\documentclass[12pt]{article}
\usepackage[left=3cm,right=3cm,top=3cm,bottom=3cm]{geometry}
\usepackage{hyperref}
\usepackage{amsmath, amsfonts, amssymb,latexsym}
\usepackage{wasysym,graphicx,stmaryrd,tikz,bclogo}
\usepackage{pgfplots}
\usepackage{multicol}
\numberwithin{equation}{section}
\makeatletter
\usetikzlibrary{arrows.meta,
                decorations.markings}
\renewenvironment{thebibliography}[1]
     {\begin{multicols}{2}[]%
      \@mkboth{\MakeUppercase\refname}{\MakeUppercase\refname}%
      \list{\@biblabel{\@arabic\c@enumiv}}%
           {\settowidth\labelwidth{\@biblabel{#1}}%
            \leftmargin\labelwidth
            \advance\leftmargin\labelsep
            \@openbib@code
            \usecounter{enumiv}%
            \let\p@enumiv\@empty
            \renewcommand\theenumiv{\@arabic\c@enumiv}}%
      \sloppy
      \clubpenalty4000
      \@clubpenalty \clubpenalty
      \widowpenalty4000%
      \sfcode`\.\@m}
     {\def\@noitemerr
       {\@latex@warning{Empty `thebibliography' environment}}%
      \endlist\end{multicols}}
\makeatother

\usepackage{amsmath, amsfonts, amssymb,latexsym,amsthm}

\input epsf.sty
\newtheorem{Theorem}{Theorem}[section]
\newtheorem{Lemma}{Lemma}[section]
\newtheorem{Proposition}[Lemma]{Proposition}

\newtheorem{Definition}[Lemma]{Definition}

\newcommand{\BEQ}{\begin{equation}}     
\newcommand{\BEA}{\begin{eqnarray}}
\newcommand{\BD}{\begin{displaymath}}
\newcommand{\EEQ}{\end{equation}}       
\newcommand{\EEA}{\end{eqnarray}}
\newcommand{\ED}{\end{displaymath}}

\newcommand{\del}{\delta}

\newcommand{\R}{\mathbb{R}}

\newcommand{\eop}{\hfill $\Box$}        
\newcommand{\Medskip}{\medskip\noindent}
\newcommand{\Bigskip}{\bigskip\noindent}

\newcommand{\half}{{1\over 2}}          

\newcommand{\Compl}{{\mathrm{Compl}}}
\newcommand{\Link}{{\mathrm{Link}}}

\begin{document}

\title
{\bf Autocatalytic cores in the diluted regime: classification and properties}
\vskip -2cm
\date\today
\vskip -2cm
\maketitle

\vskip -2cm
\begin{center}
{\bf Praneet Nandan$^a$, Philippe Nghe$^a$ and J\'er\'emie Unterberger$^b$}
\end{center}


\centerline {\small $^a$Laboratory of Biophysics and Evolution,\footnote{UMR CNRS-ESPCI 8231 Chemistry, Biology and Innovation} ESPCI Paris-PSL,} 
\centerline{10 Rue Vauquelin, 75005 Paris, France}
\centerline {\small $^b$Institut Elie Cartan,\footnote{Laboratoire 
associ\'e au CNRS UMR 7502.} Universit\'e de Lorraine,} 
\centerline{ B.P. 239, 
F -- 54506 Vand{\oe}uvre-l\`es-Nancy Cedex, France}
\centerline{jeremie.unterberger@univ-lorraine.fr}

\vspace{2mm}
\begin{quote}

\renewcommand{\baselinestretch}{1.0}
\footnotesize
{
Autocatalysis underlies the ability of chemical and biochemical systems to replicate. 
Autocatalysis was recently defined stoichiometrically for reaction networks; five types of minimal autocatalytic networks, termed autocatalytic cores were identified. 
A necessary and sufficient stoichiometric criterion was later established for dynamical autocatalysis in diluted regimes, ensuring a positive growth rate of autocatalytic species starting from infinitesimal concentrations, given that degradation rates are sufficiently low. 
Here, we show that minimal autocatalytic networks in the dynamical sense, in the diluted regime, follow the same classification as autocatalytic cores in the stoichiometric sense.
We further prove the uniqueness of the stationary regimes of autocatalytic cores, with and without degradation, for all types, except types II with three catalytic loops or more.\\
These results indicate that the stationary point is likely to be robust under perturbation at low concentrations. More complex behaviours are likely to arise by additional non-linear couplings between cores.
}
\end{quote}

\vspace{4mm}
\noindent



\newpage

\tableofcontents


\section{Introduction}


The ability of living systems to reproduce themselves is based on the chemical mechanism of autocatalysis. 
In extant biology, autocatalysis arises via DNA replication, coupled to a highly complex network of biochemical reactions that ensure metabolism, gene expression, membrane production, etc.
However, autocatalysis must have been present in the early stages of life, possibly even during so-called chemical evolution (\cite{Kauff}\cite{lan}\cite{oparin}), in a more rudimentary form.
Indeed, self-reproduction is a necessary property for Darwinian evolution, which was itself required for elaborate biochemical enzymes to appear. 
\\

Only recently, autocatalysis has been formally defined at the level of reaction networks, where multiple species altogether participate in the process of self-amplification (\cite{Andersen2021}\cite{BloLacNgh}). 
Blokhuis et al. (\cite{BloLacNgh}) defined general stoichiometric conditions for autocatalysis in reaction networks.
They deduced five classes of minimal network topologies, called autocatalytic cores, meaning that any autocatalytic network must contain a subnetwork as an autocatalytic core.
These conditions allow algorithms to systematically detect autocatalysis in large reaction networks, such as metabolisms (\cite{Arya2022}\cite{Peng2022}).
\\

Autocatalysis must also fulfill dynamical conditions, where every autocatalytic species increases in concentration over time (\cite{Kamikura2024}\cite{Lin2020}\cite{Srinivas2024}).
Indeed, reactions that degrade autocatalysts into side products may lead to the extinction of the overall network.
Blokhuis et al. further provided viability conditions, based on stochastic processes, where dynamical viability is computed on a case-by-case basis using branching processes, assuming that autocatalysts are initially absent, or present in a few copy numbers.
Reference (\cite{NgheUnt1}) proved a general stoichiometric criterion for dynamical autocatalysis in the diluted regime, given small enough degradation reactions. 
The notion of diluted regime, used in the present work and formally defined in the next section, is that of infinitesimal initial concentrations.
This regime is realized at the onset of autocatalytic amplification. Thus it characterizes the ability of autocatalysis to start spontaneously, or from a perturbation wherein chemical species are transiently injected into the reaction vessel (e.g. a meteorite landing).
In this regime, the probability of two species to meet becomes negligible compared to one-to-one and one-to-many reactions. 
The general, rate-independent criterion for dynamical autocatalysis then boils down to having a strongly connected chemical network containing at least a one-to-many reaction.
\\

However, the relationship between stoichiometry and dynamics of autocatalytic networks remains an essentially open question.
The first question is that of minimal autocatalytic cores in the dynamical sense. In the diluted regime, dynamical cores must contain autocatalytic cores. The subset of the latter that only contain one-to-one and one-to-many reactions, and which obey the criterion for diluted dynamical autocatalysis, are thus expected to be dynamical autocatalytic cores. In the present work, we enumerate minimal structures obeying diluted dynamical autocatalysis.
A second question is that of multistationarity. While chemical evolution requires multiple growth states \cite{Ameta2022,Peng2020} and autocatalytic cores are a possible candidate to trigger dynamical instabilities \cite{Vassena2024}, it remains unknown whether autocatalysis is prone to complex dynamics such as multistationarity.
Here, we show that autocatalytic cores have a single stationary state, with or without degradation reactions. This result builds on deficiency theory augmented by an analysis of the behaviour of solutions when continuously varying degradation rates. 
\\

\medskip

Here is an outline of the article. In section 2, we recall former results on stoichiometric and dynamical autocatalysis and provide useful definitions. The classification theorem for dynamical autocatalytic cores is stated in section 3. In Section 4 we state and give the proof for the uniqueness of the stationary states for these cores in the absence of degradation. Section 5 contains our main results for the stationary states of these cores with degradation, along with Theorem 5.1 which plays a vital part in the proof. We discuss our results in Section 6 and technical details are postponed to Section 7

\section{Background results and definitions}

\subsection{Autocatalytic cores}

We state definitions used for the stoichiometric classification of minimal autocatalytic networks, called cores, as introduced in reference \cite{BloLacNgh}.
A reaction network is defined by its graph $G=({\cal X},{\cal R})$, with ${\cal X}=\{x_1,\ldots,x_n\}$
the set of species, and ${\cal R}$ = set of reactions. 
All reactions are assumed to be autonomous, meaning that they have at least one reactant and one product, and non-ambiguous, meaning that the same species is either a reactant or a product, but never both (\cite{BloLacNgh}). The latter condition is not restrictive, as it only requires the addition of intermediate steps with species complexes (e.g. the catalyst-substrate complex formation step during catalysis).

\noindent Consequently, reactions are of the general form
\BEQ R \text{ (reversible) }\, : \qquad s_1 x_1+\ldots+s_k x_k \leftrightarrows s'_1 x'_1+\ldots+s'_{k'} x'_{k'} 
\label{eq:1}
\EEQ
\BEQ R_i \text{ (irreversible) }\, : \qquad s_1 x_1+\ldots+s_k x_k \to s'_1 x'_1+\ldots+s'_{k'} x'_{k'} 
\label{eq:1.1}
\EEQ
with $k,k'\ge 1$, all species $x_1,\ldots,x_k,x'_1,\ldots,x'_{k'}$
distinct,  and $s_1,\ldots,s_k; s'_1,\ldots,s'_{k'}\in \mathbb{N}_{>0}$
integer stoichiometric coefficients. We also define for irreversible reaction $R_i$ (\ref{eq:1.1})\\
\BEQ \bar{R_i} \text{ (irreversible) }\, : \qquad s'_1 x'_1+\ldots+s'_{k'} x'_{k'}\to s_1 x_1+\ldots+s_k x_k  
\label{eq:1.2}
\EEQ
where the reactants and the products exchange places. Note that for reversible reactions, $\bar{R}\simeq R$.\\
Choosing an ordering 
${\cal R}=\{R^{(1)},\ldots,R^{(m)}\}$ of ${\cal R}$, the columns
$\left(\begin{array}{c} 
-s^{(j)}_1 \\ \vdots \\ -s^{(j)}_k \\ \hline (s^{(j)})'_1 \\ \vdots \\ (s^{(j)})'_{k'}
\end{array}\right)
$  formed of stoichiometric coefficients of $R^{(1)},\ldots, R^{(j)},\ldots, R^{(m)}$ form
an $n\times m$ matrix $\mathbb S$ called {\em stoechiometric matrix}.

\Medskip {\em Non-degenerate networks.} A network is {\em non-degenerate} if and only if rk(${\mathbb S})=|{\cal X}|$.  

\Medskip \textbf{Note:} For the purpose of this paper, our non-degeneracy assumption excludes conserved quantities. But this choice is made keeping in mind that autocatalytic cores (defined later) do not have any conservation laws. 

\Medskip {\em Mass-action dynamics.} Choose direct, resp. reverse non-negative rates 
$k_+^{(j)}$, resp. $k_-^{(j)}$ (for irreversible reaction, one of the rates is 0) for the reaction 
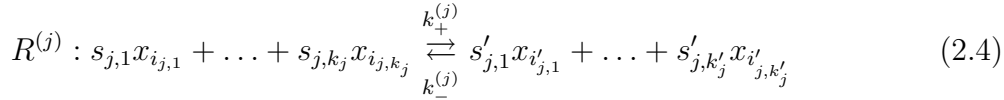
\BEQ R^{(j)}: s_{j,1} x_{i_{j,1}}+\ldots+s_{j,k_j} x_{i_{j,k_j}} \overset{k^{(j)}_+}{
\underset{k^{(j)}_-}{\rightleftarrows}} s'_{j,1} x_{i'_{j,1}}+\ldots+s'_{j,k'_j} x_{i'_{j,k'_j}} \label{eq:Rj}
\EEQ
 Let 
\BEQ C:= ([x_i])_{x_i\in {\cal X}} 
\EEQ
be the vector of species concentrations.  $C$ is an element of $\R^{{\cal X}}$, spanned by the canonical basis $(e_{\ell})$ defined by $(e_{\ell})_x=\del_{\ell,x}$.  Then the associated mass-action dynamics can be written in terms of the stoichiometric matrix as
the product (matrix $\times$ vector),
\BEQ \frac{dC}{dt} = {\mathbb S} J(C),   \label{eq:dX/dt=SJ} \EEQ
where the {\bf current} $J=J(C)$ is the vector with components
\BEQ 
J^{(j)}(C)= k^{(j)}_+ \prod_{\ell=1}^{k_j}
([x_{i_{j,\ell}}])^{s_{j,\ell}} - k^{(j)}_- \prod_{\ell'=1}^{k'_j} ([x_{i'_{j,\ell'}}])^{s'_{j,\ell'}}  \qquad  (j=1,\ldots,m).  \label{eq:current}
\EEQ 
 Alternatively, treating separately direct and reverse
reactions,  $\frac{dC}{dt}$ may be written
in terms of {\bf flows}; namely, the direct reaction $R^{(j)}$ is
characterized by the two stoichiometric vectors $(y_{R^{(j)}},y'_{R^{(j)}})$, with 
\BEQ y_{R^{(j)}}:= s_{j,1} e_{i_{j,1}} + \ldots + s_{j,k_j} e_{i_{j,k_j}},
\qquad y'_{R^{(j)}}:= s'_{j,1} e_{i'_{j,1}} + \ldots + s'_{j,k_j} e_{i'_{j,k_j}}   
\EEQ
called respectively {\bf reactant composition} and {\bf product
composition}, and by the direct flow
\BEQ \phi_{y_{R^{(j)}} \to y'_{R^{(j)}}} (C) := k^{(j)}_+ \prod_{\ell=1}^{k_j}
([x_{i_{j,\ell}}])^{s_{j,\ell}} \EEQ
whereas the associated reverse reaction is characterized by 
the permuted pair $(y'_{R^{(j)}},y_{R^{(j)}})$ and by the  reverse
flow
\BEQ \phi_{y'_{R^{(j)}} \to y_{R^{(j)}}} (C) := k^{(j)}_- \prod_{\ell'=1}^{k'_j}
([x_{i'_{j,\ell'}}])^{s'_{j,\ell}} \EEQ 
Note that   currents
\BEQ J^{(j)}(C) =  \phi_{y_{R^{(j)}} \to y'_{R^{(j)}}} (C) - 
\phi_{y'_{R^{(j)}} \to y_{R^{(j)}}} (C)   \label{eq:antisymmetry}
\EEQ
are obtained by 'antisymmetrizing' flows. Combining the above
equations, one gets the useful formula
\BEQ \frac{dC}{dt} = \sum_{y\to y'} \phi_{y\to y'}(C) (y'-y)
\label{eq:Cphiy'-y}
 \EEQ
where the sum ranges over all direct or reverse reactions $R$
characterized by their stoichiometric pair $(y,y')=(y_R,y'_R)$.

\Bigskip {\em Positive vectors.} Let $v\in\R^p$ $(p=n$ or $m)$   be a vector, then 
we say that $v$ is positive $(v>0)$ if all its components are positive,
i.e. $v_i>0$ for all $i$. 

\Bigskip We assume mass action kinetics, as except for the case of bound or precipitating species, all reactions can be broken down into elementary steps where these kinetics hold. Furthermore, as we prove our results without any assumptions on the rate constants, they are not dependent on the change in reaction rates resulting from this decomposition into elementary steps. In an evolutionary context without reactions involving ions, it is an adequate assumption for studying autocatalytic cycles.

\begin{Definition}[stoichiometrically autocatalytic network] (see Blokhuis et al. \cite{BloLacNgh}) A reaction network
$G=({\cal X},{\cal R})$  is stoichiometrically autocatalytic if and only if there exists a positive column
vector $c=\left(\begin{array}{c} c^{(1)} \\ \vdots \\ c^{(m)} 
\end{array}\right)$ such that ${\mathbb S}c>0$.
\end{Definition}

In reference \cite{NgheUnt1}, the distinction is made between {\em stoichiometrical autocatalysis} and {\em dynamical autocatalysis}. While the former is a condition where the stoichiometrical balance for all autocatalytic species is positive, Dynamical autocatalysis is the property for the system to grow exponentially from very low concentration (with a strictly positive Lyapunov exponent).
\begin{Definition}[dynamically autocatalytic network] (see \cite{NgheUnt1}) A reaction network
$G=({\cal X},{\cal R})$  is dynamically autocatalytic if and only if the underlying (mass action) dynamical system has a strictly positive largest Lyapunov exponent with a corresponding non-negative Lyapunov eigenvector.
\end{Definition}

\begin{Definition}[restricted network]  \label{def:restricted}
Let $G=({\cal X},{\cal R})$ be a network, and 
${\cal X}'\subset {\cal X}$. Then 
\BEQ G\Big|_{{\cal X}'}:=({\cal X}',{\cal R}\Big|_{{\cal X}'}) \EEQ
called {\em restriction of $G$ to ${\cal X}'$}, is the subnetwork of $G$
obtained by restricting the set of species to ${\cal X}'$ and
{\bf chemostatting} species not in ${\cal X}'$, i.e. by

\begin{enumerate}
\item removing reactions having either all reactants or all products
(or both) in ${\cal X}\setminus {\cal X}'$  (as non-autonomous after
chemostatting);
\item in all other reactions, replacing all stoichiometric coefficients
of species in  ${\cal X}\setminus {\cal X}'$ by $0$.
\end{enumerate}

\end{Definition}

Alternatively, one may also restrict the reaction set ${\cal R}$ to ${\cal R}'\subset
{\cal R}$; the result is simply denoted $({\cal X}, {\cal R}')$. 
The two restriction operations can be composed in either direction.

\begin{Definition}[minimality, autocatalytic cores]
An autocatalytic network (stoichiometric or dynamic) $G=({\cal X},{\cal R})$ is minimal if
 \BEQ \Big( {\cal R}'\subseteq {\cal R}, {\cal X}'\subseteq {\cal X},
 ({\cal X},{\cal R}')\Big|_{{\cal X}'}\ {\mathrm{autocatalytic}}
 \Big) \Longrightarrow \Big( {\cal R}'= {\cal R}, {\cal X}'={\cal X}
 \Big).
 \EEQ
 
A minimal autocatalytic network is called an stoichiometric/dynamic (resp.) {\em autocatalytic
core}.  
\end{Definition}

\begin{Definition}[Reversible extension]
     The {\em reversible extension} of a network $G=({\cal X},
{\cal R})$ 
is the network $\overleftrightarrow{G}=({\cal X},\overleftrightarrow{\cal R})$ obtained from $G$
by making reversible {\em all} reactions in ${\cal R}$, i.e.
\BEQ (R \text{ (reversible) } \in {\cal{R}} ) \Rightarrow (R \in  \overleftrightarrow{\cal{R}}) 
\EEQ
\BEQ (R_i \text{ (irreversible) } \in {\cal{R}}) \Rightarrow (R \text{ (reversible) }\in  \overleftrightarrow{\cal{R}}) .
\EEQ
(See reactions ((\ref{eq:1})) and ((\ref{eq:1.1}))
\end{Definition}

\subsection{Diluted regime}

Dynamical autocatalysis --- the ability of autocatalytic reaction networks to kinetically (and exponentially) amplify their species --- has been studied in the context of the \textit{diluted regime} \cite{NgheUnt1}, i.e. infinitesimally low concentrations. 
It is possible to translate the notion of dilution in terms of the topology of $G$ because when species concentrations are small enough, reactions involving multiple reactants become negligible as compared to those comprising a single reactant. 
In other terms, many-to-one or many-to-many reactions are negligible as compared to one-to-one or one-to-many reactions, following the definitions:

\Medskip
-- one-to-one (1--1) reactions have stoichiometry $(-1;1)$, i.e. 
they are of the form $x\to x'$ or $x\leftrightarrows x'$, with $x\not=x'\in{\cal X}$; They can be reversible or irreversible.

\noindent -- one-to-many reactions are irreversible and have stoichiometry $(-1;s'_1,\ldots,s'_{m'})$
with $s'_1,\ldots,s'_{m'}=1,2,\ldots$, $\sum_i s'_i\ge 2$, i.e. they are
of the form $x\to s'_1 x'_1 +\ldots+ s'_{m'} x_{m'}$.

\noindent-- many-to-one reactions are also irreversible and are those obtained by reversing 
one-to-many reactions, i.e. $s'_1 x'_1 +\ldots+ s'_{m'} x_{m'} \to x$.

\noindent-- many-to-many reactions can be both reversible or irreversible and have stoichiometry $(-s'_1,\ldots,-s'_{m'}; s_1,\ldots,s_m)$ with $\sum_i s'_i\ge 2$ and $\sum_i s_i\ge 2$.

\Medskip 
Stoichiometric coefficients $s'_i,s_i$ play  little role in the
discussion, therefore we more simply refer to $m,m'$ alone: 

\noindent-- a {\em simple} reaction has $m'=m=1$, i.e. it is of the form
$s_1 x_1\to or \leftrightarrows s'_1 x'_1$, with $s_1,s'_1=1,2,\ldots$ arbitrary;

\noindent-- a {\em multiple product} reaction has $m'\ge 2$, i.e. it is of 
the form $s_1 x_1+\ldots+s_m x_m \to s'_1 x'_1+\ldots+s'_{m'} x'_{m'}$
with $m'\ge 2$.

\begin{Definition}[diluted reaction networks]  \label{def:reversible}
A reaction network is {\em diluted} if it contains only 
simple reactions, and possibly some irreversible one-to-many
reactions (i.e. multiple product reaction where m=1, single reactant).
\end{Definition}

Reversible extensions of diluted autocatalytic cores will be used in section 4, where we show that they possess a single unique stationary state.

\subsection{(Top) property}

\Medskip {\em Classes.} Consider the stoechiometric matrix $\mathbb S$.
Now consider the directed graph $G({\mathbb S})$ obtained by connecting reactants to products participating in the same (reversible or irreversible) reaction. 
$G$ is in fact the graph of the \textit{split reactions}:

\Medskip {\em Split reactions.}  The arrow $x \to x'$ is a split reaction iff ${\cal R}$ contains a reaction $x\to s'x'+s'_2 s'_2+\ldots+
s'_{k'}x'_{k'}$ or $x\leftrightarrows s'x'+s'_2 s'_2+\ldots+
s'_{k'}x'_{k'}$, i.e. if it is an edge of $G({\mathbb S})$.\\

We colloquially refer to $G$ as the \textit{split graph}, as it consists in splitting one-to-many reactions into as many edges as products.
The strongly connected components of $G({\mathbb S})$ are called {\em classes}.

\Medskip {\em Class graph.} Contracting all species within a given class to a point
results in a directed acyclic graph (called: {\em class graph}).
This graph defines a partial order wherein 
${\cal C}\prec {\cal C}'$ if ${\cal C}$ is 'upstream' from ${\cal C}'$, i.e. if there exists a split reaction $x\to x'$
with $x\in {\cal C},x'\in {\cal C}'$. 

\Medskip {\em Minimal classes.} ${\cal C}$
have the property that no  class is located upstream from them, i.e.
there is no split reaction $x'\to x$ with
$x'\not\in {\cal C},x\in {\cal C}$.
 
\Medskip \textit{An example.} Take the reaction network with the reactions
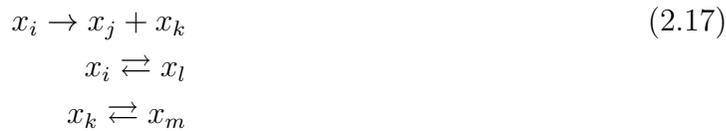
\BEA x_i \to x_j + x_k \\\nonumber
x_i \rightleftarrows x_l \\ \nonumber
x_k\rightleftarrows x_m
\EEA
The split graph is\\
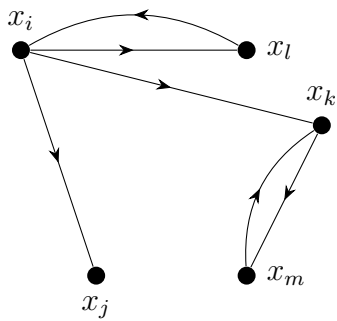
\begin{center}
    \begin{tikzpicture}[
       decoration = {markings,
                     mark=at position .5 with {\arrow{Stealth[length=2mm]}}},
       dot/.style = {circle, fill, inner sep=2.4pt, node contents={},
                     label=#1},
every edge/.style = {draw, postaction=decorate}
                        ]
\node (i) at (0,5) [dot=$x_i$];
\node (j) at (1,2) [dot=below:$x_j$];
\node (k) at (4,4) [dot=$x_k$];
\node (l) at (3,5) [dot=right:$x_l$];
\node (m) at (3,2) [dot=right:$x_m$];
\path   (i) edge (j)    (k) edge (m)    
        (i) edge (k)    (i) edge (l)
        (m) edge[bend left] (k)    (l) edge[bend right] (i);
    \end{tikzpicture}
\end{center}
The corresponding classes are ${\cal C}_i=\{x_i,x_l\},{\cal C}_j=\{x_j\},{\cal C}_k=\{x_k,x_m\}$, and the Class graph is shown below. Note that ${\cal C}_i$ is the only minimal class, and is "upstream" of ${\cal C}_j,{\cal C}_k$\\
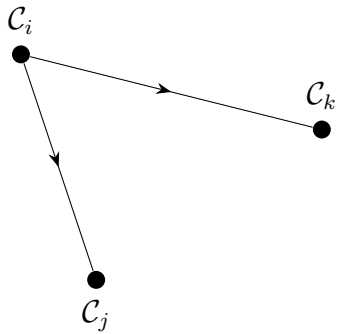
\begin{center}
    \begin{tikzpicture}[
       decoration = {markings,
                     mark=at position .5 with {\arrow{Stealth[length=2mm]}}},
       dot/.style = {circle, fill, inner sep=2.4pt, node contents={},
                     label=#1},
every edge/.style = {draw, postaction=decorate}
                        ]
\node (i) at (0,5) [dot=${\cal C}_i$];
\node (j) at (1,2) [dot=below:${\cal C}_j$];
\node (k) at (4,4) [dot=${\cal C}_k$];
\path   (i) edge (j)    (i) edge (k);
    \end{tikzpicture}
\end{center}
\Medskip {\em Restricted reactions.}  Let ${\cal C}$ be a class. If $R \, :\, x\to s'_1 x'_1+\cdots + s'_{k'} x'_{k'}$ is a reaction such that $x\in{\cal C}$, and 
$\{i=1,\ldots,k' \ |\ x'_i\in {\cal C}\}=\{1,\ldots,j'\}$ for
some $j'\ge 1$, then  the {\bf ${\cal C}$-truncated reaction} is
 \BEQ R\Big|_{{\cal C}}\, :\, x\to s'_1 x'_1+\cdots + s'_{j'} x'_{j'}.
 \EEQ
Formally, it is obtained by chemostatting (keeping constant the concentration of) species which
do not belong to $\cal C$.
\begin{Definition}[Irreducible Network]
     A reaction network is irreducible iff the Stoichiometric Matrix ${\mathbb S}$ is not similar via a permutation to a block upper triangular matrix. It is equivalent to saying that the split graph G is strongly connected.
\end{Definition}

We now recall from (Nghe-Unterberger \cite{NgheUnt1}) the (Top) property (signifying the topological nature of the property) and result, which stoichiometrically characterize dynamical autocatalytic networks in the diluted regime:

\begin{Definition}[(Top) property for autocatalysis]
A diluted reaction network $G$ satisfies (Top) if all minimal classes
of the adjacency graph $G({\mathbb S})$ contain at least one internal
one-to-many reaction. 
\end{Definition}

\begin{Proposition}   \label{prop:Top}
A diluted network 
 $G=({\cal X},{\cal R})$ is (stoichiometrically) autocatalytic  if and only if it 
satisfies (Top). Also if it satisfies (Top), it is dynamically autocatalytic (Theorem 3.1 in \cite{NgheUnt1})
\end{Proposition}

\Medskip {\bf Remark.} An irreducible, diluted network $G=({\cal X},{\cal R})$ has 
$|{\cal R}|\ge |{\cal X}|$ (for each $x\in {\cal X}$, there must
exist at least one reaction $x\to \cdots$). Therefore, such a network
is {\em non-degenerate} (i.e. rk$({\mathbb S})\ge |{\cal X}|$) if and
only if ${\mathbb S}$ is {\em full-rank} in which case rk$({\mathbb S})= |{\cal X}|$.

It was shown in \cite{NgheUnt1} that the stoichiometric property (Top) implies dynamical autocatalysis. In the appendix, we give a graph-theoretic way to obtain the same classification of autocatalytic cores into the five types for diluted systems using the \textbf{Top} property for dynamical autocatalysis as described in \cite{NgheUnt1}.\\

\section{Classification of dynamical autocatalytic cores}
\label{section:main}


Stoichiometric autocatalytic cores were classified in (Blokhuis et al.
\cite{BloLacNgh}). Using the diluted regime assumption (i.e.
dismissing many-to-one or many-to-many reactions) and our Proposition \ref{prop:Top}, it can again be proved graph-theoretically that we get only five types of cores if we consider dilute dynamical autocatalytic cores, which is included in the Appendix \ref{section:prrofcl}. We state the result, and then discuss
how it relates to the one obtained in \cite{BloLacNgh}:

\begin{Definition}[diluted autocatalytic core]
An open reaction network $G=({\cal X},{\cal R})$ is a {\em diluted
autocatalytic core} if and only if it is an autocatalytic core (stoichiometric or dynamic) composed uniquely of one-to-one or
one-to-many reactions (i.e. diluted).
\end{Definition}
\begin{Theorem} (see \cite{BloLacNgh})  \label{th:main}
Let  $G=({\cal X},{\cal R})$ be a diluted dynamicac autocatalytic core. Then $G$ is of one of the following types 
(generally speaking, possible products of  one-to-many reactions are indicated by
a bold line, see details below):

\begin{itemize}
\item[(i)] Type I: all species lie along a cycle ${\cal C}=\{x_1\to x_2
\to \ldots\to x_n\to x_{n+1}=x_1\}$, all reactions are simple
reactions along $\cal C$ in the same orientation, at least one of them is one-to-many,

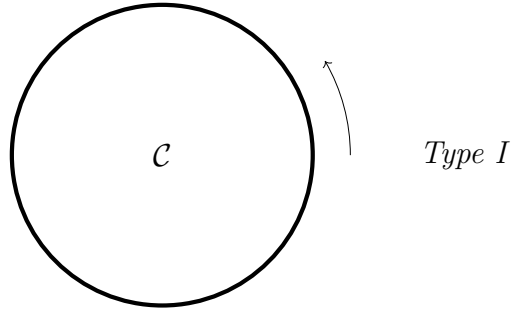
\begin{center}
\begin{tikzpicture}
\draw[ultra thick](0,0) circle(2);  \draw[->,ultra thin](2.5,0) arc(0:30:2.5 and 2.5);
\draw(0,0) node {${\cal C}$};
\draw(4,0) node {Type I};
\end{tikzpicture}
\end{center}


\Bigskip


\item[ii)] Type $II_{\ell}$ ($\ell\ge 1$): all species lie along a cycle ${\cal C}=\{x_1\to x_2\to \cdots\to x_n\to x_1\}$, back-branches $i_j\to \sigma_j$, $i=1,\ldots,\ell$ 
$(1=i_1\le\sigma_2<i_2\le \sigma_3<i_3\le \ldots\le \sigma_1<n+1)$, encoding split
reactions originated from the multiple reactions $x_{i_j}\to s_{i_j}  x_{i_j+1} + 
x_{\sigma_j}$,  span
disjoint arks along $\cal C$,

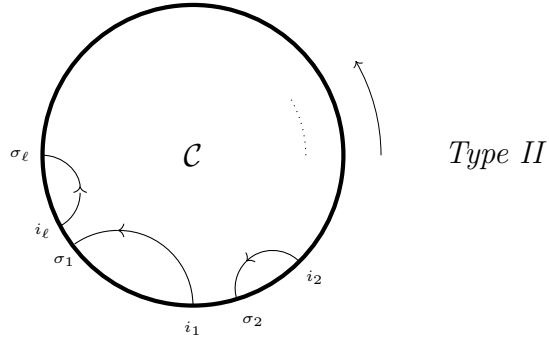
\begin{center}
\begin{tikzpicture}
\draw(0,0) node {${\cal C}$};
\draw[ultra thick](0,0) circle(2);  \draw[->,ultra thin](2.5,0) arc(0:30:2.5 and 2.5);
\draw[->](0,-2) arc(0:90:1 and 1);  \draw(0,-2.3) node {\tiny $i_1$};
\draw(-1,-1) arc(90:125:1 and 1);
\draw(-1.71,-1.41) node {\tiny $\sigma_1$};

\draw[->](1.414,-1.414) arc(45:135:0.5 and 0.5);
\draw(1.414-0.707,-1.414) arc(135:200:0.5 and 0.5);
\draw(1.62,-1.62) node {\tiny $i_2$};
\draw(0.8,-2.2) node {\tiny $\sigma_2$};

\draw[dotted] (1.5,0) arc(0:30:1.5 and 1.5);

\draw[-<](-2,0) arc(90:0:0.5 and 0.5);
\draw(-1.5,-0.5) arc(0:-60:0.5 and 0.5);
\draw(-2.3,0) node {\tiny $\sigma_{\ell}$};
\draw(-2,-1) node {\tiny $i_{\ell}$};

\draw(4,0) node {Type II};
\end{tikzpicture}
\end{center}


\Bigskip Note that split reactions along ${\cal C}$ have 
arbitrary stoichiometry $x_i\to s_i x_{i+1}$ as in Type I,
whereas  back-branches
have stoechiometric coefficient 1. 


\item[(iii)] Type III, all species lie either on first cycle $\cal C$ or
on an  ear ${\cal O}= {\cal C}'\setminus ({\cal C}\cap {\cal C}')$
added to ${\cal C}$;  ${\cal C}'\cap{\cal C}= \overrightarrow{(uv)}$
(possibly, $u=v$); reactions are simple reactions along cycles, plus a
multiple reaction $v\to x+x'$ with reactant $v$ and products $x\in {\cal C}\setminus  ({\cal C}\cap {\cal C}')$, $x'\in {\cal C}' \setminus({\cal C}\cap {\cal C}')$,

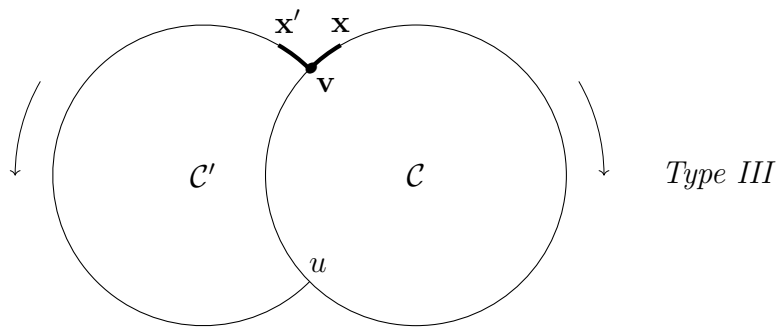
\begin{center}
\begin{tikzpicture}
\draw(0,0) circle(2);  \draw(0,0) node {${\cal C}$};
\draw(-4*0.707,0) node {${\cal C}'$};
  \draw[<-,ultra thin](2.5,0) arc(0:30:2.5 and 2.5);
\draw(-4*0.707+1.414 ,1.414) arc(45:315:2 and 2);
\draw[<-,ultra thin] (-4*0.707-2.5,0) arc(180:150: 2.5 and 2.5);

\draw[ultra thick](-2*0.707,2*0.707) arc(135:120:2 and 2);
\draw[ultra thick](-2*0.707,2*0.707) arc(45:60:2 and 2);
\draw[ultra thick](-1.2,1.2) node {$\mathbf v$};
\draw(-2*0.707,2*0.707) node {\textbullet};
\draw(-1.3,-1.2) node {$u$};
\draw[ultra thick](-1,2) node {$\mathbf x$};
\draw[ultra thick](-1.7,2.05) node {$\mathbf x'$};

\draw(4,0) node {Type III};
\end{tikzpicture}
\end{center}

\Bigskip {\em Remark.}  By hypothesis, $x,x'\not=u$. However, if $x=u$ or $x'=u$, then Type III degenerates to Type II.
If both $x=u$ and $x'=u$, then it further degenerates to Type I.


\Bigskip


\item[(iv)] Type IV, same as Type III, but with an extra multiple
reaction $w\to u+x'$ with reactant $w\in {\cal C}\setminus({\cal C}\cap {\cal C}')$
(the case $x=w$ is not excluded) and products $u$ and $x'$,

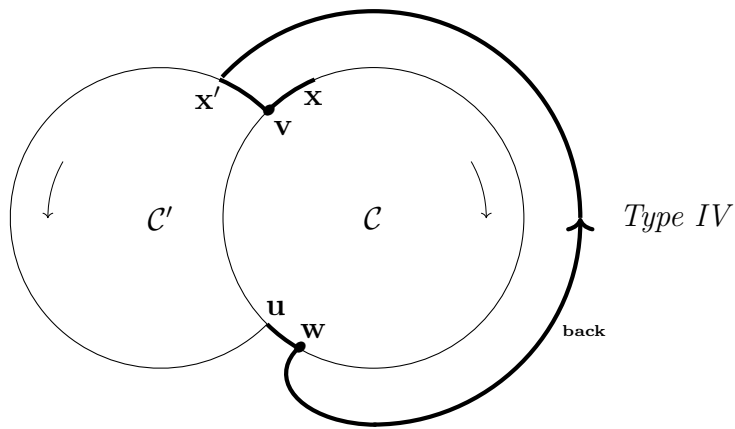
\begin{center}
\begin{tikzpicture}
\draw(0,0) circle(2);  \draw(0,0) node {${\cal C}$};
\draw(-4*0.707,0) node {${\cal C}'$};
  \draw[<-,ultra thin](1.5,0) arc(0:30:1.5 and 1.5);
\draw(-4*0.707+1.414 ,1.414) arc(45:315:2 and 2);
\draw[<-,ultra thin] (-4*0.707-1.5,0) arc(180:150: 1.5 and 1.5);

\draw[ultra thick](-2*0.707,2*0.707) arc(135:113:2 and 2);
\draw[ultra thick](-2*0.707,2*0.707) arc(45:67:2 and 2);
\draw[ultra thick](-1.2,1.2) node {$\mathbf v$};
\draw(-2*0.707,2*0.707) node {\textbullet};
\draw(-1.3,-1.2) node {$\mathbf u$};
\draw[ultra thick](-0.8,1.6) node {$\mathbf x$};
\draw[ultra thick](-2.2,1.6) node {$\mathbf x'$};

\draw[ultra thick](-2*0.5,-2*0.866) arc(-120:-135:2 and 2);

\draw(-0.8,-1.5) node {$\mathbf w$};
\draw(-2*0.5,-2*0.866) node {\textbullet};
\draw[ultra thick](-2*0.5,-2*0.866) arc(150:270:1.2 and 0.68);
\draw[->,ultra thick](0,-2.75) arc(270:360:2.75 and 2.75);
\draw[ultra thick](2.75,0) arc(360:497:2.75 and 2.75);

\draw(2.8,-1.5) node {\bf \tiny  back};

\draw(4,0) node {Type IV};
\end{tikzpicture}
\end{center}


\Bigskip The mention 'back' on the edge $w\to x'$ refers to the {\em supplementary back-branch} $w\to x'$.  


\Bigskip

\item[(v)] Type V, same as Type IV with an extra multiple
reaction $w'\to u+x$ with reactant $w'\in {\cal C}'\setminus({\cal C}\cap {\cal C}')$ (the case $w'=x'$ is not excluded)
and products $u$ and $x$,

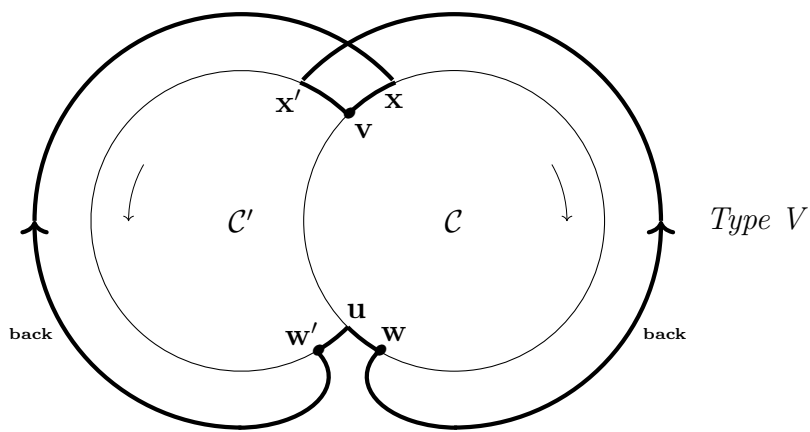
\begin{center}
\begin{tikzpicture}
\draw(0,0) circle(2);  \draw(0,0) node {${\cal C}$};
\draw(-4*0.707,0) node {${\cal C}'$};
  \draw[<-,ultra thin](1.5,0) arc(0:30:1.5 and 1.5);
\draw(-4*0.707+1.414 ,1.414) arc(45:315:2 and 2);
\draw[<-,ultra thin] (-4*0.707-1.5,0) arc(180:150: 1.5 and 1.5);

\draw[ultra thick](-2*0.707,2*0.707) arc(135:113:2 and 2);
\draw[ultra thick](-2*0.707,2*0.707) arc(45:67:2 and 2);
\draw[ultra thick](-1.2,1.2) node {$\mathbf v$};
\draw(-2*0.707,2*0.707) node {\textbullet};
\draw(-1.3,-1.2) node {$\mathbf u$};
\draw[ultra thick](-0.8,1.6) node {$\mathbf x$};
\draw[ultra thick](-2.2,1.6) node {$\mathbf x'$};

\draw[ultra thick](-2*0.5,-2*0.866) arc(-120:-135:2 and 2);
\draw[ultra thick](-4*0.707 + 2*0.5,-2*0.866) arc(300:315:2 and 2);

\draw(-0.8,-1.5) node {$\mathbf w$};
\draw(-2*0.5,-2*0.866) node {\textbullet};
\draw[ultra thick](-2*0.5,-2*0.866) arc(150:270:1.2 and 0.68);
\draw[->,ultra thick](0,-2.75) arc(270:360:2.75 and 2.75);
\draw[ultra thick](2.75,0) arc(360:497:2.75 and 2.75);

\draw(-4*0.707 + 0.8,-1.5) node {$\mathbf w'$};
\draw(-4*0.707 + 2*0.5,-2*0.866) node {\textbullet};
\draw[ultra thick](-4*0.707 + 2*0.5,-2*0.866) arc(30:-90:1.2 and 0.68);
\draw[->,ultra thick](-4*0.707 + 0,-2.75) arc(-90:-180:2.75 and 2.75);
\draw[ultra thick](-4*0.707 - 2.75,0) arc(-180:180-497:2.75 and 2.75);

\draw(2.8,-1.5) node {\bf \tiny  back};
\draw(-4*0.707-2.8,-1.5) node {\bf \tiny  back};

\draw(4,0) node {Type V};
\end{tikzpicture}
\end{center}

\Medskip The two mentions 'back' refer to the supplementary back-branches $w\to x'$ and $w'\to x$.

\end{itemize}

\noindent Furthermore, all simple reactions in Types III,IV,V have
stoichiometric coefficients 1.

\Medskip  In particular, $G$ is always irreducible.
\end{Theorem}

\Bigskip  The more complete classification obtained in 
\cite{BloLacNgh}  without the dilute regime assumption is 
almost exactly the same (definition of Types I-V is copied
from \cite{BloLacNgh}, see Suppl. Information, Fig. S1, with $u,v$, 
see Types III-IV-V,
playing the same role to help identification). The difference comes from the fact that
reactions of the type $sx\to s'_1 x'_1+\ldots+s'_{k'} x'_{k'}$ with
$s\ge 2$ arise in general. Graphs are topologically the same,
but minimality is reflected in supplementary 
vanishing, resp. non-vanishing conditions on the determinant of the
stoichiometric matrix restricted to 
sub-cycles, the so-called weight-symmetric, resp. weight-asymmetric
condition in Theorem 1 of \cite{BloLacNgh}. It is proved early on
in that paper (Proposition 2) that reactions with several different
reactants are not found in autocatalytic cores. \\

\medskip


\section{Uniqueness of stationary states (without degradation)}\label{nd}

We prove in this, and the next section, the uniqueness of stationary states for the reversible extensions of the dynamical autocatalytic cores. This section covers the proof in the absence of degradation rates. We give an argument based on deficiency theory (\cite{Fei87}, \cite{Fei95}, \cite{Fei19}). Further in Appendix \ref{sec:exp}, we give the explicit calculation of the Lyapunov function for the asymptotic stability, and the explicit form of the stationary state for each type of autocatalytic core is computed in terms of the rate constants. We also note how our work connects to the {\em Global Attractor Conjecture.}\\ 
The extension to cores with degradation, on the other hand, relies on tracking the solutions with changing degradation rates and inspecting degenerate stationary states, which is covered in the next section. The end of this section connects the two and the existing literature on inheritance. 

\noindent Recall from (\ref{eq:dX/dt=SJ}) that the dynamics is expressed
by the equation $\frac{dC}{dt}= {\mathbb S}J(C)$. 

\Medskip {\em Stationary state.} A concentration vector $C$ is a stationary
state if and only if ${\mathbb S} J(C)=0$.

\Medskip {\em Equilibrium states.} We say that a concentration vector $C$ is an {\bf equilibrium state}
if all currents $J^{(j)}(C)$, $j=1,\ldots,m$ vanish. It follows immediately from
(\ref{eq:dX/dt=SJ}) that equilibrium states are stationary states.
The reverse is not true in general.

\begin{Proposition}   \label{prop:eqstqte}
For diluted autocatalytic cores, stationary states are equilibrium states.
\end{Proposition}
\noindent \textbf{Proof.} It was proved in \cite{BloLacNgh} that all autocatalytic cores have an invertible
stoichiometric matrix $\mathbb S$. Thus, specifically for autocatalytic cores, the stationary condition ${\mathbb S}J(C)=0$ implies that $J(C)=0$; in other words, stationary states are equilibrium states.

\Bigskip  Open reaction networks equipped with mass-action rates do not
always have a unique stationary state; they may have none or several, 
or also admit periodic orbits. Diluted autocatalytic cores are specific in this respect. As the concentrations of the species which are part of the core increase, the reverse reactions of the irreversible reactions in the core start to affect the dynamics of the system.
Thus to be precise, we
consider the reversible extension of a diluted autocatalytic core $\overleftrightarrow{G}=({\cal X},\overleftrightarrow{\cal R})$, and choose direct, resp.
reverse rates $k_+^{R}$, resp. $k_-^{R}$ for all (direct)
reactions $R\in {\cal R}$. Then the following result holds:

\begin{Theorem}[uniqueness of stationary states]  \label{th:stat}
Let $G=({\cal X},{\cal R})$ be a diluted autocatalytic core, 
$\overleftrightarrow{G} = ({\cal X},\overleftrightarrow{\cal R})$ its reversible extension (see Definition \ref{def:reversible}), and 
$(k_+^R)_{R\in {\cal R}}$, resp. $(k_-^R)_{R\in {\cal R}}$ be arbitrary, positive direct, resp. reverse reaction rates. Then the corresponding
reaction network admits a unique positive stationary state $C_{stat}$, which is an equilibrium state. Furthermore, 
for any initial concentration vector $C_0>0$,  the solution of the
dynamical system $\frac{dC}{dt}={\mathbb S}J(C)$, giving the
time evolution of the concentration vector, converges to $C_{stat}$
as $t\to\infty$. 
\end{Theorem} 
The second part of the theorem concerning {\em Persistence} is addressed in Subsection \ref{subs:per}.\\
\Medskip  Let us first review some definitions. Recall that
$(e_x)_{x\in {\cal X}}$ is the standard basis of $\R^{\cal X}$. {\em Complexes}
are the elements in the vector space $\R^{\cal X}$ of the form
$y_R = s_1 e_{x_1} + \ldots + s_k e_{x_k}$ or $y'_R = s'_1 e_{x'_1} +\ldots +
s'_{k'} e_{x'_{k'}}$,  where the reaction 
\BEQ R:\qquad s_1 x_1 + \ldots + s_k x_k \to s'_1 x'_1 + \ldots
+s'_{k'} x'_{k'} \label{eq:general-R}
\EEQ
 ranges
over the reaction set $\cal R$. A reaction $R$ as above defines an
equivalence $y_R\sim y'_R$ in the set of complexes. Completing the
set of equivalences  $(y_R\sim y'_R,\ R\in {\cal R})$ yields a set
of equivalence classes called {\em linkage classes}. Then
 the {\em deficiency index} of a network $G$
is by definition
\BEQ \del = |\Compl|-|\Link|-s,
\EEQ
where: $\Compl$ is the set of complexes; $\Link$ is the set of 
linkage classes; and $s$ is the rank of the stoichiometric matrix.

\begin{Definition}[Weakly reversible]
    If every pair of complexes in a linkage class can be strongly connected (in both directions) through some sequence of reactions $\in\mathcal{R}$, the network is called weakly reversible.
\end{Definition}
Note that all five types of diluted autocatalytic cores are both, weakly reversible and irreducible.

\Medskip  The Deficiency Zero Theorem  Theorem proved by M. Feinberg (see
Feinberg \cite{Fei87}, or \cite{Fei95}, Theorem C.1 (ii)) states in particular that {\em non-degenerate weakly reversible
 networks with deficiency index $\del=0$ admit
a unique positive stationary state $C_{stat}$}, and that  this state is  {\em asymptotically stable}. (For degenerate networks, one typically
finds that the set of stationary solutions is a union of 
hyperplanes of positive dimension, whose points are indexed by
stoichiometric compatibility classes, or elements in 
(rk$( {\mathbb S}))^{\perp}$, see \cite{Fei95}, Definition 3.3).  The asymptotic stability is based on the existence of an explicit Lyapunov function for the dynamical system,
\BEQ h(C) := \sum_{x\in {\cal X}} C_{stat,x} \Big(  \frac{C_x}{C_{stat,x}}
(\ln(\frac{C_x}{C_{stat,x}})-1) + 1 \Big), \label{eq:h}
\EEQ
 namely, $h$ is strictly decreasing on integral curves. If all reactions are 1--1, then 
$\sum C_x(t)$ is a constant by probability conservation (which may
be chosen equal to 1),  $h(C)$ is the relative entropy of the 
probability measure $C$ w.r. to the stationary probability measure
$C_{stat}$. (This is a well-known standard result in Markov theory,
see e.g. (Schnakenberg \cite{Sch}, section V) in the context of
master equation systems.)

\subsection{Deficiency zero argument}

\Bigskip Consider the reversible extension of a diluted autocatalytic core. It can be checked
that, for all types, $s=|{\cal X}|$ (the stoichiometric matrix has full rank). We prove the following:

\begin{Lemma}\label{lemma:4.1}
The deficiency index of all diluted autocatalytic cores is zero. 
\end{Lemma} 

Using Feinberg's results, this Lemma implies Theorem
\ref{th:stat}.  

\Medskip {\bf Proof.}
 The previous
classification allows an easy computation of the
deficiency index.

\medskip
Type I: in case of a unique replication reaction $x_n\to s_n x_1$, one
gets:  $\Compl=\{e_{x_1},\ldots,e_{x_n},s_n e_{x_1}\}$,  $|\Compl|=n+1$,
$s=n$,  $|\Link|=1$ (all complexes are in the same linkage class), so that $\del=0$.    If  $x_k\to s_k x_{k+1}$ $(1\le k\le n-1)$ with $s_k\not=1$ is also a replication reaction, then the two linkage classes are

  
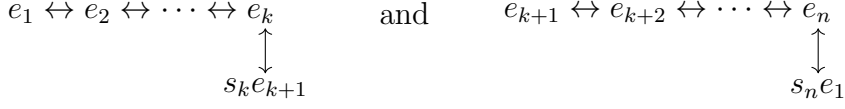
\begin{tikzpicture} 
\draw(0,0) node {$e_1\leftrightarrow e_2\leftrightarrow \cdots
\leftrightarrow e_k$}; \draw[<->](1.65,-0.2)--(1.65,-0.8); \draw(1.65,-1)
node {$s_k e_{k+1}$}; \draw(3.5,0) node {and};
\begin{scope}[shift={(7,0)}]
\draw(0,0) node {$e_{k+1}\leftrightarrow e_{k+2}\leftrightarrow \cdots
\leftrightarrow e_n$}; \draw[<->](2,-0.2)--(2,-0.8); \draw(2,-1)
node {$s_n e_1$}; 
\end{scope}
 \end{tikzpicture}

In general, any extra replication reaction increments the cardinal of
Compl and the cardinal of Link, so this has no effect on the value of  $\del$.

\medskip
Type II:  We may assume that all stoichiometric coefficients on
$\cal C$ are equal to 1 (as in the case of Type I, having replication
reactions along $\cal C$ has no effect on $\del$). Assume first that $\ell=1$, then $\Compl= \{e_2,e_3,\ldots,e_n,e_1,
e_2+ e_{\sigma_1}\}$ and $|\Link|=1$,  so that $\del=0$.
If $\ell=2$, then $\Compl$ also includes $e_{i_2+1}+e_{\sigma_2}$, and the two linkage classes are

 
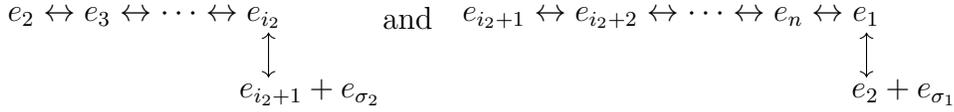
\begin{tikzpicture} 
\draw(0,0) node {$e_2\leftrightarrow e_3\leftrightarrow \cdots
\leftrightarrow e_{i_2}$}; \draw[<->](1.65,-0.2)--(1.65,-0.8); \draw(2.2,-1)
node {$e_{i_2+1}+e_{\sigma_2}$}; \draw(3.5,0) node {and};
\begin{scope}[shift={(7,0)}]
\draw(0,0) node {$e_{i_2+1}\leftrightarrow e_{i_2+2}\leftrightarrow \cdots
\leftrightarrow e_n\leftrightarrow e_1$}; \draw[<->](2.6,-0.2)--(2.6,-0.8); \draw(3.1,-1)
node {$e_2+e_{\sigma_1}$}; 
\end{scope}
 \end{tikzpicture}

In general, any back branch increments the cardinal of
Compl and the cardinal of Link, so this does not affect the value of  $\del$. 

\medskip
Type III: following the two cycles, one gets $\Compl = \{ e_u,\ldots,e_v\}\cup\{e_x,\ldots,e_u\}\cup\{e_{x'},\ldots, e_u\} \cup \{e_x + e_{x'}\}$, so 
that $|\Compl|=|{\cal X}|+1$. Since there is only one linkage class,
$\del=0$.

\medskip 
Type IV:   $|\Compl|=|{\cal X}|+2$ (due to the two 
one-to-many reactions), and correspondingly, two linkage classes

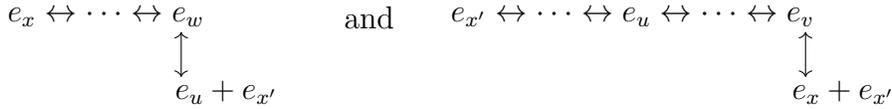
\begin{tikzpicture} 
\draw(0,0) node {$e_x\leftrightarrow \cdots
\leftrightarrow e_{w}$}; \draw[<->](1,-0.2)--(1,-0.8); \draw(1.6,-1)
node {$e_{u}+e_{x'}$}; \draw(3.5,0) node {and};
\begin{scope}[shift={(7,0)}]
\draw(0,0) node {$e_{x'}\leftrightarrow  \cdots
\leftrightarrow e_u  \leftrightarrow  \cdots
\leftrightarrow  e_v$}; \draw[<->](2.3,-0.2)--(2.3,-0.8); \draw(2.8,-1)
node {$e_x+e_{x'}$}; 
\end{scope}
 \end{tikzpicture}

\medskip Type V:  $|\Compl|=|{\cal X}|+3$ (due to the three
one-to-many reactions), and correspondingly, three linkage classes

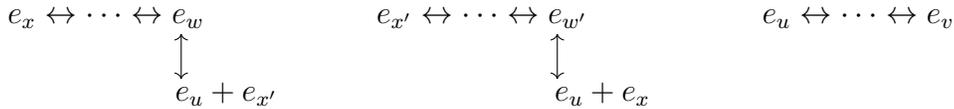
\begin{tikzpicture} 
\draw(0,0) node {$e_x\leftrightarrow \cdots
\leftrightarrow e_{w}$}; \draw[<->](1,-0.2)--(1,-0.8); \draw(1.6,-1)
node {$e_{u}+e_{x'}$}; 
\begin{scope}[shift={(5,0)}]
 \draw(0,0) node {$e_{x'}\leftrightarrow \cdots
\leftrightarrow e_{w'}$}; \draw[<->](1,-0.2)--(1,-0.8); \draw(1.6,-1)
node {$e_{u}+e_{x}$};
\end{scope}
\begin{scope}[shift={(10,0)}]
\draw(0,0) node {$e_u\leftrightarrow \cdots
\leftrightarrow e_v$};
\end{scope}
 \end{tikzpicture}\\

\subsection{Persistence, siphons and the Global attractor conjecture}\label{subs:per}
The \textit{Persistence conjecture}(\cite{Fei89}) states that trajectories of the mass action system originating in the positive orthant cannot get arbitrarily close to the boundary if the system is weakly reversible. For deficiency zero systems, this also implies the \textit{Global Attractor Conjecture} \cite{Horn74}, which in our case adds to the Deficiency zero theorem the fact that the unique stationary state is globally attracting, i.e. the solution of the dynamical system $dC/dt =
{\mathbb S} J(C)$ converges to $C_{stat}$ as $t\to\infty$ for
any positive initial condition. \\
\medskip Our case is relevant to Remark 6.4 in \cite{Deshpande2014} in which Siphons are defined.
\begin{Definition}[Siphon] See \cite{Deshpande2014}
Let $G=({\cal X},{\cal R})$ be a reaction network. A set $T\subseteq{\cal X}$ of species is a siphon iff for every reaction $(y_R\to y'_R,\ R\in {\cal R})$, if the complex $y'$ contains a species from $T$, then the complex $y$ contains a species from $T$ 
\end{Definition}
In the case of reversible extensions of dilute autocatalytic cores, the minimal siphon is ${\cal X}$ itself, and it is also self-replicable and drainable (there exists sequence of reactions that increase, decrease resp. the amount of all species). In such a case, persistence is the self-replicable nature "dominating" the drainability, but the notion of this "domination" is still vague in the literature, and one of the key necessities towards the {\em Global attractor conjecture}.

\medskip On the other hand, our case is simpler for the five core types. Lemma \ref{nntb} in the following section proves that we cannot have a boundary steady state which is not all 0, and Appendix section \ref{app:vanish} shows that near the zero stationary state (the onle one on the boundary), the total concentration of the system can only increase. Hence, it is unstable and trajectories cannot converge to it, proving the \textit{Global attractor conjecture} for reversible extensions of autocatalytic cores. The trajectories stay bounded in the positive space, (also shown in \cite{august2010} for weakly reversible networks) and must converge to the global attractor. 

\medskip

\noindent\textbf{Remark.} \textit{(Deficiency with degradation)}\\
Consider a core $G=({\cal X},{\cal R})$. 
We now assume that all species can be degradated, i.e. all
degradation rates $a_x, x\in {\cal X}$ are $>0$. This changes totally
the count in the deficiency index, since all 
linkage classes defined for zero degradation are now connected to
the zero complex, whence $|$Link$|$=1, whereas the rank $s$ of the
stoichiometric matrix is left unchanged. Recall that Complexes are elements in the vector space $\R^{\cal X}$; the zero complex is the zero vector. Even though including it in the network changes the complexes and the linkage classes, it does not affect the rank of its reaction vectors (See Chapter 3 of \cite{Fei19}). Thus, if $|{\cal R}_{>1}|$
is the number of one-to-many reactions (note that all product
composition $y_R, R\in {\cal R}_{>1}$ are distinct),
\BEQ \del = (|{\cal X}| + |{\cal R}_{>1}| + 1) - 1 - |{\cal X}| =   |{\cal R}_{>1}|.
\EEQ
Thus: $\del\ge 1$ always; the extended Deficiency One Theorem (see
Theorem A.1 in \cite{Fei95}), implying (for cores) uniqueness (but
not existence) of stationary states, applies if and only if $\del=1$;
and finally, $\del=1$ only in the following cases: 

\begin{center}
Type I (minimal); Type II$_{\ell=1}$ (minimal); Type III
\end{center}

where minimal means: minimal stoechiometric coefficients (all equal to
0 or 1, except in the case of Type I, where a single replication
reaction $B_n\to mB_1$, $m\ge 2$ is allowed). 

\medskip There exist other works investigating how degradation rates affect steady states, but they have to do with inheriting the multistationary nature of the subnetworks (\cite{banaji2018}\cite{capa2019}\cite{craciun2006}\cite{Joshi2012}). Our case differs in two ways. The inheritance of monostationary nature requires proving that no new stationary states can arise for some set of reaction rates for the old and the added reactions, while for multistationarity a proof that at least two stationary states get 'inherited' for some set of rates suffices. The second is that, unlike settings where inflows and outflows are added together (\cite{craciun2006} for example), we only add an outflow for each species. The special form of autocatalytic cores makes it so that it does not exactly fit the assumptions of existing theorems that do address monostationarity either directly or as absence of multistationarity. Thus, a new investigation is warranted.

\section{Uniqueness of states with degradation rates}
First, we prove two short, but important lemmas, \ref{lemma:nnt} and \ref{nntb}, that form a part of the hypothesis for Theorem \ref{bigboy}. Then we state the main result for the uniqueness of the stationary states of cores in the presence of degradation and demonstrate the proof. Then we demonstrate that the assumption of non-degenerate zero stationary state at vanishing degradation is valid for the cores. Finally the last sub-section gives an example of a multi-stationary network with two one-to-many reactions.

\subsection{No non-zero stationary state for large degradation rates} 


Let $(\alpha_x)_{x\in {\cal X}}$ be the degradation rates.
A direct reaction $R$ is of the form $x\to \cdots$, therefore $y_R=e_x$;
by abuse of notation, we write $k_{x\to y'}$ instead of $k_{e_x\to y'}$.
Conversely, a reverse reaction $R$ is of the form $\cdots \to x'$,
therefore $y'_R=e_{x'}$, and we write then $k_{y\to x'}$ instead of
$k_{y\to e_{x'}}$. One-one reactions $R$ are such that $||y||_1:=
\sum_x y_x = 1$ and $||y'||_1=1$; all other reactions have either
$||y||_1>1$ or $||y'||_1>1$ (but not both).   An expression
like $\sum_{x\to y'} f(x,y')$ is to be understood as the sum of
the values $f(x,y')$ for all reactions $R:e_x\to y'$ with $x$ fixed.

\begin{Lemma}\label{lemma:nnt}
Let $K:=\max_{x\to y' \ |\ ||y'||_1>1}   k_{x\to y'}$ and 
$M:= \max_x \sum_{x\to y'} (||y'||_1-1)$. Then, if
\BEQ \forall x\in {\cal X}\, \qquad \alpha_x\ge KM \EEQ
there is no strictly positive stationary state. 
\end{Lemma}

\noindent{\bf Proof.} Let  $C>0$. Then 
\BEQ \langle 1, \frac{dC}{dt}\rangle = \sum_x  [x] \Big( \sum_{x\to y'}  k_{x\to y'}
\langle 1,y'-e_x\rangle - \alpha_x\Big) +  \sum_{x'} \sum_{y\to x'} k_{y\to x'} [y] \langle 1,e_{x'}-y\rangle. \EEQ
The second sum is $<0$ since $\langle 1,e_{x'}-y\rangle = -(||y'||_1-1)\le 0$, more precisely $<0$ for all one-to-many reactions. 
The factor between parentheses in the first term is 
$\sum_{x\to y'} k_{x\to y'} (||y'||-1) - \alpha_x \le KM-\alpha_x\le 0$. 
\hfill \eop

\subsection{No stationary state on the boundary}
The $positive$ $orthant$ is defined as the space $\mathbb{R}_+=\{x$ $|$ $[x_i]>0$ $\forall$ $x_i\in\mathcal{X}\}$ 
\begin{Lemma}\label{nntb}
    For any of the 5 types of cores ($G=(\mathcal{X},\mathcal{R})$, with or without degradation), there cannot be a stationary state with some concentrations 0 but not all (i.e. on the "boundary" of the positive orthant)
\end{Lemma}
\noindent\textbf{Proof.} Let the set of species with vanishing concentrations be $\mathcal{B}\subseteq\mathcal{X}$. Note that for any species $x_i\in\mathcal{X}$, the possible types of outgoing reactions for any of the cores (for some other species $x_j,x_k\in\mathcal{X}$) are
\begin{itemize}
    \item $x_i \rightarrow \emptyset/x_j/x_j +x_k$ 
    \item $x_i+x_j\rightarrow x_k$
    \item $2x_i\rightarrow x_j$
\end{itemize}
No other type of reaction contributes to the decrease in the concentration of $x_i$, and hence consequently, in the mass action equation for $d[x_i]/dt$, all other terms are positive. The latter takes the general form \\
\begin{equation}\label{eq:st}
    \frac{d[x_i]}{dt}=-[x_i](\sum_{X_j\in\mathcal{X}}s_j[x_j]+s)+f([x_j]_{X_j\neq X_i})
\end{equation}
where some $s_j,s$ may be 0 but some are positive, and $f$ is a polynomial of the rest of the concentrations with all coefficients positive.\\
Now, consider $x_i\in\mathcal{B}$, i.e. $[x_i]=0$ at the stationary state. We get from (\ref{eq:st}) that $f=0$, which in turn implies that for any species $x_j$, which is the reactant of a reaction with $x_i$ as a product (or one of), $[x_j]=0$ and thus $x_j\in\mathcal{B}$. Now, for each autocatalytic core $G$, we define a Directed Graph $G'=(\mathcal{X},E')$, where 
\begin{itemize}
    \item Edge $\{x_i,x_j\},\{x_j,x_i\}\in E'$ if $x_i\rightleftarrows x_j\in \mathcal{R}$
    \item Edge $\{x_j,x_i\},\{x_k,x_i\}\in E'$ if $x_i\rightleftarrows x_j+x_k\in \mathcal{R}$
\end{itemize}
Note that $\{x_i,x_j\}\in E'$ and $x_i\in\cal B$ $\implies x_j\in\cal B$. Since $G'$ defined for each core is a weakly connected digraph, we get $\mathcal{B}=\mathcal{X}$, the zero stationary state.\\

\subsection{General uniqueness of stationary states}

\begin{Theorem}\label{bigboy}
Take a chemical reaction network of $n$ species, $G=({\cal X},{\cal R})$ and consider $G'=({\cal X},{\cal R'})$ to be the extension of G with added non-zero degradation reactions for each species $x\in{\cal X}$. Let $[x]$ denote the concentration of species $x$ and $C$ is the vector of concentrations. Let the network satisfy mass action kinetics with the following assumptions:
\begin{enumerate}
    \item $G$ is a weakly reversible deficiency 0 network
    \item Neither $G$ nor $G'$ has any stationary state on the boundary of the positive orthant except the zero state
    \item There exists a continuous function $m_b(a,k)$, such that any positive stationary state of $G'$ with kinetic rates $k$ and degradation rates $a$ satisfies $m_b(a,k)\geq\max\{[x],x\in{\cal X}\}$
    \item There are no degenerate non-zero stationary states for $G'$ for all choices of parameters
    \item The Jacobian (defined below) of the mass action equations of $G'$ at the zero state is 0 for exactly one value of $\alpha\geq0$, if we take degradation rates of the form $\alpha.a_i$ for each species $x_i\in{\cal X}$, $a_i,\alpha\in\mathbb{R}^+$
\end{enumerate}
Then, the network $G'$ has at most one non-zero stationary state for any values of kinetic and degradation rates. 
\end{Theorem}
\noindent{\bf Proof.} By Assumption 1 and deficiency theory, $G$ has only one non-zero stationary state, which is asymptotically stable. Assume that there exists a set of rates $k$ and degradation values $a$ for which $G'$ has at least two positive stationary states. We consider the degradation rates in a variable form: $(\alpha.a)$,$\alpha\in\mathbb{R}^+$ and prove contradiction with the hypotheses by letting $\alpha$ vary.

\medskip

Let $f(\alpha.a,C)$ denote the mass action equations of the system for these fixed kinetics.\\
Let $S(\alpha)$ denote the set of solutions in $C$ of the mass action equations $f(\alpha.a,C)$ with $C\geq0$. Let $S^*(\alpha)=$$S(\alpha)\cap\{C>0\}$.The Jacobian matrix $$
\textbf{J}(\alpha.a,C)=\left[ 
\begin{array}{cccc} 
\frac{\partial \textbf{f}}{\partial x_1} & \frac{\partial \textbf{f}}{\partial x_2} & \dots & \frac{\partial \textbf{f}}{\partial x_n}
\end{array} 
\right]
$$ is the (square in this case) matrix of all the first-order partial derivatives of the equations, and its determinant is referred to as the Jacobian Determinant. The matrix is invertible if the determinant is non-zero. A state $C$ is called degenerate if $C\in S(\alpha)$ and det $J(\alpha.a,C)=0$.

\medskip

Take $M$ as the bound defined by Lemma \ref{lemma:nnt}. At $\alpha=1$ there are at least two non-zero stationary states, and at $\alpha=M$ there are none. Also by Assumption 5, there are no degenerate stationary states apart possibly from the origin for $0<\alpha\leq M$. 

\medskip

Considering the bound on the solutions based on assumption 3, the solutions for any $\alpha\leq M$ are bounded by $\max\{m_b(\alpha.a',k)$ $|$ $0\leq\alpha\leq M\}$ and cannot diverge. Call this compact subset $\mathcal{V} \subset \mathbb{R}^{n}$. Let $\mathcal{V}^+=\mathcal{V}\cap\mathbb{R}^{n+}$

\medskip

We start with a short Lemma,
\begin{Lemma}\label{lemma:max}
    (Maximal solution of C) Let $C_{ss}\in S^*(\alpha^o)$ be non-degenerate,\\
    Define $I_{max}=\uplus \{I$ $|$ $\alpha^o\in I, \exists$ $ C^o_{st} (\alpha):I\rightarrow\mathcal{V}$ satisfying conditions (*),(**),(***)\},\\
    (*) $C^o_{st}(\alpha^o)=C_{ss}$\\
    (**) $\forall \alpha \in I$, $C^o_{st}(\alpha)\in S^*(\alpha)$\\
    (***) $\forall \alpha \in I$, det $J(\alpha,C^o_{st}(\alpha))\neq0$,\\
    Then there exists a unique $C_{st}(\alpha):I_{max}\rightarrow\mathcal{V}$ such that it satisfies (*),(**),(***) for $I_{max}$
\end{Lemma}
\noindent\textbf{Proof.} If $C_{ss}$ is a stationary state of the system $G'$ at $\alpha^o$ with an invertible Jacobian matrix, then, by the implicit function theorem, there exists an open interval $I^o$ around $\alpha^o$ and a unique continuous function $C_{st}:I^o\rightarrow\mathcal{V^+}$ such that $C_{st}(\alpha^o)=C_{ss}$ and $C_{st}(\alpha)$ $\in S(\alpha)$ for all $\alpha\in I^o$. Thus $I_{max}$ is non-empty.

\medskip

If $I^a$ and $I^b$ are two intervals  with continuous functions $C_{st}^a(\alpha)$ and $C_{st}^b(\alpha)$ on the intervals (respectively) satisfying (*),(**),(***), with $C_{st}^a(\alpha^o)=C_{st}^b(\alpha^o)=C_{ss}$, \\
Let $I'=\{\alpha$ $|$ $C_{st}^a(\alpha)=C_{st}^b(\alpha)\}$, then $I'$ is open in $I^a\cap I^b$ because of (***), and it is closed by continuity; thus (by connectedness) $I'=I^a\cap I^b$.\\ 
Setting $C_{st}(\alpha)=$ $\begin{cases}
    C_{st}^a(\alpha) \text{ if } \alpha \in I^a \\
     C_{st}^b(\alpha) \text{ if } \alpha \in I^b
\end{cases}$, we get a function continuous on the set $I^a\cup I^b$ \\
Gluing all such intervals together gives a unique continuous function $C_{st}$ over $I_{max}$.\\ $\square$

\Bigskip

As there are at least two distinct stationary states at $\alpha=1$, select any two distinct states $C_{ss}^1$ and $C_{ss}^2$, and define as in the Lemma \ref{lemma:max}, the intervals $I_{max}^1$ and $I_{max}^2$ and continuous functions $C^1_{st}:I^1_{max}\rightarrow\mathcal{V}$ and $C^2_{st}:I^2_{max}\rightarrow\mathcal{V}$, such that $C^i_{st}(1)=C^i_{ss}$ for $i=1,2$. We exhibit a contradiction by considering the behaviour of functions $C^i_{st},i=1,2$ at the end points of $I'=I_{max}^1\cap I_{max}^2=(\alpha_{min},\alpha_{max})$.

\medskip

Let $\alpha'=\alpha_{min}$ or $\alpha_{max}$; by compactness there exists a sequence $\alpha_n$ in $I'$ converging to $\alpha'$, such that the sequence ($C_{st}^1(\alpha_n)$,$C_{st}^2(\alpha_n)$) converge in $\mathcal{V}^2$. Let ($C_{st}^1(\alpha')$,$C_{st}^2(\alpha')$)=$\lim_{n\rightarrow\infty}(C_{st}^1(\alpha_n),C_{st}^2(\alpha_n))$.

\medskip
Consider first $\alpha'=\alpha_{max}$; if $C_{st}^1(\alpha_{max})\neq$$C_{st}^2(\alpha_{max})\neq 0$, by Assumption 4 we can apply the implicit function theorem again to extend the interval $I'$, contradicting the maximality of $\alpha_{max}$. Thus either $C_{st}^1(\alpha_{max})=0$ and det $J(\alpha_{max},C_{st}^1(\alpha_{max}))=0$, or $C_{st}^2(\alpha_{max})=0$ and det $J(\alpha_{max},C_{st}^2(\alpha_{max}))=0$. This implies that det $J(\alpha_{max},0)=0$

\medskip
Now consider $\alpha'=\alpha_{min}$. If $\alpha_{min}>0$, we get again det $J(\alpha_{min},0)=0$ which is contradictory with Assumption 5 as $\alpha_{min}\neq\alpha_{max}$; thus $\alpha_{min}=0$. Also note that $C_{st}^1(\epsilon),C_{st}^2(\epsilon)\neq0$ for $0<\epsilon<\alpha_{max}$\\

The origin is non-degenerate (by Assumption 5 again) and the bulk stationary state (call it $C_{\delta}$) is also non-degenerate by Theorem 15.2.2 from \cite{Fei19}.\\
If $C_{st}^{1,2}(\alpha_{min})$=$C_{st}^{1,2}(0)=C_{\delta}$, it violates the uniqueness of the continuous function we would get by applying the implicit function theorem at $C_{\delta}$. If $C_{st}^{1/2}(0)=0$, along with the constant function $C(\alpha)=0$, it violates the uniqueness of the implicit function at $C=0$.

\smallskip

Thus the existence of the two stationary states $C^1_{ss}\not=C^2_{ss}$ at $\alpha=1$ is contradictory with the hypotheses of the Theorem.

\subsubsection*{The zero stationary state for vanishing degradation rates}\label{subs:van}

The Jacobian of a core at vanishing degradation and zero stationary state ($J(0.a,0)$) is identical to the Jacobian of the dynamical core as classified in Section \ref{section:main}, but with all one-one reactions made reversible. Denote the latter chemical reaction network by $G=(\mathcal{X},\mathcal{R})$. \\
We use the child selection formalism developed in \cite{Vassena2020} to determine the properties of the Jacobian of this system. 

\medskip

For any reaction $r\in\mathcal{R}$, let $\bar{r}$ denote the reverse reaction, with reactant and product species interchanged.\\
The reaction network of $n$ species, $G$, has a form where for all species $x\in\mathcal{X}$ there exists either 1 or 2 outgoing reactions with a single reactant (which is $x$). $t$ reactions in $\mathcal{R}$ are one-to-many reactions, and all the other reactions are in the form of $(n-t)$ reversible pairs of 1-1 reactions, $r$ and $\bar{r}$  (total $2n-2t+t$ reactions). Furthermore, there exists a species $x\in\mathcal{X}$, with only one outgoing reaction which is also a one-one reaction.
 
\medskip

A child selection is an injective map \textbf{Cs: $\mathcal{X}\to\mathcal{R}$} which associates to every species $x\in\mathcal{X}$ a reaction $r\in\mathcal{R}$ such that $x$ is an reactant species of reaction $r$. Then for the Jacobian matrix $J$,
\begin{equation}
    det\text{ }J=\sum_{Cs} det\text{ }S^{Cs} \prod_{x\in\mathcal{X}} \frac{\partial k_{Cs(x)}}{\partial x}
\end{equation}
$S^{Cs}$ is the $n\times n$ matrix whose $i^{th}$ column is the $Cs(i)^{th}$ column of the stoichiometric matrix $S$, $k_{Cs(x)}$ is the mass action rate of reaction $Cs(x)$, and the sum runs over all possible child selections $Cs$.

\medskip

In our particular case, any child selection maps the $n$ species to $n$ reactions out of $2n-t$. If $Cs(\mathcal{X})=\{Cs(x): x\in \mathcal{X}\}$ contains both $r$ and $\bar{r}$, then $det$ $S^{Cs}=0$.\\
A child selection with a non-zero contribution to $det$ $J$ must have all $t$ one-to-many reactions and $(n-t)$ 1-1 reactions while avoiding reversible pairs. 

\medskip

When all the one-to-many reactions are included, there is only one assignment of the 1-1 reactions possible, and it is exactly the form of the reactions for the core in Section \ref{section:main}. The contribution for this child selection is non-zero, as $S^{Cs}$ is irreversible (by a case-by-case check) by the hypothesis of Section \ref{section:main}, and all partial derivatives of the mass action rates for these reactions are positive constants. Thus the Jacobian determinant is non-zero. This is shown explicitly for each type in Appendix \ref{app:vanish}.

\noindent It is easy to see that the only possibility assigns to each $x$ the reaction outgoing from $x$ as depicted in the irreversible graphs of section 3.
\medskip

\begin{Theorem}
    \label{smallboy}
    All autocatalytic cores (apart from Type II with $\ell>2$), with strictly non-zero degradation, can have at most one positive stationary state
\end{Theorem}

\noindent\textbf{Proof.} Type I and Type III can be explicitly solved to get the uniqueness of stationary states, as demonstrated in Appendix \ref{sec:onethree}\\
For the rest of the cores, they satisfy all the conditions of Theorem \ref{bigboy}, and hence can have at most one non-zero stationary state. Section \ref{nd} gives condition 1, \ref{nntb} gives condition 2, \ref{subs:van} gives condition 5, and conditions 3,4 are proven for each case in Appendix \ref{sec:rest}.

\medskip
\noindent\textbf{Remark. } \textit{Some degradations zero, some non-zero}\\
In the case of some degradation rates vanishing but not all, uniqueness still holds for all types except type V (See Appendix \ref{sec:onethree} and \ref{app:some}).


\subsection{Multistationarity for interaction of multiple cores}

We end this section with a example of how we can have multistationarity. This example is a slightly modified version of Example 2.3 in \cite{Joshi2023}. Species $X_3$ and $X_4$ for a Type I core, and it has been coupled to a fork with $X_1$ and $X_2$. The inputs to $X_1$ and $X_2$ can again be viewed as a coupling to an out reaction (degradation) of an autocatalytic core at equilibrium. 

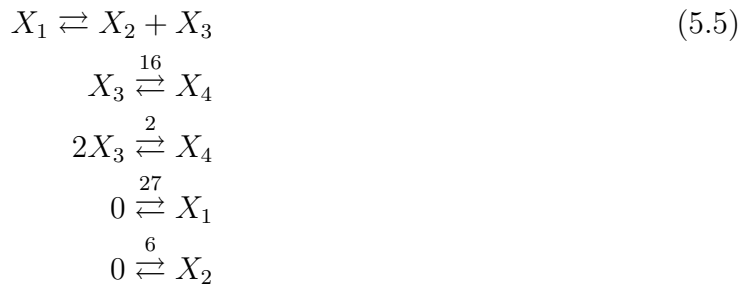
\BEA
X_1 \rightleftarrows X_2 +X_3 \\ \nonumber
X_3 \overset{16}{\rightleftarrows} X_4 \\ \nonumber
2X_3 \overset{2}{\rightleftarrows} X_4 \\ \nonumber
0 \overset{27}{\rightleftarrows} X_1 \\ \nonumber
0 \overset{6}{\rightleftarrows} X_2 \\ \nonumber
\EEA
 \Bigskip

 Fix the unspecified rates as 1, this system admits three distinct positive stationary concentrations \\
 \medskip\\
 $[x_1]=18,[x_2]=15,[x_3]=2,[x_4]=20$\\
 $[x_1]=13,[x_2]=20,[x_3]=1,[x_4]=9$\\
 $[x_1]=21,[x_2]=12,[x_3]=3,[x_4]=33$\\
 \medskip\\
 Example 2.6 from \cite{Joshi2023} is also a multistationary system which involves a non-linear coupling reaction between two autocatalytic reactions. 

\newpage

\section{Discussion}
In this work, we characterized minimal autocatalytic networks, called autocatalytic cores.
Our two main results are that:
(i) Stoichiometric autocatalytic cores are identical to dynamical autocatalytic cores in the diluted regime.
(ii) Dynamical autocatalytic cores have a single stationary state. 
\\

A reaction is stoichiometrically autocatalytic if there exists a set of reaction rates such that every species of the network is positively produced \cite{BloLacNgh}, provided that every reaction has at least one reactant and one product, a condition called \textit{autonomy}. Equivalently, the image of the stoichiometric matrix, which has plus and minus signs in every column, intersects the strictly positive orthant.
Minimal stoichiometries allowing autocatalysis, or stoichiometric autocatalytic cores, were classified into five types \cite{BloLacNgh}, which differ by the number of one-to-many reactions they possess and the entanglement of their internal catalytic cycles. In this regard, the Type II cores are special as they are the only type where the number of catalytic cycles $i$ varies, leading to subcategories denoted $II_i$.
For all types, each category or sub-category comprises an infinity of variants generated by the addition of intermediate 1-to-1 reaction steps. 
By definition of minimality, any autocatalytic system must contain an autocatalytic core. 
Thus, possessing an autocatalytic core is a necessary condition for autocatalysis, which has been used to detect autocatalysis in real chemistries \cite{Peng2022}.
We should also mention that more relaxed and more restricted stoichiometric criteria have been proposed in the literature \cite{Hordijk2018,Andersen2021,Deshpande2014}, with Type II corresponding to minimal autocatalytic sets as defined in \cite{Steel2018}.
\\

The present work deals with dynamical autocatalysis, following up on Unterberger $\&$ Nghe \cite{NgheUnt1}, as autocatalysis is ultimately a dynamical property \cite{Lin2020,Horvath2021}. Stoichiometric autocatalysis does a priori not guarantee dynamical autocatalysis, because the required reaction vectors may not be compatible with kinetic laws (\cite{Lin2020}), or degradation rates are too high as compared to the production rate of species, thus turning the zero zero-concentration state into an attractor.
The kinetic viability of autocatalytic networks has been studied on a case-by-case basis \cite{BloLacNgh}, but there is currently no general result characterizing dynamical autocatalysis for arbitrary degradation rates of each species.
However, in the limit regime of diluted reaction networks, reference \cite{NgheUnt1} shows that there exist finite values of degradation rates for which networks with the topological property (Top) possess a positive Lyapunov exponent, meaning a positive exponential growth rate.
Colloquially, (Top) is equivalent to requiring that the strongly connected components which have no inflows from other components, possess at least one one-to-many reaction.
One may then consider minimal networks allowing dynamical autocatalysis in the diluted regimes, here called \textit{dynamical autocatalytic cores}, which corresponds to minimal networks verifying (Top).
We showed here that dynamical cores are a subset of stoichiometric cores.
This guarantees that autocatalytic systems capable of exponential growth starting from zero or small concentrations (given small degradation rates) are exactly those detected using stoichiometric cores.
\\

The eigenvector associated with the Lyapunov exponent, up to normalization, gives the fraction of each species during exponential growth, called the \textit{chemical composition}.
In practice, this regime may be realized at intermediate times in large closed chemical reactors, where the growth mode associated with the Lyapunov exponent becomes dominant compared to all other modes, but the accumulation of products does not yet cause visible non-linearities. 
This type of behaviour is well-known in microbiology, where growth in the bioreactor is typically characterized by three phases leading to a sigmoidal profile (called 'Monod' growth): an initial phase that appears as a delay ('lag' phase), a stationary exponential growth phase ('log' phase), and a saturation ('stationary' phase).
The growth mode associated with the Lyapunov exponent can also be observed in open reactors, according to several modalities.
One is the so-called CSTR, for a continuously stirred chemical reactor, where reactants consumed by the reaction are injected at a constant flux, the reactor is actively mixed, and some of the solution is extracted at an equal rate in order to maintain the total volume constant.
CSTRs can be approximated by serial dilution protocols, where a finite fraction is extracted from a reactor after a given incubation time, and then injected in fresh solution at regular time intervals. The composition approaches the Lyapunov eigenvector one as the incubation time gets smaller \cite{Blokhuis2018}. 
Yet another modality is the chemostated open reactor, meaning that the concentration of reactants is maintained fixed, and the dilution rate is matched with the growth rate. Although theoretically convenient, this type of reactor requires semi-permeable boundaries. This situation is in fact realized in living cells, as these are equipped with elaborated import pumps and osmotic regulation mechanisms that maintain the total concentration constant. However, it was shown recently that this type of mechanism could be realized in the absence of evolved bio-molecules, by transport phenomena and osmosis in a rudimentary multi-phase system containing autocatalytic reactions \cite{Lu2024}. 
\\ 

As products accumulate, many-to-one and many-to-many reactions start to participate in the dynamics.
Reverse reactions should then be considered, resulting in non-linear kinetics, making it non-obvious whether the Lyapunov mode is the only stationary growth regime.
Multistationarity is of particular interest for emergent properties in reaction networks \cite{Novichkov2021}, notably in the context of the origin of life, where multiple growth modes in competition would enable rudimentary forms of Darwinian evolution \cite{Ameta2022, Adamski2020, Pavlinova2022}.
Here, multistationarity is understood in the composition space (the fraction of each species), which in a chemostated system corresponds to actual multistationarity in concentration, as well as in homeostatic systems that grow while maintaining a constant total concentration (as cells during exponential growth) \cite{Sughiyama2022}.
Formally, the reverse reactions lead us to consider what we termed the \textit{reversible extension} of the network.  
In the absence of degradation, deficiency theory provides a direct argument showing the uniqueness of a stationary composition \cite{Fei19}.
However, this theory is not directly applicable when adding arbitrary degradation rates.
By using direct computations of stationary regimes and interpolating the behaviour between large degradation rates and zero degradation, we could show the uniqueness of the stationary regime for almost all possible rate constants of the mass action kinetics.
This result was obtained for Types I, Type II$_{1,2}$, and Types III to V.
We could not establish the same result for Types II$_>2$; however numerical tests suggest that the same result applies. Thus, it remains an open question to address the uniqueness of the stationary state for these types, which correspond to minimal networks in the theory of collective autocatalytic sets. 
\\

Autocatalytic motifs were shown to be among the possible stoichiometries prone to trigger instabilities \cite{Vassena2024}, which in our context means that the zero concentration state is an unstable stable solution, but the bulk solution (where all species of the autocatalytic core have positive concentrations) is the only possible stable one. The fact that a stationary state is unique implies that phenomena richer than growth must proceed through an interplay between different Types of minimal cores and other types of negative/positive feedbacks, or more complicated kinetics that cannot be simplified into mass-action type equations. This uniqueness also means that autocatalytic cores alone lead to growth that is robust to perturbations, which is a potential advantage for primordial metabolism \cite{Ameta2021}.

\newpage

\numberwithin{equation}{subsection}
\section{Appendix}
The first subsection gives practical examples of the cores we get in the classification from metabolic networks. Section 8.2 is the definition of a cycle as used in Section \ref{section:prrofcl}. \ref{sec:exp} gives the form of the unique positive stationary state for each core in terms of the rate constants. \ref{app:vanish} show for each core, the explicit form of the Jacobian determinant at the zero state for vanishing degradations, as proved in Subsection \ref{subs:van}. The last two subsections contain the computations necessary to prove the uniqueness of the non-zero stationary state for each type.\\  
\subsection{Examples of autocatalytic cores from metabolism}\label{sec:examp}
\Bigskip {\em Example 1. Reverse Krebs cycle} (see e.g. (Smith-Morowitz
\cite{SmiMor}), Fig. 4.1). It is presented as Type II in \cite{BloLacNgh}; it may actually be seen as a degenerate case of Type III (see above) with $x=u$.  The path from $u$ to $v$ is the main
anabolic path leading from Oxaloacetate (4 carbon atoms) to Citrate (6), through 
Succinate and OxaloSuccinate, by a succession
of enzymatic reactions (reducative carboxylations, hydratations/dehydratations, reductions). The fragmentation reaction $v\to x+x'$
is the retro-aldolization Citrate $\longrightarrow$ Oxaloacetate +
Acetate. Then the path from $x'$ to $u$ is the anabolic path from
Acetate (2) to Oxalocetate, which is homologous to the subpath of the
main anabolic path leading from Succinate to OxaloSuccinate. 

\Bigskip {\em  Example 2. Glyoxylate cycle}, see. e.g. \cite{Mor}, 
Fig. 13.27. The glyoxylate cycle, 
a less-known metabolic cycle used by plants, bacteria, protists and
fungi to convert Acetyl-CoA to Succinate for
glucid biosynthesis,
is of Type III  if Acetate is chemostatted. The path from $u$ to $v$ is the main anabolic path
leading  (through OxaloAcetate) from Malate (4 carbon atomes) to IsoCitrate (6), by a succession of enzymatic
reactions (oxydations, aldolizations). The fragmentation reaction
$v\to x+x'$ is the retro-aldolization IsoCitrate $\longrightarrow$
Glyoxylate + Succinate. Then, the addition of Acetate by aldolization
leads from Glyoxylate (2) back to Malate, and (on the other branch),
an oxydation of Succinate (4) to Fumarate (4), followed by hydratation, leads also back to Malate. Note that the second branch is not degenerate
as in Example 1, but is not an anabolic path either (the number of
carbon atoms is preserved).

\subsubsection*{Other Metabolic pathways}
\Bigskip
\begin{itemize}
    \item \textit{Type III}
    \Bigskip

-- an anabolic path from a shorter molecule $u$ to a longer one $v$
along ${\cal C}\cap {\cal C}'$;

-- followed by a fragmentation (e.g. retroaldolization) from 
$v$ into  short molecules $x$ and $x'$;

-- and two anabolic paths, one from $x$ along ${\cal C}$, one from $x'$
along ${\cal C}'$, leading both
back to $u$.  

\Bigskip

\item \textit{Type IV}
\Bigskip

-- a path of anabolic reactions and skeleton rearrangements
leading from  a short molecule $x'$ to $u$ along ${\cal C}'$ , and then from $u$ to a
long molecule $v$  (possibly, $v=u$) along ${\cal C}\cap {\cal C}'$;

-- a fragmentation (retroaldolization, etc.) of $v$ into a shorter
molecule $x$ and 
directly back into $x'$;

--  and a  path of anabolic reactions leading along ${\cal C}$ from $x$ to a long
molecule $w$ (possibly, $x=w$), which is then fragmented back into $u$ and into $x'$.

Pictorially, with molecule sizes increasing upwards,

\begin{center}
\begin{tikzpicture}[scale=1.35]
\draw(0,-1.1) node {$x'$};
\draw[->](0,-0.8)--(0,-0.2);  \draw(-0.2,-0.5) node {${\cal C}'$};
\draw(0,0) node {$u$};
\draw[->,dashed](0,0.2)--(0,0.8);\draw(-0.5,0.5) node {${\cal C}\cap {\cal C}'$};
\draw(0,1) node {$v$};
\draw[->,ultra thick](0,1.2)--(0,1.4);
\draw[ultra thick](0,1.4)--(0,1.6);
\draw[->, ultra thick](0,1.6)--(-1,1.6)--(-1,-1)--(-0.2,-1);

\draw[->,ultra thick](0,1.6)--(2.5,1.6)--(2.5,-1); \draw(2.5,-1.2) node {$x$};
\draw[dashed,->](2.3,-1.2)--(1.3,-1.2)--(1.3,-0.7);
\draw(1.8,-1) node {${\cal C}$};
 \draw(1.3,-0.5) node {$w$};
\draw[ultra thick, ->](1.1,-0.5)--(0.9,-0.5); \draw[ultra thick] (0.9,-0.5)--(0.7,-0.5);
\draw[->, ultra thick](0.7,-0.5)--(0.15,-0.15); \draw[->, ultra thick](0.7,-0.5)--(0.15,-0.85);
\draw(0.7,-0.8) node {\tiny back};

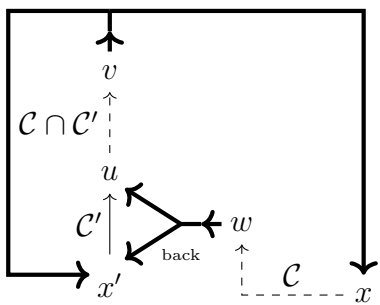
\draw(5,0) node {Pictogram (Type IV)};
\end{tikzpicture}
\end{center}

Paths from $u$ to $v$ and from $x$ to $w$ are depicted by a dashed line because they
may have length 0 (case $u=v$ or  $x=w$). All other paths (plain lines) have
length $\ge 1$.

\end{itemize}

\subsection{Further Definitions}

\begin{Definition}[cycle] \label{def:cycle}
Let $G=({\cal X},{\cal R})$ be a network, $n\ge 2$,  and $x_1,\ldots,x_n\in 
{\cal X}$. Then ${\cal C}=\{x_1\to x_2\to \cdots\to x_n\to x_1\}$
is a cycle iff there exist split reactions $x_1\to x_2, x_2\to x_3,\ldots,x_{n-1}\to x_n,
x_n\to x_1$ going along the cycle. See Figure below.  By abuse of notation,
we write $x_{n+1}=x_1$.

A {\em split reaction along the cycle} is any split reaction $x_i\to x_{i+1}$
coming from a reaction in $\cal R$.

We write $i_1\prec i_2\prec\ldots \prec i_k$ if $x_{i_1}\not=x_{i_2}\not=
\ldots\not=x_{i_{k-1}}$ and $x_{i_2}\not=x_{i_3}\not=\ldots\not=x_{i_k}$  (the case $i_1=1, i_k=n+1$, with the identification $x_{n+1}\equiv x_1$,  is not excluded) and, starting from 
$i_1$ and going round the cycle once,  one successively  encounters 
$i_2,\ldots,i_k$. (If $i_1=1$ this simply means $i_1=1<i_2<\ldots <
i_k\le n+1$). Because the circle closes back onto itself, this is not
an order relation (e.g. $1\prec n$ and $n\prec 1$). If, for some
$j=1,\ldots,k-1$, $|i_{j+1}-i_j|\not=1,n-1$ (i.e. if $i_{j+1},i_j$ are
not neighbors), we write more precisely: $i_j \prec\prec i_{j+1}$. 
\end{Definition}

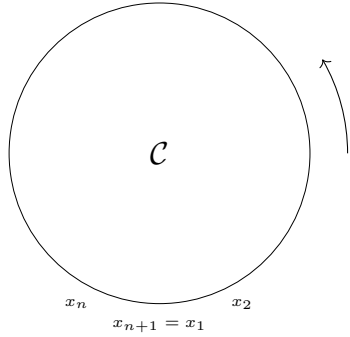
\Medskip
\begin{center}
\begin{tikzpicture}
\draw[ultra thin](0,0) circle(2);
  \draw[->](2.5,0) arc(0:30:2.5 and 2.5);
\draw(0,-2.3) node {\tiny $x_{n+1}=x_1$}; 
\draw(1.1,-2.) node {\tiny $x_2$};
\draw(-1.1,-2) node {\tiny $x_n$};  
\draw(0,0) node {${\cal C}$};

\end{tikzpicture}
\end{center}


\subsection{Proof of Classification Theorem}\label{section:prrofcl}


Let $G=({\cal X},{\cal R})$ be an autocatalytic core composed uniquely of one-to-one or
one-to-many reactions. In particular, by Proposition \ref{prop:Top},
it verifies (Top).
{\em We first prove by absurd that  $G$ must be irreducible}. Namely, assume
$G$ contains two strongly connected components ${\cal C}_1,{\cal C}_2$
with ${\cal C}_1\to {\cal C}_2$ and ${\cal C}_1$ minimal, i.e. such
that $\not\!\exists {\cal C},\  {\cal C}\to {\cal C}_1$.  (Top) implies
the existence of a one-to-many reaction $R: X\to s_1 x_1+\cdots + 
s_k x_k$ with $X,x_1,\ldots,x_j\in {\cal C}_1$ $(1\le j\le k)$;
$x_{\ell}\not\in {\cal C}_1$ if $\ell>j$; and $s_1+\ldots+s_j\ge 2$.
Then the irreducible restricted network $G'=({\cal C}_1, {\cal R}\Big|_{{\cal C}_1})$
(see Appendix) includes the one-to-many reaction $R\Big|_{{\cal C}_1}: X\to s_1 x_1 + \ldots + s_j x_j$, hence $G'$ satisfies (Top).
Therefore, $G$ cannot be minimal.

\Bigskip This has the following significant consequence: {\em verifying
(Top) for an autocatalytic core reduces to the trivial task of checking that
the reaction set ${\cal R}$ contains a one-to-many reaction}. This
is the key principle of the proof.

\Bigskip Going further, we consider mutually exclusive cases.
We will often make use of the following remark.  Let $G=({\cal X},{\cal R})$ be an irreducible autocatalytic
 network, ${\cal R}'\subset {\cal R}$ and ${\cal X}'\subset {\cal X}$
such that $({\cal X}',{\cal R}')\not=({\cal X},{\cal R})$, i.e.
either ${\cal R}'\subsetneq {\cal R}$ or ${\cal X}'\subsetneq {\cal X}$ (or both).
 Then the restricted network $G'=({\cal X},{\cal R}')\Big|_{{\cal X}'}$
is autocatalytic if it is irreducible and contains a one-to-many
reaction. This gives an easy argument to prove that $G$ is not minimal.  

\Medskip A key notion in the classification is that of a cycle, 
see Definition \ref{def:cycle}  in Appendix. Briefly said, a cycle
\BEQ {\cal C}:\qquad x_1 \to x_2 \to \cdots\to x_n\to x_1 \EEQ
is a succession of split reactions $x_i\to x_{i+1}$ ($1\le i\le n-1$), $x_n\to x_1$ coming from reactions in $\cal R$. The
(index) ordering $1\prec 2\prec \cdots\prec n\prec 1$ is defined
along the cycle. It is invariant under cyclic permutations
so that $k\prec \cdots\prec n\prec 1\prec \cdots\prec k$
defines the same cyclic order.


\subsubsection{Case 1 (simple reactions)}  \label{subsection:case1}


  {\bf We assume here that all reactions are simple}(that is, either a one-to-one reaction or a one-to-many reaction with a single type of product species).
By (Top), at least one of them is one-to-many, $R_1: x_1 \to sx_2$ 
with $s\ge 2$. Since $G$ is irreducible, one can find a cycle 
\BEQ {\cal C}:\qquad x_1 \to x_2 \to \cdots\to x_n\to x_1, \EEQ
see again Definition \ref{def:cycle}. Call $R_i: x_i \to s_i x_{i+1}$, $i=1,\ldots,n$
the reactions along the cycle. Then $G'=({\cal C}, \{R_1,\ldots,R_n\})
\subset G$ satisfies (Top), so (by minimality) $G=G'$ is of Type I.


\subsubsection{Case 2 (one cycle)}  \label{subsection:case2}


 Barring Case 1., $G$ contains at least one one-to-many
reaction 
\BEQ R_1 : \qquad x_1 \to s_2 x_2 + s' x' + \cdots \label{eq:R_1}
 \EEQ
  {\bf We assume here that  there exists a cycle ${\cal C}:x_1\to
x_2 \to \cdots\to x_n\to x_1$ containing all species.}

\Medskip It is easy to see that $R_1$ has exactly two different products; indeed, 
assuming (by absurd) $R_1: \ x_1 \to s_2 x_2 + s' x' + s'' x'' +\cdots$,
with $x_1\prec x_2 \prec x'\prec x''$,  the restricted network
$G|_{{\cal X}'}$, ${\cal X}'=\{x'\to \cdots \to x_1\} \subsetneq {\cal X}$ (in blue)
would also be irreducible and satisfy (Top), see Figure (the dotted
line stands for possible extra split reactions $x_1\to\cdots$ coming
from $R_1$),

\medskip
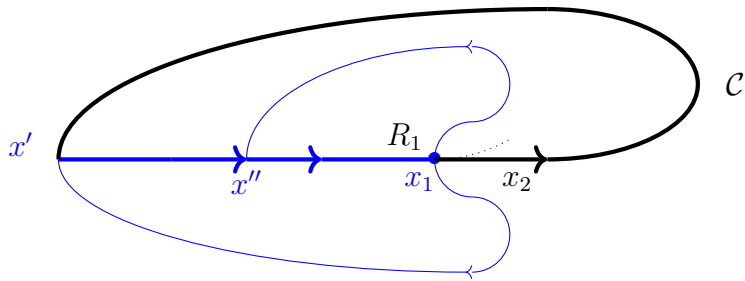
\begin{center}
\begin{tikzpicture}
\draw[blue](-0.5,0.2) node {$x'$};
\draw[ultra thick,blue](0,0)--(1.5,0);
\draw[->,ultra thick,blue](1.5,0)--(2.5,0);

\draw[blue](2.5,-0.3) node {$x''$};

\draw[->,ultra thick,blue](2.5,0)--(3.5,0); 
\draw[ultra thick,blue](3.5,0)--(5,0);

\draw[blue](4.8,-0.3) node {$x_1$};
\draw[blue](5,0) arc(180:90:0.5 and 0.5);
\draw[blue](5,0) arc(180:270:0.5 and 0.5);
\draw[blue](5.5,0.5) arc(-90:90:0.5 and 0.5);
\draw[>-,blue](5.5,1.5) arc(90:180:3 and 1.5);

\draw[dotted](5,0) arc(-90:-60:2 and 2);

\draw[blue](5,0) node {\textbullet}; \draw(4.6,0.3) node {$R_1$};
\draw[blue](5.5,-0.5) arc(90:-90:0.5 and 0.5);
\draw[>-,blue](5.5,-1.5) arc(-90:-180:5.5 and 1.5);

\draw[->,ultra thick](5,0)--(6.5,0);  \draw(6.1,-0.3) node {$x_2$};
\draw[ultra thick](6.5,0) arc(-90:90:2 and 1);
\draw[ultra thick](6.5,2) arc(90:180:6.5 and 2);

\draw[ultra thick](9,1) node {$\cal C$};
\end{tikzpicture}
\end{center}
\medskip

\noindent For the same reason, any reaction with $\ge 3$ different products is
excluded.

\Medskip

We note certain reactions that become redundant in this framework and can be removed without breaking (Top)
\begin{itemize}
    \item {\bf short-cuts}, i.e. simple reactions $x_i\to sx_j$ not
along the cycle, i.e. such that $|j-i|\not =1,n-1$
\item $x_i\to s'' x_{i+1}$
become redundant in presence of multiple reactions $x_i \to sx_{i+1} + s'x'$
\item multiple reactions $x\to s'x'+s'' x''$, such 
that neither of its split reactions $x\to x'$ nor $x\to x''$ is  along the
cycle
\item reaction $x_i \to s_1 x_{i+1} + s'_1 x'_1$ is redundant in the presence of 
$x_i \to s_2 x_{i+1} + s'_2 x'_2$ (i.e. with different stoichiometric
coefficients, $s_1\not=s_2$ or $s'_1\not=
s'_2$, or different products, $x'_1\not=x'_2$) 
\end{itemize}

On the other hand, split reactions $x_i\to x_{i+1}$ may either
be simple reactions, or a piece  of (i.e. a split reaction coming
from) a multiple reaction. Denote
by $I_{\ell}=\{i_1,\ldots,i_{\ell}\}$ the set of indices $i$ such that 
$G$ contains a one-to-many reaction $R_i: x_i\to sx+ s' x' $
with $x=x_{i+1}$. Then  the multiple reactions of $G$ belong to
the closed list,
\BEQ R_j : \qquad x_{i_j}\to s_{i_j} x_{i_j+1} + s'_j x_{\sigma_j},
\qquad j=1,\ldots,\ell  \label{eq:Rj}
\EEQ
with  $\sigma_j\in \{1,\ldots,n\}$, $\sigma_j\not=i_j,i_j+1$. Denote by $\Sigma_{\ell}$ the set of these $\sigma_j$. See 
illustrating Figure 
corresponding to Type II in the Theorem in \S \ref{section:main}. The stoichiometric coefficient $s'_j$ attached
to the side-branch $j=1,\ldots,\ell$, see (\ref{eq:Rj}), is equal to 1 (otherwise
$G\Big|_{{\cal C}_j}$ restricted to the subcycle 
${\cal C}_j: x_{\sigma_j}\to x_{\sigma_j+1}\to\cdots \to x_{i_j}\to x_{\sigma_j}$
would be irreducible and satisfy (Top)).  

\Medskip We include $R_1,R_2,\ldots,R_{\ell}$ one after the another,
and discuss by induction on $j\ge 2$ the various possibilities of inserting
$i_j,\sigma_j$ along the $2(j-1)$ arks defined by the previously defined
indices $I_{j-1}\uplus \Sigma_{j-1}:=\{i_1,\sigma_1,i_2,\sigma_2,\ldots,i_{j-1},\sigma_{j-1}\}$. Namely --  starting from $i_1$, following along the cycle, and considering
all indices in $I_{j-1}\uplus \Sigma_{j-1}$ -- we get
 \BEQ
  i_1=\tau_1\prec
\cdots \prec \tau_{2(j-1)}\prec i_1
\EEQ
 with $\{\tau_1,\ldots,\tau_{2(j-1)}\}=I_{j-1}\uplus \Sigma_{j-1}$. The $2(j-1)$ arks
$(\square_i)_{1\le i\le 2(j-1)}$ are then 
\BEQ \square_i=\{k\ |\ 
\tau_i\prec k \prec \tau_{i+1}\} \ \  (1\le i<2(j-1)), \qquad 
\square_{2(j-1)}=\{k\ |\ \tau_{2(j-1)}\prec k\prec \tau_1\}.
\EEQ   
Note that some of them are possibly empty.


\bigskip

\Medskip
If $\ell=1$, we get a type II cycle with $\ell$=1.

\Medskip {\em Remark.} Note that the indices of the reactions $R_j$, $j=\{1,2,...m\}$ are arbitrarily assigned; interchanging them does not change the type of cycle. This simple remark will be used many times.


\subsubsection{Two one-to-many reactions}


We denote by 
\BEQ \square_1=\{i\not=i_1,\sigma_1\ |\ i_1\prec 
i\prec \sigma_1\} 
\EEQ
 the set of indices found along the ark from
$i_1$ to $\sigma_1$, and similarly 
\BEQ \square_2=\{i\not=i_1,\sigma_1\ |\ \sigma_1\prec 
i\prec i_1\}
\EEQ
 the complementary set. Thus
 \BEQ i_1 \prec  \, \square_1 \prec \sigma_1  \, \prec \square_2 \EEQ
 by which it is meant that $i_1\prec i \prec \sigma_1 \prec i'$
 for all $i\in \square_1,i'\in \square_2$. 
 
\Medskip Adding a second one-to-many reaction is equivalent to choosing $i_2$ and $\sigma_2$ in $\square_1\uplus \square_2$, or in other words, to inserting $i_2,\sigma_2$ along the two corresponding arks. Exhausting the different possibilities -- with, in particular, $i_2,\sigma_2$ inserted along the same ark or not --, we find the 
following cases:

\begin{itemize}
\label{threetwoone}
\item[(i)] (nested violation, $N_1$ case)  $i_1 \prec i_2 \prec \sigma_2
\prec \sigma_1 $ (both $i_2,\sigma_2$ inserted along $\square_1$)

\Medskip
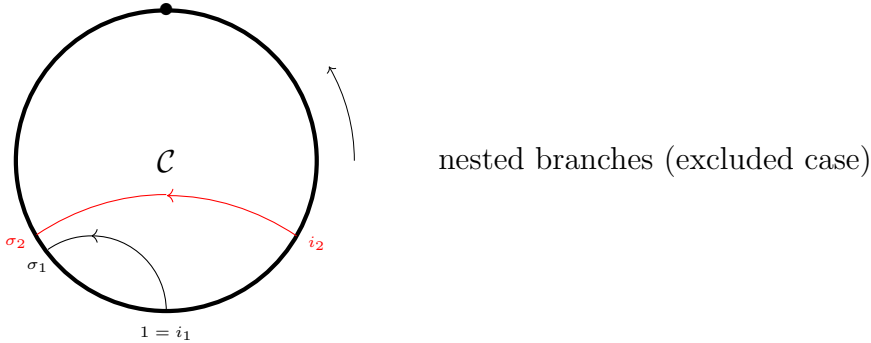
\begin{figure}[h]
\centering
\begin{tikzpicture}
\draw(0,0) node {${\cal C}$};
\draw[ultra thick](0,0) circle(2);  \draw[->,ultra thin](2.5,0) arc(0:30:2.5 and 2.5);
\draw[->](0,-2) arc(0:90:1 and 1);  \draw(0,-2.3) node {\tiny $1=i_1$};
\draw(-1,-1) arc(90:125:1 and 1);
\draw(-1.71,-1.41) node {\tiny $\sigma_1$};

\draw[->,red](2*0.866,-2*0.5) arc(60:90:3.45 and 4);
\draw[red](0,-0.45) arc(90:120:3.45 and 4);
\draw[red](2,-1.1) node {\tiny $i_2$};
\draw[red](-2,-1.1) node {\tiny $\sigma_2$};

\draw(0,2) node {\textbullet}; 

\draw(6.5,0) node {nested branches (excluded case)};
\end{tikzpicture}
\caption{Invalid nested branches}
\label{nested}
\end{figure}

\Medskip
Actually, $\sigma_2\not=i_2+1$ (as follows from (\ref{eq:Rj})) so that, more precisely, $i_1 \prec i_2 \prec\prec \sigma_2
\prec \sigma_1 \prec 1$. The black dot along the ark from $i_2$ to
$\sigma_2$ indicates any of the species in the non-empty set of species 
${\cal X}_{ext}$ indexed by $i_2+1,\ldots,
\sigma_2-1$. Similarly, the notation $\prec\prec$ denotes the presence of at least one species between $i_2$ and $\sigma_2$, 
so that the newly formed ark ${\cal X}_{ext}=\{i_2+1\prec \cdots\prec \sigma_2-1\}$ is not empty.  

This is contradictory with minimality since the restricted network
$G\Big|_{{\cal X} \setminus {\cal X}_{ext}}$ is strongly connected, and contains the one-to-many reaction $R_1$.

\item[(ii)] $i_1\prec \sigma_1\prec \sigma_2 \prec i_2 $ or
$i_1\prec \sigma_1\prec i_2\prec \sigma_2$
 (both $i_2,\sigma_2$ inserted along $\square_2$) is a nested violation
case, as we can remove any species between $i_1$ and $\sigma_1$ and still get a strongly connected network with a one-to-many reaction ($R_2$).

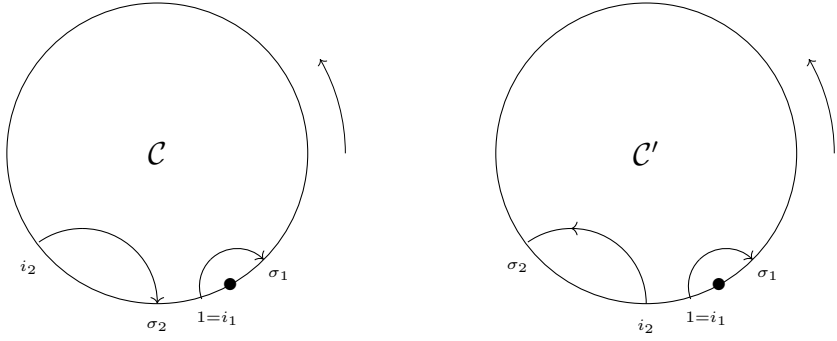
\begin{center} 
\begin{tikzpicture}

\draw(0,0) node {${\cal C}$};
\draw(0,0) circle(2);  \draw[->,ultra thin](2.5,0) arc(0:30:2.5 and 2.5);
\draw[<-](0,-2) arc(0:90:1 and 1);  \draw(0,-2.3) node {\tiny $\sigma_2$};
\draw(-1,-1) arc(90:125:1 and 1);
\draw(-1.71,-1.51) node {\tiny $i_2$};

\draw[<-](1.414,-1.414) arc(45:135:0.5 and 0.5);
\draw(1.414-0.707,-1.414) arc(135:200:0.5 and 0.5);
\draw(1.62,-1.62) node {\tiny $\sigma_1$};
\draw(0.8,-2.2) node {\tiny 1=$i_1$};

\draw(0.9682,-1.75) node {\textbullet}; 

\draw(6.5,0) node {${\cal C}'$};
\draw(6.5,0) circle(2);  \draw[->,ultra thin](9,0) arc(0:30:2.5 and 2.5);
\draw[->](6.5,-2) arc(0:90:1 and 1);  \draw(6.5,-2.3) node {\tiny $i_2$};
\draw(5.5,-1) arc(90:125:1 and 1);
\draw(4.79,-1.51) node {\tiny $\sigma_2$};

\draw[<-](7.914,-1.414) arc(45:135:0.5 and 0.5);
\draw(7.914-0.707,-1.414) arc(135:200:0.5 and 0.5);
\draw(8.12,-1.62) node {\tiny $\sigma_1$};
\draw(7.3,-2.2) node {\tiny 1=$i_1$};

\draw(7.4682,-1.75) node {\textbullet}; 
\end{tikzpicture}
\end{center}

\textbf{Whenever we get a form like $i_j \prec i_k \prec \sigma_k \prec \sigma_j$ (corresponding to one arc along the cycle) or $i_j  \prec \sigma_j \prec i_k \prec \sigma_k$ (corresponding to two arcs along the cycle), we would henceforth refer to it as a nested violation or nested branches ($N_1$ and $N_2$ respectively).} \\  


\item[(iii)] $i_1\prec \sigma_2 \prec i_2 \prec \sigma_1$ ($i_2$, $\sigma_2$ permuted compared to (i)) yields a Type II core
with $\ell$=2.\\

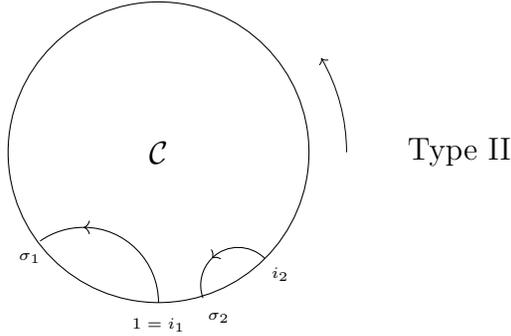
\begin{figure} [h]
\centering
\begin{tikzpicture}
\draw(0,0) node {${\cal C}$};
\draw(0,0) circle(2);  \draw[->,ultra thin](2.5,0) arc(0:30:2.5 and 2.5);
\draw[->](0,-2) arc(0:90:1 and 1);  \draw(0,-2.3) node {\tiny $1=i_1$};
\draw(-1,-1) arc(90:125:1 and 1);
\draw(-1.71,-1.41) node {\tiny $\sigma_1$};

\draw[->](1.414,-1.414) arc(45:135:0.5 and 0.5);
\draw(1.414-0.707,-1.414) arc(135:200:0.5 and 0.5);
\draw(1.62,-1.62) node {\tiny $i_2$};
\draw(0.8,-2.2) node {\tiny $\sigma_2$};

\draw(4,0) node {Type II};
\end{tikzpicture}
\caption{A cycle of Type II with $\ell$=2}
\label{Type2p}
\end{figure}

\item[(iv)] (entangled violation $E$ case)  $i_1\prec i_2 \prec\prec \sigma_1
\prec \sigma_2$ ($i_2$ inserted along $\square_1$, $\sigma_2$ along $\square_2$,  assuming further that  $\sigma_1\not=
i_2+1$) 
  We then reach the same conclusion as in (i) with ${\cal X}_{ext}$ indexed by $i_2+1,\ldots,\sigma_1-1$: the restricted network
$G\Big|_{{\cal X} \setminus {\cal X}_{ext}}$ is strongly connected, and contains the one-to-many reaction $R_1$, therefore $G$ is 
not minimal. 

\Medskip By extension, 
for any $j\ge 2$, we shall call {\bf entangled violation} (of 
minimality) denoted by $E$, any pair configuration such that, for some pair
of indices $1\le a\not=b\le j$,
 \BEQ i_a\prec i_b\prec\prec \sigma_a\prec \sigma_b \qquad
 {\mathrm{(entangled\  violation)}}
 \EEQ

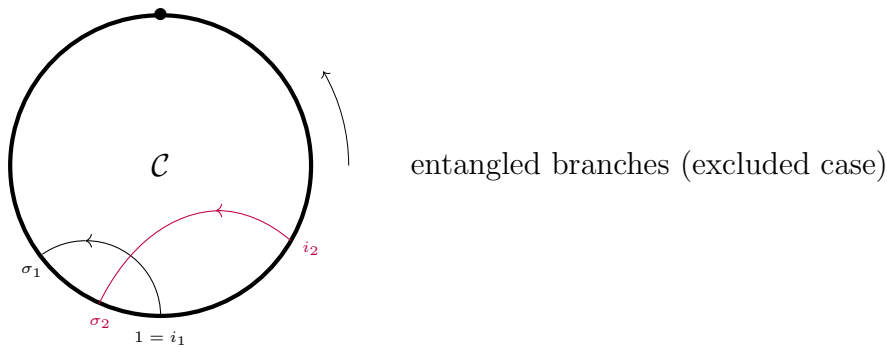
\Medskip

\begin{center}
\begin{tikzpicture}
\draw(0,0) node {${\cal C}$};
\draw[ultra thick](0,0) circle(2);  \draw[->,ultra thin](2.5,0) arc(0:30:2.5 and 2.5);
\draw[->](0,-2) arc(0:90:1 and 1);  \draw(0,-2.3) node {\tiny $1=i_1$};
\draw(-1,-1) arc(90:125:1 and 1);
\draw(-1.71,-1.41) node {\tiny $\sigma_1$};

\draw[->,purple](2*0.866,-2*0.5) arc(60:90:2 and 3);
\draw[purple](0.8,-0.6) arc(90:144:2 and 3);
\draw[purple](2,-1.1) node {\tiny $i_2$};
\draw[purple](-0.8,-2.1) node {\tiny $\sigma_2$};

\draw(0,2) node {\textbullet}; 

\draw(6.5,0) node {entangled branches (excluded case)};
\end{tikzpicture}
\end{center}


\Medskip {\em Remark.} Nested configurations are sensitive to
relative pair orientations, e.g. exchanging $i_2$ and $\sigma_2$ turns
two nested branches (Fig. \ref{nested} above) into a Type II core with
$\ell=2$ (Fig. \ref{Type2p}). So are entangled configurations, but note
that 
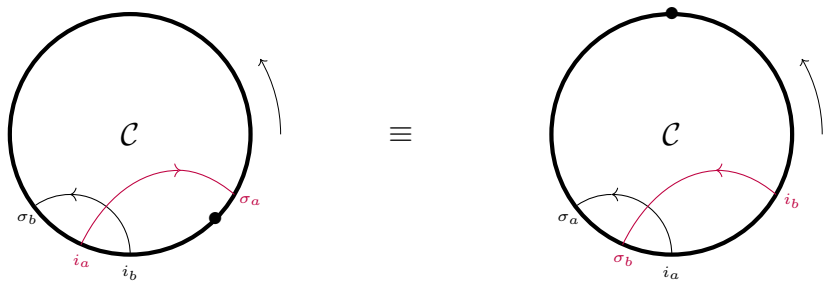
\Medskip

\begin{center}
\begin{tikzpicture}[scale=0.8]
\draw(0,0) node {${\cal C}$};
\draw[ultra thick](0,0) circle(2);  \draw[->,ultra thin](2.5,0) arc(0:30:2.5 and 2.5);
\draw[->](0,-2) arc(0:90:1 and 1);  \draw(0,-2.3) node {\tiny $i_b$};
\draw(-1,-1) arc(90:125:1 and 1);
\draw(-1.71,-1.41) node {\tiny $\sigma_b$};

\draw[-<,purple](2*0.866,-2*0.5) arc(60:90:2 and 3);
\draw[purple](0.8,-0.6) arc(90:144:2 and 3);
\draw[purple](2,-1.1) node {\tiny $\sigma_a$};
\draw[purple](-0.8,-2.1) node {\tiny $i_a$};

\draw(1.414,-1.414) node {\textbullet}; 

\draw(4.5,0) node {$\equiv$};

\begin{scope}[shift={(9,0)}]
\draw(0,0) node {${\cal C}$};
\draw[ultra thick](0,0) circle(2);  \draw[->,ultra thin](2.5,0) arc(0:30:2.5 and 2.5);
\draw[->](0,-2) arc(0:90:1 and 1);  \draw(0,-2.3) node {\tiny $i_a$};
\draw(-1,-1) arc(90:125:1 and 1);
\draw(-1.71,-1.41) node {\tiny $\sigma_a$};

\draw[->,purple](2*0.866,-2*0.5) arc(60:90:2 and 3);
\draw[purple](0.8,-0.6) arc(90:144:2 and 3);
\draw[purple](2,-1.1) node {\tiny $i_b$};
\draw[purple](-0.8,-2.1) node {\tiny $\sigma_b$};

\draw(0,2) node {\textbullet}; 
\end{scope}
\end{tikzpicture}
\end{center}

\Medskip By moving around $\sigma_a$ along the cycle, relative
pair orientations of entangled pairs seem reversed (but the black dot, indicating 
that the ark from $i_b$ to $\sigma_a$ is not empty, is not at the
same location). 

\item[(v)]   $i_1\prec i_2 \prec \sigma_1
\prec \sigma_2$ as in (iv), but  (special case) $\sigma_1=i_2+1$.  This
is a  Type IV core (see Fig.  in \S \ref{section:main} (iv)),
with the identification $\{i_1\prec i_2\prec \sigma_1=i_2+1
\prec \sigma_2\prec i_1\} \leftrightarrow \{v\prec w \prec x'
\prec u\prec v\}$ (the cycle containing all species is uncovered by dropping the
edges $v\to x'$ and $w\to u$).

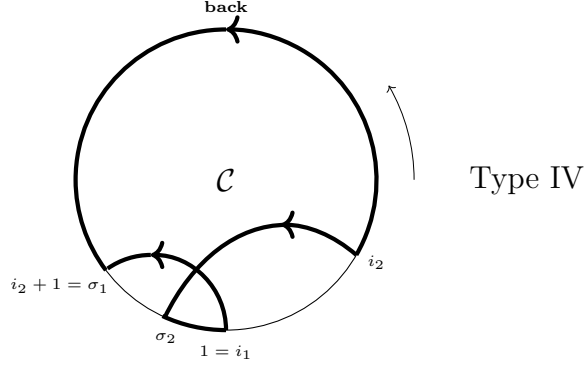
\Medskip

\begin{center}
\begin{tikzpicture}
\draw(0,0) node {${\cal C}$};
\draw(0,0) circle(2);
\draw[ultra thick,->](1.732,-2*0.5) arc(-30:90:2 and 2);
\draw[ultra thick](0,2) arc(90:217:2 and 2);
  \draw[->,ultra thin](2.5,0) arc(0:30:2.5 and 2.5);
\draw[->, ultra thick](0,-2) arc(0:90:1 and 1);  \draw(0,-2.3) node {\tiny $1=i_1$};
\draw[ultra thick](-1,-1) arc(90:125:1 and 1);
\draw(-2.21,-1.41) node {\tiny $i_2+1=\sigma_1$};
\draw[ultra thick](0,-2) arc(-90:-115:2 and 2);

\draw[->,ultra thick](2*0.866,-2*0.5) arc(60:90:2 and 3);
\draw[ultra thick](0.8,-0.6) arc(90:144:2 and 3);
\draw(2,-1.1) node {\tiny $i_2$};
\draw(-0.8,-2.1) node {\tiny $\sigma_2$};

\draw(0,2.3) node {\bf \tiny back};

\draw(4,0) node {Type IV};
\end{tikzpicture}
\end{center}

\item[(vi)] $i_1\prec \sigma_2\prec \sigma_1 \prec i_2 $  ($i_2$ inserted along $\square_2$, $\sigma_2$ along $\square_1$) Exchanging indices 1,2 yields back cases (iv)-(v).

\end{itemize}

\medskip

In the following, we add inductively pairs of indices 
$(i_j,\sigma_j)$, $j=3,4,\ldots$   For $j=2$, we got only
two cores (Types II and IV), all other possibilities being 
non-minimal autocatalytic. Clearly,  adding another multiple reaction $(j=3)$ to a non-minimal autocatalytic network cannot
yield a minimal autocatalytic network, so that we may content ourselves with starting from a Type II or Type IV core and
adding a pair of indices $(i_3,\sigma_3)$, discarding systematically nested and entangled violations. 

\Medskip It proves convenient to postpone the discussion 
of the relatively simple case when $i_j$ and $\sigma_j$ are located along the same ark (see $\S \ref{ark}$  below). So we assume
for now that they belong to two different arks (a condition
stipulated simply as 'different arks').


\subsubsection{Three one-to-many reactions, different arks}
\label{subsubsection:3-one-to-many-different-arks}

\label{threere}
We  start either from a Type II or a Type IV core and add
a couple of indices $(i_3,\sigma_3)$ located on two different arks.

\begin{figure}
\centering
\begin{tikzpicture}
\draw(0,0) node {${\cal C}$};
\draw(0,0) circle(2);  \draw[->,ultra thin](2.5,0) arc(0:30:2.5 and 2.5);

\draw[red](0,-2) arc(270:286:2);
\draw[red](0.3,-1.6) node {$\square_1$};
\draw[blue](0.67,-1.9) arc(286:317:1.7);
\draw[blue](1.35,-2.05) node {$\square_2$};
\draw[red](1.40,-1.40) arc(-45:217:2);
\draw[red](0,2.3) node {$\square_3$};
\draw[blue](-1.55,-1.25) arc(218:268:2);
\draw[blue](-1.3,-2.0) node {$\square_4$};

\draw[->](0,-2) arc(0:90:1 and 1);  \draw(0,-2.3) node {\tiny $1=i_1$};
\draw(-1,-1) arc(90:125:1 and 1);
\draw(-1.71,-1.41) node {\tiny $\sigma_1$};

\draw[->](1.414,-1.414) arc(45:135:0.5 and 0.5);
\draw(1.414-0.707,-1.414) arc(135:200:0.5 and 0.5);
\draw(1.62,-1.62) node {\tiny $i_2$};
\draw(0.8,-2.2) node {\tiny $\sigma_2$};

\draw(4,0) node {Type II};
\end{tikzpicture}
\caption{A cycle of Type II with $\ell$=2 and arks}
\label{Type2}
\end{figure}
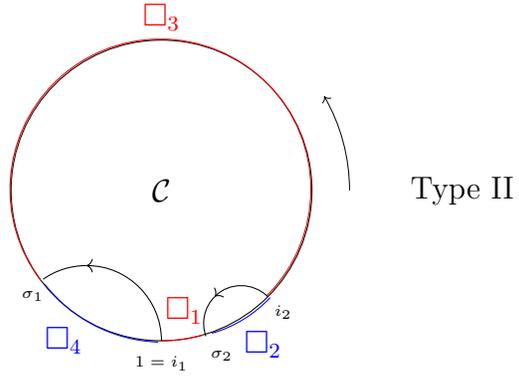

\Medskip \textbf{A. Starting from Type $II_2$ (see Fig. \ref{Type2}).} 
The ordering along the circle is: 
\BEQ i_1 \prec \square_1 \prec \sigma_2 \prec \square_2 \prec i_2 \prec \square_3 \prec \sigma_1\prec \square_4 \prec i_1. 
\EEQ
We discuss the various possibilities, to conclude that
no new cores can be formed:

\begin{itemize}
\item[(i)] ($i_3\in \square_1$) Then necessarily  $\sigma_3\in \square_4$ (otherwise $\sigma_3\in \square_2$ or $\square_3$, and   -- discarding the pair $(i_2,\sigma_2)$ -- one gets nested
branches).  But then the pairs of indices $(i_1,\sigma_1),(i_3,\sigma_3)$  make an entangled violation. Similarly if $i_3\in \square_3$, by exchanging $(i_1,\sigma_1)$ and $(i_2,\sigma_2)$.  

\item[(ii)] ($i_3\in\square_2$) Then $\sigma_3\in \square_1$ (if $\sigma_3\in \square_3$, $(i_1,\sigma_1)$ and $(i_3,\sigma_3)$ form nested branches and if $\sigma_3\in\square_4$, the same pair forms an entangled violation). But then
the pairs of indices $(i_2,\sigma_2)$ and $(i_3,\sigma_3)$ make an entangled violation. Similarly if $i_3\in \square_4$, by exchanging $(i_1,\sigma_1)$ and $(i_2,\sigma_2)$. 

\end{itemize}

\Bigskip

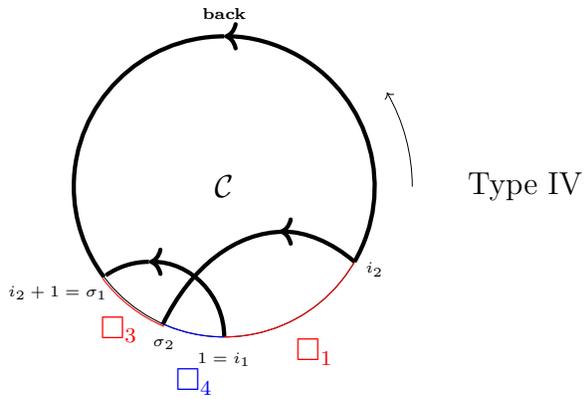
\begin{figure}
\centering
\begin{tikzpicture}
\draw(0,0) node {${\cal C}$};
\draw(0,0) circle(2);
\draw[ultra thick,->](1.732,-2*0.5) arc(-30:90:2 and 2);
\draw[ultra thick](0,2) arc(90:217:2 and 2);
  \draw[->,ultra thin](2.5,0) arc(0:30:2.5 and 2.5);
\draw[->, ultra thick](0,-2) arc(0:90:1 and 1);  \draw(0,-2.3) node {\tiny $1=i_1$};
\draw[ultra thick](-1,-1) arc(90:125:1 and 1);
\draw(-2.21,-1.41) node {\tiny $i_2+1=\sigma_1$};

\draw[blue](0,-2) arc(-90:-115:2 and 2);
\draw[red](1.2,-2.2) node {$\square_1$};
\draw[red](-0.8,-1.86) arc(-113:-143:2);
\draw[red](-1.4,-1.9) node {$\square_3$};
\draw[red](0,-2) arc(-90:-30:2);
\draw[blue](-0.4,-2.6) node {$\square_4$};

\draw[->,ultra thick](2*0.866,-2*0.5) arc(60:90:2 and 3);
\draw[ultra thick](0.8,-0.6) arc(90:144:2 and 3);
\draw(2,-1.1) node {\tiny $i_2$};
\draw(-0.8,-2.1) node {\tiny $\sigma_2$};

\draw(0,2.3) node {\bf \tiny back};

\draw(4,0) node {Type IV};
\end{tikzpicture}
\caption{A cycle of Type IV with arks}
\label{Type4}
\end{figure}


\Medskip \textbf{B. Starting from Type IV  (see Fig. \ref{Type4}).} Since $\sigma_1=i_2+1$,  $\square_2=\emptyset$, and the
ordering is now
\BEQ i_1 \prec \square_1 \prec i_2 \prec \sigma_1 \prec \square_3 \prec \sigma_2\prec \square_4 \prec i_1
\EEQ

\Medskip Discussing the various possibilities, we prove that only
Type V cores can be formed this way:

\begin{itemize}
\item[(i)]  ($i_3\in \square_1$)  If $\sigma_3\in \square_4$, the pairs $(i_2,\sigma_2)$ and $(i_3,\sigma_3)$ form nested branches. On the other hand, if $\sigma_3\in \square_3$, the same pairs form an entangled violation.
\item[(ii)] ($i_3\in\square_3$) If $\sigma_3\in\square_4$, then   pairs $(i_1,\sigma_1)$ and $(i_3,\sigma_3)$ form a nested violation ($N_2$). Hence
necessarily  $\sigma_3 \in\square_1$, so that
\BEQ i_1 \prec \sigma_3 \prec i_2 \prec \sigma_1 \prec i_3 \prec \sigma_2. 
\EEQ
 Here there is no outright violation of any case, but we note that the pairs $(i_2,\sigma_2)$ and $(i_3,\sigma_3)$, and also the pairs $(i_1,\sigma_1)$ and $(i_3,\sigma_3)$, take the form of Section 3.2.1 cases (iv)-(v). This means that we must have $\sigma_1=i_2+1$ (from the 1,2 pairs), $\sigma_3=i_1+1$ (from the 3,1 pairs) and $\sigma_2=i_3+1$ (from the 2,3 pairs). This is a Type V core, see  Fig. in \S \ref{section:main} (v), with the identification $\{i_1\prec 
 \sigma_3=i_1+1\prec i_2 \prec \sigma_1=i_2+1 \prec i_3\prec \sigma_2=i_3+1\prec i_1\} \leftrightarrow \{v\prec x \prec w \prec x'\prec w' \prec u \prec v\}$. The cycle connecting all species
is uncovered by dropping the edges $v\to x'$, $w\to u$ and the
backbranch $w'\to x$.
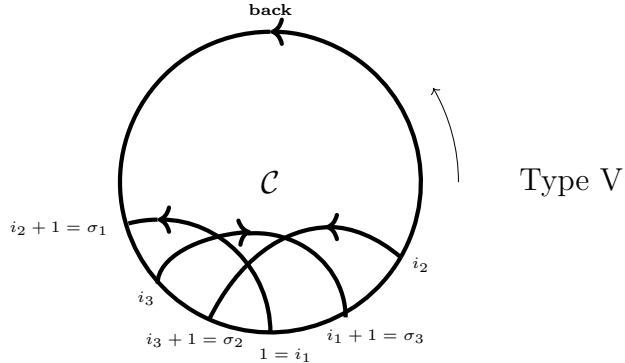
\begin{center}
\begin{tikzpicture}
\draw(0,0) node {${\cal C}$};
\draw(0,0) circle(2);
\draw[ultra thick,->](1.732,-2*0.5) arc(-30:90:2 and 2);
\draw[ultra thick](0,2) arc(90:198:2 and 2);
  \draw[->,ultra thin](2.5,0) arc(0:30:2.5 and 2.5);
\draw[->, ultra thick](0,-2) arc(0:90:1.5 and 1.5);  \draw(0.2,-2.3) node {\tiny $1=i_1$};

\draw[ultra thick](1,-1.8) arc(0:95:1.2 and 1.12); \draw(1.4,-2.05) node {\tiny $i_1+1=\sigma_3$};
\draw[->, ultra thick](-1.5,-1.35) arc(180:100:1.5 and 0.7);\draw(-1.65,-1.55) node {\tiny $i_3$};

\draw[ultra thick](-1.5,-0.5) arc(90:105:1.5 and 1.5);
\draw(-2.81,-0.61) node {\tiny $i_2+1=\sigma_1$};
\draw[ultra thick](1.732,-1) arc(-30:-162:2 and 2);

\draw[->,ultra thick](2*0.866,-2*0.5) arc(60:90:2 and 3);
\draw[ultra thick](0.8,-0.6) arc(90:144:2 and 3);
\draw(2,-1.1) node {\tiny $i_2$};
\draw(-1,-2.1) node {\tiny $i_3+1=\sigma_2$};

\draw(0,2.3) node {\bf \tiny back};

\draw(4,0) node {Type V};
\end{tikzpicture}
\end{center}

\item[(iii)] ($i_3\in \square_4$)  If $\sigma_3\in \square_1$, pairs $(i_2,\sigma_2)$ and $(i_3,\sigma_3)$ form nested branches ($N_2$).  On the other hand, if $\sigma_3\in \square_3$, these pairs form an entangled violation.
\end{itemize}


\subsubsection{Four one-to-many reactions, different arks}


Starting from the Type V core produced in \S \ref{subsubsection:3-one-to-many-different-arks} {\bf B.} (ii), we add a  new pair $(i_4,\sigma_4)$  in different arks. The ordering along the cycle is now
(see previous Figure)

\BEQ i_1 \prec  \sigma_3=i_1+1 \prec \square_2 \prec i_2 \prec \sigma_1=i_2+1 \prec \square_4 \prec i_3 \prec \sigma_2=i_3+1 
\prec \square_6 \prec i_1 \EEQ
with $\square_1, \square_3,\square_5=\emptyset$. 
Discussing the various possibilities, we find that no new core
can be formed this way:

\begin{itemize}
\item[(i)] ($i_4\in\square_2$) If  $\sigma_4\in\square_4$, the pairs $(i_4,\sigma_4)$ and $(i_3,\sigma_3)$ form a nested violation ($N_2$). If $\sigma_4\in \square_6$, then the pairs $(i_2,\sigma_2)$ and $(i_4,\sigma_4)$ form a nested violation ($N_1$). 
\item[(ii)] ($i_4\in\square_4$) If  $\sigma_4\in\square_2$, the pairs $(i_4,\sigma_4)$ and $(i_3,\sigma_3)$ form a nested violation ($N_1$). If $\sigma_4\in \square_6$, pairs $(i_1,\sigma_1)$ and $(i_4,\sigma_4)$ form a nested violation ($N_2$). 
\item[(iii)] ($i_4\in \square_6$) If $\sigma_4\in\square_2$, then pairs $(i_2,\sigma_2)$ and $(i_4,\sigma_4)$ form a nested violation ($N_2$). If $\sigma_4\in\square_4$, then pairs $(i_1,\sigma_1)$ and $(i_4,\sigma_4)$ form a nested violation ($N_1$). 

\end{itemize}


\subsubsection{Left over case (see end of \ref{threetwoone})}
\label{ark}

The left over case is obtained starting from Type II or Type IV and inserting the pair $i_j,\sigma_j$ in the same arc. This cannot be done in the ordered pair $i_j \prec \sigma_j $ ($\S \ref{threetwoone}$ $N_1$). Thus it is done in the form of the ordered pair $\sigma_j \prec i_j$ (i.e. opposite in direction to the cycle).\\ 

\medskip

For this insertion into $\square_i$, a cyclic permutation of the indices will not result in N$_1$ violation only if starting from $\square_i$, all $\sigma_k$ (k$\textless$ j) are encountered before their corresponding $i_k$.

\begin{itemize}
    \item For type IV cycles ($\S \ref{threere}$), there is no such ark that can satisfy this condition. \\
 \item For type II with $\ell=$2, where we had $i_1 \prec \square_1 \prec \sigma_2 \prec \square_2 \prec i_2 \prec \square_3 \prec \sigma_1\prec \square_4$, we can insert this pair in $\square_1$ or $\square_3$ which will both yield the same notation except for an exchange of labels for reactions 2 and 3. We will get\\
$i_1 \prec \sigma_2 \prec i_2 \prec \sigma_3 \prec i_3 \prec \sigma_1$\\
Which is a Type II reaction with $\ell$=3.
\end{itemize}

\bigskip

We can keep inserting such $\sigma_j \prec i_j$ pairs in such a notation right after any $i_k$, k$\textless$j, without violating the minimality condition. We get the general notation for the type II$_{\ell}$ cycle -\\
$i_1 \prec \sigma_2 \prec i_2 \prec \sigma_3 \prec i_3 \prec .....\prec \sigma_{\ell} \prec i_{\ell}\prec \sigma_1$\\
\bigskip\\
If on the other hand we try inserting $\sigma_{\ell+1}$ and $i_{\ell+1}$ into this general notation at separate places (for $\ell \textgreater 2$), we will always get a violation. This is shown below.\\
\medskip\\
Assume on the contrary that you could insert $\sigma_{\ell+1}$ and $i_{\ell+1}$ in a cycle of type II$_{\ell}$, but not in the form $\sigma_{\ell+1} \prec i_{\ell+1}$ (or $i_{\ell+1} \prec \sigma_{\ell+1}$, which is always a nesting violation). This implies that there are necessarily other species in between $\sigma_{\ell+1}$ and $i_{\ell+1}$ and also between $i_{\ell+1}$ and $\sigma_{\ell+1}$. We select indices k1 and k2, such that either $i_{k1}$ or $\sigma_{k1}$ (but not both) is between $\sigma_{\ell+1}$ and $i_{\ell+1}$ and either $i_{k2}$ or $\sigma_{k2}$ (but not both) is between $i_{\ell+1}$ and $\sigma_{\ell+1}$ in the notation. 

Then the loop takes the form:
\BEQ i_{k1} \prec \square_1 \prec \sigma_{k2} \prec \square_2 \prec i_{k2} \prec \square_3 \prec \sigma_{k1}\prec \square_4 \prec i_{k1}. 
\EEQ
With $i_{\ell+1}$ and $\sigma_{\ell+1}$ filling up different squares.
This is identical to $\S \ref{threere}$ Part 1 except for a change of index (there also, we had to insert an entangled branch in a Type II cycle). Since all the cases there were a violation of minimality (E violation), this case also violates minimality.\\

\Medskip
\begin{center}
\begin{tikzpicture}

\draw(-2,0) arc(-180:0:2 and 2);
\draw[dashed](-2,0) arc(-180:-210:2 and 2);
\draw[dashed](2,0) arc(0:30:2 and 2);
  \draw[->,ultra thin](2.5,0) arc(0:30:2.5 and 2.5);
\draw[->](0,-2) arc(0:90:1 and 1);  \draw(0,-2.3) node {\tiny $i_1$};
\draw(-1,-1) arc(90:125:1 and 1);
\draw(-1.71,-1.41) node {\tiny $\sigma_1$};

\draw[->](1.414,-1.414) arc(45:135:0.5 and 0.5);
\draw(1.414-0.707,-1.414) arc(135:200:0.5 and 0.5);
\draw(1.62,-1.62) node {\tiny $i_2$};
\draw(0.8,-2.2) node {\tiny $\sigma_2$};

\draw(2,-1) node {\tiny $i_3$};
\draw[dashed,purple](2*0.866,-2*0.5) arc(60:210:0.5 and 0.5);
\draw[dashed,purple](2*0.866,-2*0.5) arc(60:150:2 and 2);
\draw[dashed,red](2*0.866,-2*0.5) arc(60:180:1 and 1.25);

\begin{scope}[shift={(8,0)}]
\draw(-2,0) arc(-180:0:2 and 2);
\draw[dashed](-2,0) arc(-180:-210:2 and 2);
\draw[dashed](2,0) arc(0:30:2 and 2);
  \draw[->,ultra thin](2.5,0) arc(0:30:2.5 and 2.5);
\draw[->](0,-2) arc(0:90:1 and 1);  \draw(0,-2.3) node {\tiny $i_1$};
\draw(-1,-1) arc(90:125:1 and 1);
\draw(-1.71,-1.41) node {\tiny $\sigma_1$};

\draw[->](1.414,-1.414) arc(45:135:0.5 and 0.5);
\draw(1.414-0.707,-1.414) arc(135:200:0.5 and 0.5);
\draw(1.62,-1.62) node {\tiny $i_2$};
\draw(0.8,-2.2) node {\tiny $\sigma_2$};

\draw(-1,-2) node {\tiny $i_3$};
\draw[dashed,purple](-2*0.5,-2*0.866) arc(120:60:2 and 2);
\draw[dashed,purple](-2*0.5,-2*0.866) arc(120:45:1 and 1);
\draw[dashed,purple](-2*0.5,-2*0.866) arc(0:60:2 and 2);

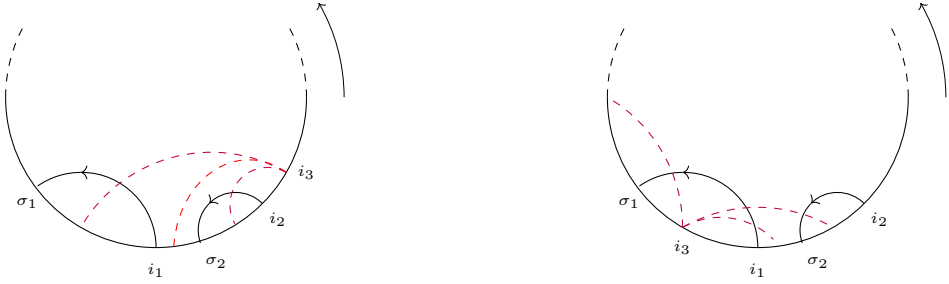
\draw(-4,-3) node {No entangled branches possible for type II $\ell \textgreater$ 1};

\end{scope}
\end{tikzpicture}
\end{center}
\bigskip
Thus any minimal, strongly connected cycle with more than three one-to-many reactions has to be of Type II, and up to the case of Three one-to-many reactions we have shown that minimality is violated for a single cycle apart from the types discussed in the classification theorem.


\subsubsection{Case 3 (two or more cycles)}


Barring Case 1 (see \S \ref{subsection:case1}), $G$ contains a multiple reaction 
\BEQ R_1:\ x_1\to s_2 x_2 +
s'_2 x'_2 + \cdots
\EEQ
 Choose a cycle ${\cal C}=\{x_1\to x_2\to \ldots\to 
x_{\ell+1}=x_1\}$ passing through the edge
$x_1\to x_2$. Barring Case 2 (see \S \ref{subsection:case2}), ${\cal C}$ does not contain all species. 
The case when ${\cal C}$ contains $x'_2$ is excluded, because the
restricted network
$({\cal C}, {\cal R}\Big|_{{\cal C}})$ is irreducible, satisfies (Top) and 
is strictly included in $G$; thus $x'_2\not\in {\cal C}$.

\Medskip Consider an ear 
\BEQ {\cal O}: \qquad v=x_1\to x'_2\to \cdots\to x'_{k'}\to 
x'_{k'+1}=u 
\EEQ
$(x'_2,\ldots,x'_{k'}\not\in {\cal C}, u\in {\cal C})$  built on $\cal C$ and starting
with the edge $x_1\to x'_2$,

\Medskip
\begin{center}
\begin{tikzpicture}
\draw(0,0) circle(2);  \draw(0,0) node {${\cal C}$};
\draw(-4*0.707,0) node {${\cal C}'$};
  \draw[<-,ultra thin](2.5,0) arc(0:30:2.5 and 2.5);
\draw(-4*0.707+1.414 ,1.414) arc(45:315:2 and 2);
\draw[<-,ultra thin] (-4*0.707-2.5,0) arc(180:150: 2.5 and 2.5);

\draw[ultra thick](-2*0.707,2*0.707) arc(135:120:2 and 2);
\draw[ultra thick](-2*0.707,2*0.707) arc(45:60:2 and 2);
\draw(-0.2,1.2) node {\tiny $v=x_1=x'_1=x_{\ell+1}$};
\draw(-2*0.707,2*0.707) node {\textbullet};
\draw(-0.2,-1.4) node {\tiny $u=x_{k+1}=x'_{k'+1}$};
\draw(-0.8,2.1) node {\tiny $x=x_2$};
\draw(-2.4,1.6) node {\tiny $x'=x'_2$};

\end{tikzpicture} 
\end{center}

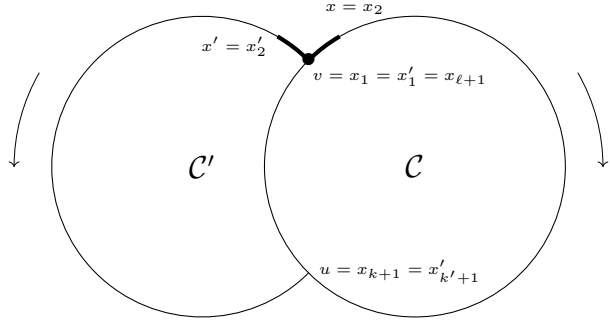
\Medskip Since $u\in {\cal C}$, there exists $k\in\{1,\ldots,\ell\}$ such that
$u=x_{k+1}$. Let $x'_1:=x_1$. Then 
\BEQ {\cal C}':\qquad v = x'_1\to x'_2\to \cdots \to x'_{k'} \to u
\to x_{k+2}\to \cdots\to x_{\ell}\to v
\EEQ
is a second cycle.
Since the restricted network $({\cal C}\cup {\cal C}', {\cal R}\Big|_{{\cal C}\cup
{\cal C}'})$ is irreducible and  satisfies (Top), 
\BEQ {\cal X}={\cal C}\cup {\cal C}' \EEQ
 by
minimality. Similarly, simple reactions along $\cal C$, resp. 
${\cal C}'$, are 1--1, otherwise $G\Big|_{{\cal C}}$, resp. 
$G\Big|_{{\cal C}'}$ would be irreducible and satisfy (Top). 
For the same reason, $s_2=1$, resp. $s'_2=1$.

\Medskip As in Case 2 (see \S \ref{subsection:case2}), one notes that simple or multiple reactions  $x\to s_1 y_1+\cdots + s_{m} y_{m}$ such that {\em none of the split reactions $x\to y_i,i=1,\ldots,m$ is one of the reactions $x_j\to x_{j+1}$, resp.  $x'_j\to x'_{j+1}$ along ${\cal C}$, resp. ${\cal C}'$}, are 
redundant, i.e. contradict minimality. Also, multiple reactions
$x\to s_1 y_1+\cdots + s_{m} y_{m}$ $(s_1+\ldots+s_m\ge 2)$  with all $x,y_1,\ldots,y_{m}\in{\cal C}$, resp. ${\cal C}'$, would contradict minimality since
$G\Big|_{{\cal C}}$, resp. $G\Big|_{{\cal C}'}$ would be irreducible
and satisfy (Top).  Thus all there remains
to do is to discuss possible {\em cycle-mixing reactions} (other
than $R_1$), namely,  
multiple reactions  with $m'\ge 2$ different products,
\BEQ a\to sb+s' c + \cdots \qquad (m' \ {\mathrm{products}}) 
\label{eq:cycle-mixing} 
\EEQ
with species $a,b,c$ not belonging to the same cycle, either 
\BEQ (1) \qquad x_j \to sx_{j+1}+ s' x'_{j'} + \cdots, \ \ j\in \{2,\ldots,k\}, \ j'\in\{2,\ldots,k'\}  
\EEQ
or
\BEQ
(2)\qquad  x'_{j'}\to s' x'_{j'+1} + s x_j + \cdots, \ \ j'\in\{2,
\ldots,k'\}, \ j\in \{2,\ldots,k\}
\EEQ 
One has excluded reactions of type (1) with $j=1$, and similarly 
reactions of type (2) with $j'=1$, because they are redundant
with $R_1$.   The two cases are clearly symmetric. One may further eliminate
 cycle-mixing reactions of type (1) with $j\in\{2,\ldots,k-1\}$; namely, if $j<k$, 
species in the set  ${\cal X}_{ext}:=\{x_{j+1},\ldots,x_k\}$
along ${\cal C}$ may be chemostatted (see Definition
\ref{def:restricted}) because the path $x_j\to x'_{j'} 
\overset{{\cal C}'}{\to} \cdots
\overset{{\cal C}'}{\to} x_{k+1}$ connects $x_j$ to $x_{k+1}$, whence the restricted network $G\Big|_{{\cal X}
\setminus {\cal X}_{ext}}$ is irreducible autocatalytic, contradicting minimality.  Similarly (still considering a reaction of type 1),
if $j'>2$, 
${\cal X}'_{ext}:=\{x'_2,\ldots,x'_{j'-1}\}$  along ${\cal C}'$ can be chemostatted because
the path $v=x_1\overset{{\cal C}'}{\to} x_2 \overset{{\cal C}'}{\to} \cdots\to x_j\to x'_{j'}$   connects $v$ to $x'_{j'}$ (the reaction $R_1$ is then turned into a simple reaction, but the
restricted network $G\Big|_{{\cal X}
\setminus {\cal X}'_{ext}}$ remains autocatalytic thanks to 
the cycle-mixing reaction (1)). All together, we see that only the cycle-mixing reaction
 \BEQ (1) \qquad  x_k\to su+ s' x'_2,  \EEQ
forming the back-branch 
\BEQ x_k\to x'_2, \EEQ
 is possible. Similarly, only the cycle-mixing reaction
\BEQ (2) \qquad x'_{k'}\to s' u + s x_2, \EEQ
forming the back-branch 
\BEQ x'_{k'}\to x_2, \EEQ
is possible (since no other products are possible, we have simultaneously proved that $m'=2$ in (\ref{eq:cycle-mixing})). As
above for the reaction $R_1$, minimality imposes $s=s'=1$ in (1)
and in (2).  This yields three different cases.

\subsubsection{No cycle-mixing reaction}

In absence of reactions (1), (2), one gets Type III. (This 
is the only case for which there is no single cycle connecting all
species, which explains why we have not found this case previously in the proof.)

\subsubsection{One cycle-mixing reaction}

Including either (1) or (2) but not both  (the two cases are equivalent by
symmetry), one gets Type IV, with its single back-branch.

\subsubsection{One cycle-mixing reaction}

Including both (1) and (2), one gets Type V, with its two
back-branches.  

\medskip

\medskip

\subsection{'Hand-made argument': Lyapunov function}
Theorem C.1 (ii) in \cite{Fei95}
is based on a general  convexity inequality (Proposition 5.3), and
on the fact that a certain relation holds in an abstract vector space
$\R^{\cal C}$ with standard basis $(\omega_y)_{y\in{\mathrm{Compl}}}$
indexed by the set of complexes, namely, 
\BEQ \sum_R \phi_{y_R\to y'_R}(C_{stat}) (\omega_{y'_R} - \omega_{y_R})=0. \label{eq:phiomega}
\EEQ
 In general, only
the stationary property, $\sum_R \phi_{y_R\to y'_R}(C_{stat})(y'_R-y_R)=0$ holds, which implies (\ref{eq:phiomega}) by applying
the linear operator 
\BEQ Y:  \R^{{\mathrm{Compl}}} \to \R^{\cal X}, \ \  \omega_y \mapsto y;
\label{eq:Y}
\EEQ
 precisely, it is proved in \cite{Fei95} that
the kernel of $Y$ restricted to span$(\omega_{y'_R}-\omega_{y_R}, 
R\in {\cal R})$ has dimension $\del$; thus (\ref{eq:phiomega}) 
follows if the deficiency index $\del$ is zero.   However, (\ref{eq:phiomega})  is actually  obvious if $C_{stat}$ is an equilibrium state; namely,
letting $(R,\bar{R})$ be any couple (direct reaction, reverse reaction),
the equilibrium property for $C_{stat}$ implies $\phi_{y_R\to y'_R}(C_{stat}) (\omega_{y'_R} - \omega_{y_R}) +\phi_{y_{\bar{R}}\to y'_{\bar{R}}}(C_{stat}) (\omega_{y'_{\bar{R}}} - \omega_{y_{\bar{R}}})  = (\phi_{y_R\to y'_R}(C_{stat}) -\phi_{y'_R\to y_R}(C_{stat}))(\omega_{y'_R} - \omega_{y_R})  =0$ by (\ref{eq:antisymmetry}). Let us prove briefly that $h$ is
 a Lyapunov function in this particular case for the sake of the reader. 
First it is well-known (and follows from strict concavity
 of the logarithm) that $h(C)\ge 0$ and vanishes only for $C=C_{stat}$, see e.g. Appendix C in \cite{Fei95}.   We let $\langle \, \cdot\, , \cdot\,  
\rangle$ be the standard scalar product in $\R^{{\cal X}}$.
The generalized relative entropy function $h$ satisfies the following
identity,
\BEQ \phi_{y\to y'}(C)= \phi_{y\to y'}(C_{stat}) \, \times\,  \prod_{i=1}^n
(C_i/C_{stat,i})^{y_i}  =   \phi_{y\to y'}(C_{stat}) \, \times\, 
e^{\langle y,\nabla h (C)\rangle}
\EEQ
  Now the convexity inequality $e^a (b-a)\le e^b - e^a,\ a,b\in\R$
  implies 
\BEA \frac{dh}{dt} &=& \langle \frac{dC}{dt}, \nabla h(C)\rangle \\
&=& \sum_{y\to y'}  \phi_{y\to y'}(C) \langle    y'-y,   \nabla h(C) \rangle
\nonumber\\
&=& \sum_{y\to y'}    \phi_{y\to y'}(C_{stat})  
e^{\langle y,\nabla h(C)\rangle} \, \langle y'-y,   \nabla h(C) \rangle
\nonumber\\
&\le & \sum_{y\to y'} \phi_{y\to y'}(C_{stat}) \, (e^{\langle y',\nabla
h(C)\rangle} - e^{\langle y,\nabla h(C)\rangle}) \nonumber\\
&=& \half \sum_{y\to y'} (\phi_{y\to y'}(C_{stat})- \phi_{y'\to y}(C_{stat})) \, (e^{\langle y',\nabla
h(C)\rangle} - e^{\langle y,\nabla h(C)\rangle}) = 0.
\EEA

The second line comes from (\ref{eq:Cphiy'-y}).  The last line comes by antisymmetrizing and using (\ref{eq:antisymmetry}), together with the
fact that $C_{stat}$ is an equilibrium state.  The inequality
is strict except if $C=C_{stat}$.

\subsection{Explicit formulas for the equilibrium state}\label{sec:exp}

We use the classification theorem (Theorem
\ref{th:main}) to give explicit formulas for 'minimal' stoichiometric
coefficients (all coefficients equal to $0$ or $1$, except
$s_n=2$ for the replication reaction $x_n\to x_1$ in Type I). 

\begin{itemize}
\item[(i)] (Type I) Setting to zero the currents associated to the reactions $x_j\overset{k_j^+}{\underset{k_j^-}{\rightleftarrows}}  x_{j+1}$, $j=1,\ldots,n-1$ along the branch
\BEQ x_1 \to x_2 \to\cdots \to
x_n
\EEQ
yields a sequence of equations
$k_1^+ C_{stat,1} = k_1^-  C_{stat,2},\ldots, k_{n-1}^+ C_{stat,n-1} = k_{n-1}^-
C_{stat,n}$ which define each $C_{stat,j}$, $j=1,\ldots,n-1$ uniquely
in terms of $C_{stat,n}$; in particular, $C_{stat,1}=(K(1\rightleftarrows n))^{-1} C_{stat,n}$, where 
\BEQ K(1\rightleftarrows n) = \frac{k_1^+ \cdots k_{n-1}^+}{k_{n-1}^-
\cdots k_1^-}
\EEQ
is the quotient (product of direct rates)/(product of reverse rates)
along the branch. Completing the loop, we get $k_n^+ C_{stat,n} = k_n^- C_{stat,1}^{2}$.
Substituting $K(1\rightleftarrows n) C_{stat,1}$ to $C_{stat,n}$ yields 
\BEQ C_{stat,1} = K(1\rightleftarrows n) \frac{k_n^+}{k_n^-}, \qquad 
C_{stat,k} = K(1\rightleftarrows k) C_{stat,1}\ \  (2\le k\le n)
\EEQ 

\item[(ii)] (Type II) Following each subcycle ${\cal C}_j: x_{\sigma_j}\to
x_{\sigma_j+1}\to 
\cdots \to x_j$ $(j=1,\ldots,\ell)$,  one gets as in (i)  $C_{stat,\sigma_j}=(K(\sigma_j\rightleftarrows j))^{-1} C_{stat,j}$. Then
the current associated to $R_j:\ x_{i_j}\to x_{i_j+1} + x_{\sigma_j}$
vanishes if and only if  $k^+_{i_j} C_{stat,i_j} = k^-_{i_j} C_{stat,i_j+1}
C_{stat,\sigma_j}$. This yields
\BEQ C_{stat,i_j+1} = \frac{k^+_{i_j}}{k^-_{i_j}} \frac{C_{stat,i_j}}{C_{stat,\sigma_j}} = K(\sigma_j\rightleftarrows j) \frac{k^+_{i_j}}{k^-_{i_j}}
\EEQ
and more generally,
\BEQ C_{stat,m} = K(\sigma_j\rightleftarrows m), \qquad i_j \prec m\preceq i_{j+1} \EEQ
with $K(m'\rightleftarrows m)= \frac{k_{m'}^+\cdots k_m^+}{k_m^-\cdots
k_{m'}^-}$ is the quotient (product of direct rates)/(product of
reverse rates) along the subcycle $m'\to m'+1\to\cdots m$. 

\item[(iii)]  (Type III)  Removing first the one-to-many reaction $ R_v: \ v
\overset{k_v^+}{\underset{k_v^-}{\rightleftarrows}}
s_v x+s'_v x'$ (and the associated reverse reaction), one gets a star-shaped
network, with one branch starting from $u$ and ending in 
$v$, and two branches starting from $x$, resp. $x'$ and ending in $u$. Proceeding as in (i) along each of these branches, 
one gets the  relations 
\BEQ C_{stat,u}= (K(u\rightleftarrows v))^{-1} C_{stat,v}, \qquad  C_{stat,x}= (K(x\rightleftarrows u))^{-1} C_{stat,u}, \qquad  C_{stat,x'}= (K(x'\rightleftarrows u))^{-1} 
C_{stat,u}.  \label{eq:eq-iii}
\EEQ
 Now, the current associated to $R_v$ vanishes if and only if
$k_v^+ C_v = k_v^- C_x C_{x'}$. Using (\ref{eq:eq-iii}) yields
\BEA && C_{stat,u} = K(u\rightleftarrows v) K(x\rightleftarrows u)
K(x'\rightleftarrows u)  \frac{k_v^+}{k_v^-} \nonumber\\
&& C_{stat,x} =  K(u\rightleftarrows v) 
K(x'\rightleftarrows u)  \frac{k_v^+}{k_v^-}, \qquad 
 C_{stat,x'} =  K(u\rightleftarrows v) 
K(x\rightleftarrows u)  \frac{k_v^+}{k_v^-} \nonumber\\
\EEA
from which the other coefficients $C_{stat,\cdot}$ are easily determined.

\item[(iv)] (Type IV) We proceed as in (iii) by removing first the one-to-many
reactions $R_v$ and  $R_w:\ w\overset{k_w^+}{\underset{k_w^-}{\rightleftarrows}}
 u+ x'$. Then we have one branch from $x'$ to $v$ through $u$ 
along ${\cal C'}$, and another branch from $x$ to $w$ along ${\cal C}$,
from which we get the relations
\BEQ  C_{stat,x'} = (K(x'\rightleftarrows v))^{-1} C_{stat,v}, \
 C_{stat,u}=  (K(u\rightleftarrows v))^{-1} C_{stat,v}, \  
  C_{stat,x}= (K(x\rightleftarrows w))^{-1} C_{stat, w}   \label{eq:eq-iv}
\EEQ

 Now, the currents associated to $R_v,R_w$ vanish if and only if
\BEQ k_v^+ C_{stat,v} = k_v^- C_{stat,x} C_{stat,x'}, \qquad k^+_w 
C_{stat,w} =
k^-_w C_{stat,u} C_{stat,x'}.
\EEQ
 Substituting (\ref{eq:eq-iv})  into the previous relations yields

\BEA && C_{stat,w} =  K(x\rightleftarrows w)
K(x'\rightleftarrows v) \frac{k^+_v}{k^-_v}, \qquad  C_{stat,x} =  
K(x'\rightleftarrows v) \frac{k^+_v}{k^-_v} \nonumber\\
&& C_{stat,v} =  K(x'\rightleftarrows v) \sqrt{ K(x\rightleftarrows w)
K(u\rightleftarrows v)  \frac{k^+_w}{k^-_w} \frac{k^+_v}{k^-_v}}
\EEA
from which the other coefficients $C_{stat,\cdot}$ are easily determined.

\item[(v)] (Type V) We proceed as in (iv) by first removing the one-to-many
reactions $R_v, R_w$ and $R_{w'}:\ w'\overset{k_{w'}^+}{\underset{k_{w'}^-}{\rightleftarrows}}
 u+ x$. Then we have one branch from $x'$ to $w'$ 
along ${\cal C'}$, one branch from $x$ to $w$ along ${\cal C}$, and
one branch from $u$ to $v$ along ${\cal C}\cap {\cal C}'$,
from which we get the relations
\BEQ  C_{stat,x'} = (K(x'\rightleftarrows w'))^{-1} C_{stat,w'}, \
 C_{stat,u}=  (K(u\rightleftarrows v))^{-1} C_{stat,v}, \  
  C_{stat,x}= (K(x\rightleftarrows w))^{-1} C_{stat, w}   \label{eq:eq-v}
\EEQ

The currents associated to $R_v,R_w, R_{w'}$ vanish if and only if
\BEQ k_v^+ C_{stat,v} = k_v^- C_{stat,x} C_{stat,x'}, \qquad k^+_w 
C_{stat,w} =
k^-_w C_{stat,u} C_{stat,x'}, \qquad k^+_{w'} 
C_{stat,w'} =
k^-_{w'} C_{stat,u} C_{stat,x}
\EEQ
 Substituting (\ref{eq:eq-v})  into the previous relations yields
\BEA && (1)\qquad  C_{stat,v} = \frac{k_v^-/k_v^+}{K(x\rightleftarrows w)
K(x'\rightleftarrows w')} \, c^*_w c^*_{w'}, \nonumber\\ 
&& (2)\qquad C_{stat,w} = \frac{k_w^-/k_w^+}{K(u\rightleftarrows v)
K(x'\rightleftarrows w')} \, c^*_v c^*_{w'}, \qquad  (3) \qquad 
 C_{stat,w'} = \frac{k_{w'}^-/k_{w'}^+}{K(u\rightleftarrows v)
K(x\rightleftarrows w)} \, c^*_v c^*_{w} \nonumber\\
\EEA
Forming quotients $\frac{(1)}{(2)},\frac{(1)}{(3)}$,
\BEQ \frac{C_{stat,v}}{C_{stat,w}} = \sqrt{ \frac{k^-_v}{k^+_v}
\frac{k^+_w}{k^-_w} \frac{K(u\rightleftarrows v)}{
K(x\rightleftarrows w)} }, \qquad  \frac{C_{stat,v}}{C_{stat,w'}} = \sqrt{ \frac{k^-_v}{k^+_v}
\frac{k^+_{w'}}{k^-_{w'}} \frac{K(u\rightleftarrows v)}{
K(x'\rightleftarrows w')} }
\EEQ
makes it possible to eliminate $C_{stat,w},C_{stat,w'}$ from (1),
giving finally
\BEQ C_{stat,v} =  K(u\rightleftarrows v) \sqrt{
K(x\rightleftarrows w) K(x'\rightleftarrows w')  \frac{k^+_w}{k^-_w}
\frac{k^+_{w'}}{k^-_{w'}} }
\EEQ
and similar formulas for $C_{stat,w},C_{stat,w'}$ by circular
permutations between $v,w,w'$.

\end{itemize}

\subsection{Non-degeneracy of the zero stationary state for vanishing degradation rates}\label{app:vanish}
Here we calculate the explicit Jacobian determinant at the zero stationary state of each core for no degradation, and show that it cannot be zero. Also assume that $C_{\delta}$ is a vector of very small non-negative concentrations, $([\delta x_i])_{x_i\in{\cal X}}$. The system near zero behaves as $\frac{dC}{dt}=J\times C_{\delta}$.
\subsubsection*{Type I}
The reactions are:
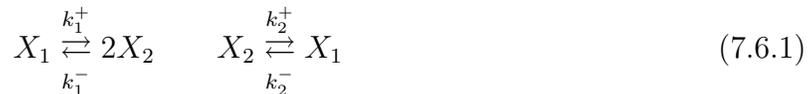
\BEA
X_1 \overset{k_1^+}{\underset{k_1^-}{\rightleftarrows}} 2X_2 \qquad X_2 \overset{k_2^+}{\underset{k_2^-}{\rightleftarrows}} X_1  
\EEA
$$
J=\left[ 
\begin{array}{cc} 
-(k_1^++k_2^-) & k_2^+ \\
2k_1^++k_2^- & -(k_2^+) 
\end{array} 
\right]
$$
We get Det $J=-k_1^+k_2^+$, which is non-zero\\ 
At $C_{\delta}$, $\left.\left(\sum_{X_i\in{\cal X}}\frac{d[x_i]}{dt}\right)\right|_{C_{\delta}}=k_1^+[\delta x_1]\geq0$. This is also true for all of the following cases of types.\\

\subsubsection*{Type III}
The reactions are:
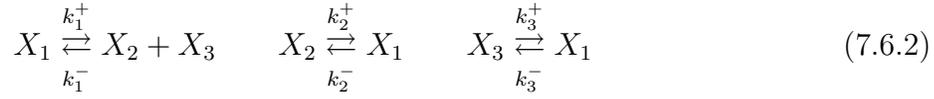
\BEA
X_1 \overset{k_1^+}{\underset{k_1^-}{\rightleftarrows}} X_2+X_3  \qquad X_2 \overset{k_2^+}{\underset{k_2^-}{\rightleftarrows}} X_1 \qquad X_3 \overset{k_3^+}{\underset{k_3^-}{\rightleftarrows}} X_1  
\EEA
$$
J=\left[ 
\begin{array}{ccc}  
-(k_1^++k_2^-+k_3^-) & k_2^+ & k_3^+ \\
k_2^-+k_1^+ & -k_2^+ & 0 \\
k_3^-+k_1^+ & 0 & -k_3^+ 
\end{array} 
\right]
$$
And we get Det $J=k_1^+k_2^+k_3^+$, non-zero\\

\subsubsection*{Type II $l=1,2$}
For one fork,
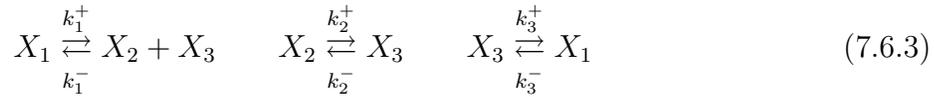
\BEA
X_1 \overset{k_1^+}{\underset{k_1^-}{\rightleftarrows}} X_2+X_3  \qquad X_2 \overset{k_2^+}{\underset{k_2^-}{\rightleftarrows}} X_3 \qquad X_3 \overset{k_3^+}{\underset{k_3^-}{\rightleftarrows}} X_1  
\EEA
$$
J=\left[ 
\begin{array}{ccc}  
-(k_1^++k_3^-) & 0 & k_3^+ \\
k_1^+ & -k_2^+ & k_2^- \\
k_3^-+k_1^+ & k_2^+ & -(k_3^++k_2^-) 
\end{array} 
\right]
$$
Again, we get a non-zero Det $J=k_1^+k_2^+k_3^+$\\

\medskip

For two forks 
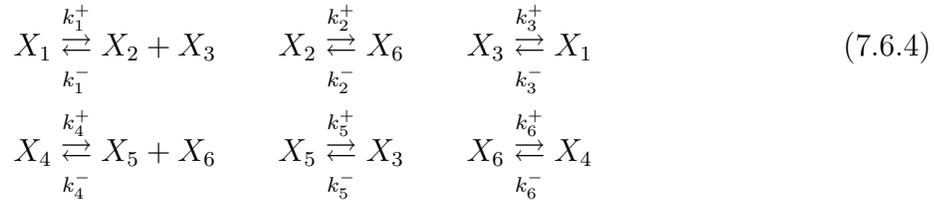
\BEA
X_1 \overset{k_1^+}{\underset{k_1^-}{\rightleftarrows}} X_2+X_3  \qquad X_2 \overset{k_2^+}{\underset{k_2^-}{\rightleftarrows}} X_6 \qquad X_3 \overset{k_3^+}{\underset{k_3^-}{\rightleftarrows}} X_1  \\ \nonumber
X_4 \overset{k_4^+}{\underset{k_4^-}{\rightleftarrows}} X_5+X_6  \qquad X_5 \overset{k_5^+}{\underset{k_5^-}{\rightleftarrows}} X_3 \qquad X_6 \overset{k_6^+}{\underset{k_6^-}{\rightleftarrows}} X_4  
\EEA
$$
J=\left[ 
\begin{array}{cccccc}  
-(k_1^++k_3^-) & 0 & k_3^+ & 0 & 0 & 0 \\
k_1^+ & -k_2^+ & 0 & 0 & 0 & k_2^- \\
k_3^-+k_1^+ & 0 & -(k_3^++k_5^-) & 0 & k_5^+ & 0 \\
0 & 0 & 0 & -(k_4^++k_6^-) & 0 & k_6^+ \\
0 & 0 & k_5^- & k_4^+ & -(k_5^+) & 0 \\
0 & k_2^+ & 0 & k_4^++k_6^- & 0 & -(k_6^++k_2^-)
\end{array} 
\right]
$$
Which yeilds Det $J=-k_1^+k_2^+k_3^+k_4^+k_5^+k_6^+$, always non-zero\\

\subsubsection*{Type IV}
\BEA
X_1 \overset{k_1^+}{\underset{k_1^-}{\rightleftarrows}} X_2+X_3  \qquad X_2 \overset{k_2^+}{\underset{k_2^-}{\rightleftarrows}} X_5 \qquad X_3 \overset{k_3^+}{\underset{k_3^-}{\rightleftarrows}} X_4  \\ \nonumber
X_5 \overset{k_5^+}{\underset{k_5^-}{\rightleftarrows}} X_4+X_3  \qquad X_4 \overset{k_4^+}{\underset{k_4^-}{\rightleftarrows}} X_1
\EEA
$$
J=\left[ 
\begin{array}{ccccc}  
-(k_1^++k_4^-) & 0 & 0 & k_4^+ & 0 \\
k_1^+ & -k_2^+ & 0 & 0 & k_2^- \\
k_1^+ & 0 & -k_3^+ & k_3^- & k_5^+ \\
k_4^- & 0 & k_3^+ & -(k_4^++k_3^-) & k_5^+ \\
0 & k_2^+ & 0 & 0 & -(k_5^++k_2^-)
\end{array} 
\right]
$$
Which gives non-zero Det $J=2k_1^+k_2^+k_3^+k_4^+k_5^+$\\

\subsubsection*{Type V}
\BEA
X_1 \overset{k_1^+}{\underset{k_1^-}{\rightleftarrows}} X_2+X_3  \qquad X_2 \overset{k_2^+}{\underset{k_2^-}{\rightleftarrows}} X_4 \qquad X_3 \overset{k_3^+}{\underset{k_3^-}{\rightleftarrows}} X_5  \\ \nonumber
X_4 \overset{k_4^+}{\underset{k_4^-}{\rightleftarrows}} X_6+X_3  \qquad X_5 \overset{k_5^+}{\underset{k_5^-}{\rightleftarrows}} X_6+X_2  \qquad X_6 \overset{k_6^+}{\underset{k_6^-}{\rightleftarrows}} X_1
\EEA
$$
J=\left[ 
\begin{array}{cccccc}  
-(k_1^++k_6^-) & 0 & 0 & 0 & 0 & k_6^+ \\
k_1^+ & -k_2^+ & 0 & k_2^- & k_5^+ & 0 \\
k_1^+ & 0 & -k_3^+ & k_4^+ & k_3^- & 0 \\
0 & k_2^+ & 0 & -(k_4^++k_2^-) & 0 & 0 \\
0 & 0 & k_3^+ & 0 & -(k_5^++k_3^-) & 0 \\
k_6^- & 0 & 0 & k_4^+ & k_5^+ & -k_6^+ 
\end{array} 
\right]
$$
Which computes to Det $J=-4k_1^+k_2^+k_3^+k_4^+k_5^+k_6^+$, non-zero\\

\subsection{General analysis of stationary states of cores (Types I and III)}\label{sec:onethree}


{\em General notations.} 
\begin{enumerate}
\item Let cyc be one of the cycles (${\cal C}$ for
Types I-II,  ${\cal C}$ or ${\cal C}'$ for Types III-V). 
Assume $x\in$cyc is not the reactant/product of a multiple reaction, then $x_+$, resp. $x_-$ is the species following, resp.
preceding $x$ along cyc.  The rate of the reactions connecting
$x_{\pm}$ to $x$ are $x_-\overset{k^+_{x_-}}{\underset{k
^-_{x_-}}{\rightleftarrows}} x $ and $x \overset{k^+_x}{\underset{k^-_{x}}{\rightleftarrows}} x_+$. 

\item Let $R:x\to sz+s' z'$ be a multiple reaction with two different
products $z,z'$. Then rates are $x\overset{\nu^+_x}{\underset{\nu^-_x}{\rightleftarrows}} sz+s' z'$.   

\end{enumerate}

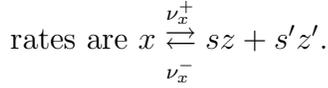
\Bigskip

\Bigskip
{\bf Type I.} The stationary equation for $B_n$ yields the 
"linearization" equation, 
\BEQ  (\nu_+ + k_{n-1}^- + a_n)[B_n] - k_{n-1}^+ [B_{n-1}] = \nu_- [B]^2 
\label{eq:54}
\EEQ
Substituting into the equation for $B=B_0$ yields
\BEQ (k_{on}+a_0) [B] - k_{off} [B_1]  - 2k^+_{n-1} [B_{n-1}]
=   -2 ( k_{n-1}^- + a_n)[B_n] \label{eq:56}
\EEQ
 a Markov-type  equation for $B$,
with Markov transitions from $B_1$ and (shunting $B_n$) from  $B_{n-1}$. 
Then the equations for $B_i,i=1,\ldots,n-1$ describe Markov
transitions to and from $B_{i\pm 1}$, including the special case
$i=n-1$,   
\BEQ (k_{n-1}^+ + k_{n-2}^- +a_{n-1}) [B_{n-1}] - k_{n-2}^+ [B_{n-2}] = k_{n-1}^- [B_n] \EEQ
All together, we get
\BEQ -\underline{M} \left([B] \  [B_1] \ \cdots\  [B_{n-1}]\right)^t = [B_n] \phi  \label{eq:58}
\EEQ
where $\underline{M}$ is a generalized irreducible Markov matrix  describing transitions
along the shortened cycle \begin{tikzpicture} 
\draw(0,0) node {
$B_0\leftrightarrow B_1\leftrightarrow \cdots \leftrightarrow B_{n-1}$};
\draw(-1.75,0.3)[<-,red] arc(180:0:1.7 and 0.3);
\draw[red](0,0.8) node {\tiny $2k^+_{n-1}$}; \end{tikzpicture} with  positive killing
rates $a_1,\ldots,a_n$, and   $\phi=\left(\begin{array}{c} \phi_1 \\
\vdots\\ \phi_n \end{array}\right)$.  The solution of (\ref{eq:58}) is unique, 
\BEQ [B] = \psi_0 [B_n], \qquad 
[B_i] = \psi_i [B_n], \qquad i=1,\ldots,n-1 \EEQ
We must assume $\psi_i>0$. Putting back these expressions for $ [B],[B_{n-1}]$
into (\ref{eq:54}) yields finally 
\BEQ \nu_- \psi_0^2 [B_n] = (\nu_+ + k_{n-1}^- +a_n)  - k^+_{n-1}
\psi_{n-1}.
\EEQ 

If $\psi_i>0$, $i=1,\ldots,n-1$ and $(\nu_+ + k_{n-1}^- +a_n) > k^+_{n-1}\psi_{n-1}$, these formulas yield
a unique stationary state.


\vskip 2cm
\noindent{\bf Type III.} The stationary equation for $A_n$ yields the 
"linearization" equation, 
\BEQ  (\nu_+ + k_{n-1}^- + a_n)[A_n] - k_{n-1}^+ [A_{n-1}] = \nu_- [B'_0][B''_0] 
\label{eq:60}
\EEQ

Proceeding as for Type I, we want to shunt $[A_n]$.  Putting aside
the equation for $A_n$, we have $|{\cal X}|-1$ Markov-type equations
for the other species; specific cases are the equations for $A_{n-1}$,
\BEQ (k^+_{n-1} + k^-_{n-2} + a_{n-1})[A_{n-1}] - k_{n-2}^+ [A_{n-2}] =  k_{n-1}^- [A_n]  \label{eq:61} 
\EEQ

and the equations for $B'_0, B''_0$, which we linearize using (\ref{eq:60}),
\BEQ (k'_{0,+}+a'_0) [B'_0] - k'_{0,-} [B'_1] - k^+_{n-1} [A_{n-1}] = 
-(k_{n-1}^- + a_n) [A_n]
\EEQ
and similarly with  $' \rightarrow\  ''$.  We thus get
a linear equation of the type
\BEQ - \underline{M} ([A_0] \ \cdots\ [A_{n-1}] \ [B'_0] \ \cdots \ 
[B'_n] \ [B''_0] \ \cdots\ [B''_n])^t = [A_n] \phi \EEQ
where  $\underline{M}$ is a generalized irreducible Markov matrix  describing transitions
along the shortened cycle

\medskip
\begin{center}
 \begin{tikzpicture} 
\draw(0,0) node {$A_0\leftrightarrow B_1\leftrightarrow \cdots \leftrightarrow A_{n-1}$};
\draw[<->](-2,0.3)--(-2,0.6); \draw(-2,0.9) node {$B'_n$};
\draw(0,0.9) node {$\leftrightarrow B'_n \leftrightarrow \cdots
\leftrightarrow B'_0$};
\draw[<->](-2,-0.3)--(-2,-0.6); \draw(-2,-0.9) node {$B''_n$};
\draw(0,-0.9) node {$\leftrightarrow B''_n \leftrightarrow \cdots
\leftrightarrow B''_0$};
\draw[->,red](1.7,0.3)--(1.5,0.6);
 \draw[red](2,0.45) node {\tiny $k^+_{n-1}$}; 
 \draw[red](2,-0.45) node {\tiny $k^+_{n-1}$};
\draw[->,red](1.7,-0.3)--(1.5,-0.6);
 \end{tikzpicture}
\end{center}
\medskip 
   with  positive killing
rates. The unique solution is of the form 
\BEQ [A_i]=\psi_i [A_n], \ 0\le i\le n-1; \qquad [B'_i]=\psi'_i
[A_n], \ 0\le i\le n'; \qquad [B''_i]=\psi''_i [A_n], \ 0\le i\le n''
\EEQ
Inserting this into  (\ref{eq:60}) yields
\BEQ \nu_- \psi'_0\psi''_0 [A_n] = (\nu_+ + k_{n-1}^-+a_n) - k^+_{n-1} 
\psi_{n-1}. \EEQ
As in the case of Type I, if all coefficients $\psi_i,\psi'_i,\psi''_i$
are $>0$ and $(\nu_+ + k_{n-1}^-+a_n) > k^+_{n-1} 
\psi_{n-1}$, these formulas define a unique stationary state.

\subsection{General analysis of stationary states of cores (Type IV,II$_2$, and V)}\label{sec:rest}

\begin{Lemma}
    Species not involved in splitting reactions can be linearly substituted at the stationary state into the mass action equations of other species without affecting the sign of coefficients in that equation
\end{Lemma}
\noindent {\bf Proof.} Consider a chunk of any Type of core of the form:\\
\medskip
$ {\rightleftarrows}X_{j-1}\overset{k_{j-1}^+}{\underset{k_{j-1}^-}{\rightleftarrows}} X_j\overset{k_{j}^+}{\underset{k_{j}^-}{\rightleftarrows}} X_{j+1}{\rightleftarrows}$\\
If the mass action system of equations are denoted by $f_i(x \in {\cal X})$ for i=$1,2...n$,\\
\begin{equation}
    f_j=-(k_j^++k_{j-1}^-+a_j)[x_j]+k_j^-[x_{j+1}]+k_{j-1}^+[x_{j-1}]
\end{equation}\\
At the stationary state, $f_j=0$ yields $[x_j]$ as a linear combination of $[x_{j-1}]$ and $[x_{j+1}]$, which can be used to write the flow between $X_{j-1}$ and $X_{j}$ as it appears in $f_{j-1}$ in terms of $[x_{j+1}]$,
\begin{equation}
    -k_{j-1}^+[x_{j-1}] +k_{j-1}^-[x_j]= -k_{j-1}^+[x_{j-1}] + \frac{k_{j-1}^-k_{j}^- [x_{j+1}]}{k_j^++k_{j-1}^-+a_j} +
\frac{k_{j-1}^-k_{j-1}^+ [x_{j-1}]}{k_j^++k_{j-1}^-+a_j}\end{equation}
where we can note that the coefficient of $[x_{j-1}]$ remains negative while the $[x_j]$ with a positive coefficient is replaced by $[x_{j+1}]$ with a positive coefficient. 
\begin{equation}
    \frac{k_{j-1}^-k_{j}^- [x_{j+1}]}{k_j^++k_{j-1}^-+a_j} -
\frac{k_{j}^+k_{j-1}^+ [x_{j-1}]}{k_j^++k_{j-1}^-+a_j}-\frac{a_{j}k_{j-1}^+ [x_{j-1}]}{k_j^++k_{j-1}^-+a_j}=\bar{k}_{j-1}^-[x_{j+1}]-\bar{k}_{j-1}^+[x_{j-1}]-\bar{a}_{j}[x_{j-1}]
\end{equation}
The latter term in $\bar{a}$ can be absorbed into the degradation of species $[x_{j-1}]$ itself, and the same rates (but with the opposite flow) appear in the equation for $[x_{j+1}]$. Thus $X_j$ can be shorted out of the network.\\

\subsubsection{Type IV}
\medskip
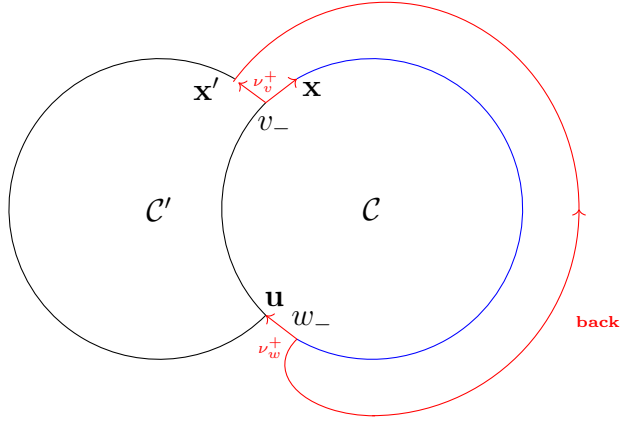
\begin{center}
\begin{tikzpicture}
  \draw(0,0) node {${\cal C}$};
\draw(-4*0.707,0) node {${\cal C}'$};
 
\draw(-4*0.707+2*0.5,1.732) arc(60:315:2 and 2);
\draw[blue](-2*0.5,1.732) arc(120:-120:2 and 2);
\draw(-2*0.707,-2*0.707) arc(-135:-225:2 and 2);

\draw(-1.3,1.1) node {$v_-$};

stationary
\draw(-1.3,-1.2) node {$\mathbf u$};
\draw[ultra thick](-0.8,1.6) node {$\mathbf x$};
\draw[ultra thick](-2.2,1.6) node {$\mathbf x'$};

\draw[->,red](-2*0.707,2*0.707)--(-2*0.5,1.732);
\draw[->,red](-2*0.707,2*0.707)--(0.05-4*0.707+2*0.5,-0.05+1.732);
\draw[red](-2*0.707,1.7) node {\tiny $\nu^+_{v}$};


\draw(-0.8,-1.5) node {$w_-$};
\draw[red](-1.35,-1.85) node {\tiny $\nu^+_{w}$};


\draw[red,->](-2*0.5,-2*0.866)--(-2*0.707,-2*0.707);
\draw[red](-2*0.5,-2*0.866) arc(150:270:1.2 and 0.68);
\draw[->,red](0,-2.75) arc(270:360:2.75 and 2.75);
\draw[red](2.75,0) arc(360:502:2.57 and 2.75);

\draw[red](3,-1.5) node {\bf \tiny  back};

\end{tikzpicture}
\end{center}
$\mathcal{X}=\{x,x',v,w,u\}$\\
We introduce two new variables $\delta$ and $\delta'$ as follows
\begin{equation}\label{eq:4-0}
    \delta=\nu_{v}^+[v]-\nu_{v}^-[x][x']  ,\delta'=\nu_{w}^+[w]-\nu_{w}^-[u][x']
\end{equation}
Then, using these variables and using the lemma on all species not involved in forks, we get the mass action equations at stationarity as
\BEA\label{eq:4-1} 
-(k_x^++a_x)[x]+\delta+k_x^-[w]=0\\ \nonumber
-(k_{x'}^++a_{x'})[x']+\delta+\delta'+k_{x'}^-[u]=0\\ \nonumber
-(k_u^++k_{x'}^-+a_u)[u]+\delta'+k_u^-[v]+k_{x'}^+[x']=0\\ \nonumber
-(k_u^-+a_v)[v]-\delta+k_u^+[u]=0\\ \nonumber
-(k_x^-+a_w)[w]-\delta'+k_x^+[x]=0
\EEA
Further, from (\ref{eq:4-1}) we also get that
\begin{equation}\label{eq:4-11}
2\delta'= a_{x'}[x']+a_u[u]+a_v[v] \qquad \delta-\delta'=a_x[x]+a_w[w]    
\end{equation}
which implies that for positive concentrations, $\delta$ and $\delta'$ are positive (i.e. the solution is in a subset of the positive quadrant)\\
\begin{Lemma}
    For type IV core with rate constants and degradation values as in (\ref{eq:4-1}), there exists a continuous function $m_b(a,k)$, such that any positive stationary state satisfies $m_b(a,k)\geq\max\{[z],z\in{\cal X}\}$
\end{Lemma}
\noindent\textbf{Proof.} The fact that $\delta$ and $\delta'$ $\geq0$ gives inequalities on the steady-state concentrations based on the rate constants from equation (\ref{eq:4-1}),
\BEA
\frac{k_x^++a_x}{k_x^-}[x]\geq[w] \qquad  \frac{k_{x'}^++a_{x'}}{k_{x'}^-}[x']\geq[u] \\ \nonumber
\frac{k_u^++k_{x'}^-+a_u}{k_u^-}[u]\geq [v] \qquad \frac{k_u^++k_{x'}^-+a_u}{k_{x'}^+}[u]\geq [x']
\EEA
As $\delta\geq 0$,
\begin{equation}
    \left(\frac{k_{x'}^++a_{x'}}{k_{x'}^-}\right)\left(\frac{k_u^++k_{x'}^-+a_u}{k_u^-}\right)\nu_{v}^+ [x'] \geq \nu_{v}^+[v] \geq \nu_{v}^-[x][x']
\end{equation}
which for any non-zero stationary state gives
\begin{equation}
    \left(\frac{k_{x'}^++a_{x'}}{k_{x'}^-}\right)\left(\frac{k_u^++k_{x'}^-+a_u}{k_u^-}\right)\frac{\nu_{v}^+}{\nu_{v}^-}\geq[x] \text{  (which also bounds $[w]$)}
\end{equation}
$\delta'\geq0$ gives,
\begin{equation}
     \nu_{w}^+[w]\geq\nu_{w}^-[u][x']\geq\nu_{w}^-\left(\frac{k_{x'}^-}{k_{x'}^++a_{x'}}\right)[u]^2 
\end{equation}
And with the bound for $[w]$, we get an upper bound for $[u],[v]$ and $[x']$ as well. These bounds are the smallest when all the degradation is 0.

\Bigskip

These equations can then be used to write down each of the concentrations $[x],[x'],[u],[v]$ and $[w]$ as a linear combination of $\delta$ and $\delta'$ with coefficients formed as expressions in the rate constants.  

\BEA\label{eq:4-00}
&&[x]=\frac{(a_w+k_x^-)\delta-k_x^-\delta'}{a_wa_x+a_xk_x^-+a_wk_x^+} \qquad [w]=\frac{-(a_x+k_x^+)\delta'+k_x^+\delta}{a_wa_x+a_xk_x^-+a_wk_x^+}\\ \nonumber
&&[u]=\frac{((a_v+k_u^-)a_{x'}+2(a_v+k_u^-)k_{x'}^+)\delta'+(-a_{x'}k_u^-+a_vk_{x'}^+)\delta}{(a_v+k_u^-)a_{x'}k_{x'}^-+(a_ua_v+a_uk_u^-+a_vk_u^+)a_{x'}+(a_ua_v+a_uk_u^-+a_vk_u^+)k_{x'}^+}\\ \nonumber
&&[v]=\frac{(a_{x'}k_u^++2k_u^+k_{x'}^+)\delta'-(a_ua_{x'}+a_{x'}k_u^++a_{x'}k_{x'}^-+a_uk_{x'}^+)\delta}{(a_v+k_u^-)a_{x'}k_{x'}^-+(a_ua_v+a_uk_u^-+a_vk_u^+)a_{x'}+(a_ua_v+a_uk_u^-+a_vk_u^+)k_{x'}^+}\\ \nonumber
&&[x']=\frac{(2a_vk_{x'}^-+2k_u^-k_{x'}^-+a_ua_v+a_uk_u^-+a_vk_u^+)\delta'+(a_ua_v+a_uk_u^-+a_vk_u^++a_vk_{x'}^-)\delta}{(a_v+k_u^-)a_{x'}k_{x'}^-+(a_ua_v+a_uk_u^-+a_vk_u^+)a_{x'}+(a_ua_v+a_uk_u^-+a_vk_u^+)k_{x'}^+}
\EEA
\medskip
The condition that the rate constants are all positive $(>0)$ directly give a strict sign to the coefficients of $\delta$ and $\delta'$ in the expressions of concentrations except for the coefficient of $\delta$ in the expression of $[u]$. Let $\epsilon=(-a_{x'}k_u^-+a_vk_{x'}^+)$. We can then replace these expressions in equation (\ref{eq:4-0}) and taking care only for the sign, we get two new equations:-
\medskip
\BEA
0=A\delta'-B\delta-C\delta\delta'+D\delta'^2-E\delta^2 \nonumber \\
0=-A'\delta'+B'\delta+\textcolor{red}{C'}\delta\delta'-D'\delta'^2+\textcolor{red}{E'}\delta^2 
\label{eq:4-2}
\EEA
\medskip\\
where $A,B,C,D,E$ and $A',B',C',D',E'$ are functions of the rates, and for positive rate constants, all of them have positive values except $C'$ and $E'$. For these two,\\
$\epsilon>0$ $\Leftrightarrow$ $E'<0$ $\Rightarrow$ $C'<0$\\
\medskip\\
Also important is the fact that since we want the concentrations to be non-negative, 
\BEA
0\geq -\nu_{v}^-[x][x'] \nonumber \\
0\geq -\nu_{w}^-[u][x'] 
\EEA
\BEA
0\geq -C\delta\delta'+D\delta'^2-E\delta^2 \nonumber \\
0\geq +\textcolor{red}{C'}\delta\delta'-D'\delta'^2+\textcolor{red}{E'}\delta^2 
\label{eq:4-3}
\EEA

The stationary states of the type 4 core correspond to the solutions of equation (\ref{eq:4-2}) in the positive quadrant following the inequalities (\ref{eq:4-3}). \\
\medskip\\



\medskip

\begin{Lemma}
    There is no degenerate stationary state in the positive quadrant for (\ref{eq:4-2})
\end{Lemma}
\noindent {\bf Proof.} Assume that (\ref{eq:4-2}) has a degenerate stationary state in the positive quadrant (not at origin) satisfying the given inequality, \\
The Jacobian Matrix of the system at this point is 
$$
J=\left[ 
\begin{array}{cc} 
 -B-2E\delta-C\delta' & A+2D\delta'-C\delta  \\
 B'+2\textcolor{red}{E'}\delta+\textcolor{red}{C'}\delta' & -A'-2D'\delta'+\textcolor{red}{C'}\delta 
\end{array} 
\right]
$$
The state being degenerate implies that the columns of this matrix are linearly dependent, so assume there exists $\mu$ such that: Column 1 + Column 2 $\times \mu\frac{\delta'}{\delta}$ $=0$. In terms of expressions, this gives
\BEA
0= -B\delta-2E\delta^2-C\delta'\delta+\mu(A\delta'+2D\delta'^2-C\delta\delta') \nonumber \\
0=B'\delta+\textcolor{red}{C'}\delta\delta'+2\textcolor{red}{E'}\delta^2+\mu(-A'\delta'-2D'\delta'^2+\textcolor{red}{C'}\delta\delta') 
\EEA
Using equation (\ref{eq:4-2}) gives
\BEA
A\delta'-B\delta=(\mu-1)(A\delta'+2D\delta'^2-C\delta\delta')=(\mu-1)(B\delta+D\delta'^2+E\delta^2)\nonumber\\
B'\delta-A'\delta'=(\mu-1)(-A'\delta'-2D'\delta'^2+\textcolor{red}{C'}\delta\delta')=(\mu-1)(-B'\delta-\textcolor{red}{E'}\delta^2-D'\delta'^2)
\EEA
From the inequality (\ref{eq:4-3}), we have that the LHS of both equations are positive. As we have assumed the solution to be in the positive quadrant, the RHS of the first equation being positive implies that $\mu>1$.\\
We immediately have $\mu<1$ if \textcolor{red}{$E'$} is positive. If it is negative, \textcolor{red}{$C'$} is negative as well and we get $\mu<1$ in both cases. Thus we end up in a contradiction.  

\Bigskip

\begin{Lemma}
    For degradation of the form $a'=\{\alpha a_x, \alpha a_u, \alpha a_w, \alpha a_v, \alpha a_{x'}\}$, the stationary state at the origin has Jacobian determinant 0 for only one value of $\alpha\in\mathbb{R}^+$ 
\end{Lemma} 
\noindent {\bf Proof.} The origin is always a solution for the hyperbolae for any value of degradation rates. The Jacobian matrix at this stationary state is 
$$
J=\left[ 
\begin{array}{cc} 
 -B & A  \\
 B' & -A' 
\end{array} 
\right]
$$
Assume fixed reaction constants ($\{$\textbf{$k$},\textbf{$\nu$}$\}$), $a=\{a_x, a_u, a_w, a_v, a_{x'}\}$, and let us consider the same system with degradation rates of the form ($a'=\{\alpha a_x, \alpha a_u, \alpha a_w, \alpha a_v, \alpha a_{x'}\}$) where $\alpha>0$.\\

\medskip
We formulate the Jacobian determinant at origin as a function of $\alpha$.\\
Using equation (\ref{eq:4-00}), 
$$
J=\left[ 
\begin{array}{cc} 
 -(\alpha \, g(k,\nu,a)+ \alpha \, h(k,\nu,a)+ \alpha^2 p(k,\nu,a) + \alpha^3 q(k,\nu,a)) & f(k,\nu)+\alpha \, g(k,\nu,a)  \\
 f'(k,\nu) & -(f'(k,\nu)+\alpha \, g'(k,\nu,a)+ \alpha^2 h'(k,\nu,a)) 
\end{array} 
\right]
$$
where $f,g,h,p,q$ and $f',g',h'$ are functions of the rate constants and degradation rates which have positive values for positive rates. \\
\begin{equation}
    Det\: J= F(\alpha,k,\nu,a)-f(k,\nu)f'(k,\nu)
\end{equation}
where $F(\alpha,k,\nu,a)$ is, in fact, a degree 5 polynomial in $\alpha$ with all positive coefficients which is a strictly increasing function of $\alpha$. For fixed rates and varying $\alpha>0$, $Det\: J=0$ for only one value of $\alpha$.\\

\Bigskip

\medskip

Lemmas 8.3-5 imply that the Type IV cycle satisfies the assumptions of Theorem \ref{bigboy} if the zero stationary state is non-degenerate for vanishing degradation rates.

\subsubsection{Type II, $\ell$=2}

\begin{figure}
\centering
\begin{tikzpicture}
\draw(0,0) node {${\cal C}$};
\draw(0,0) circle(2);  \draw[->,ultra thin](2.5,0) arc(0:30:2.5 and 2.5);

\draw[red](0,-2) arc(270:286:2);
\draw[blue](0.67,-1.9) arc(286:317:1.7);
\draw[red](1.40,-1.40) arc(-45:217:2);
\draw[blue](-1.55,-1.25) arc(218:268:2);

\draw(0.3,-1.6) node {$y_3$};
\draw(0,2.3) node {$x_3$};

\draw[->](0,-2) arc(0:90:1 and 1);  \draw(0,-2.3) node { $y_2$};
\draw(-1,-1) arc(90:125:1 and 1);
\draw(-1.71,-1.41) node {$y_1$};

\draw[->](1.414,-1.414) arc(45:135:0.5 and 0.5);
\draw(1.414-0.707,-1.414) arc(135:200:0.5 and 0.5);
\draw(1.62,-1.62) node {$x_2$};
\draw(0.8,-2.2) node {$x_1$};

\draw(4,0) node {Type II};
\end{tikzpicture}
\end{figure}

$\mathcal{X}=\{x_1,x_2,x_3,y_1,y_2,y_3\}$\\
We again introduce two new variables $\delta$ and $\delta'$ as follows
\begin{equation}\label{eq:2-0}
    \delta=\nu_{1}^+[x_2]-\nu_{1}^-[x_3][x_1]  ,\delta'=\nu_{2}^+[y_2]-\nu_{2}^-[y_3][y_1]
\end{equation}
Then, using these variables and using the lemma on all species not involved in forks, we get the mass action equations at stationary state as
\BEA\label{eq:2-1}
-(k_{x_1}^++k_{y_3}^-+a_{x_1})[{x_1}]+\delta+k_{x_1}^-[{x_2}]+k_{y_3}^+[{y_3}]=0\\ \nonumber
-(k_{x_1}^-+a_{x_2})[x_2]-\delta+k_{x_1}^+[x_1]=0\\ \nonumber
-(k_{x_3}^++a_{x_3})[{x_3}]+\delta+k_{x_3}^-[{y_1}]=0\\ \nonumber
-(k_{y_1}^++k_{x_3}^-+a_{y_1})[{y_1}]+\delta'+k_{y_1}^-[{y_2}]+k_{x_3}^+[{x_3}]=0\\ \nonumber
-(k_{y_1}^-+a_{y_2})[y_2]-\delta'+k_{y_1}^+[y_1]=0\\ \nonumber
-(k_{y_3}^++a_{y_3})[{y_3}]+\delta'+k_{y_3}^-[{x_1}]=0
\EEA
Further, from (\ref{eq:2-1}) we also get that
\begin{equation}\label{eq:2-22}
\delta= a_{x_3}[x_3]+a_{y_1}[y_1]+a_{y_2}[y_2] \qquad \delta'= a_{y_3}[y_3]+a_{x_1}[x_1]+a_{x_2}[x_2]    
\end{equation}
So again, we are restricted to the positive quadrant in the $\delta,\delta'$ space.\\

\Medskip

\begin{Lemma}
    For type II, $l=2$ core with rate constants and degradation values as in (\ref{eq:2-1}), there exists a continuous function $m_b(a,k)$, such that any positive stationary state satisfies $m_b(a,k)\geq\max\{[z],z\in{\cal X}\}$
\end{Lemma}
\noindent {\bf Proof.} The inequalities after putting $\delta,\delta'\geq0$ in (\ref{eq:2-1}) are
\BEA
\frac{k_{x_1}^++k_{y_3}^-+a_{x_1}}{k_{x_1}^-}[x_1]\geq[x_2] \qquad \frac{k_{y_3}^++a_{y_3}}{k_{y_3}^-}[y_3]\geq[x_1] \\ \nonumber
\frac{k_{y_1}^++k_{x_3}^-+a_{y_1}}{k_{y_1}^-}[y_1]\geq[y_2] \qquad \frac{k_{x_3}^++a_{x_3}}{k_{x_3}^-}[x_3]\geq[y_1] 
\EEA
And we get a bound for $[y_3]$ and $[x_3]$ from $\delta,\delta'\geq0$ which bound everything\\
\BEA
    \left(\frac{k_{x_1}^++k_{y_3}^-+a_{x_1}}{k_{x_1}^-}\right)\nu_{1}^+[x_1]\geq\nu_{1}^+[x_2]\geq\nu_{1}^-[x_3][x_1] \\ \nonumber
    \left(\frac{k_{y_1}^++k_{x_3}^-+a_{y_1}}{k_{y_1}^-}\right)\nu_{2}^+[y_1]\geq\nu_{2}^+[y_2]\geq\nu_{1}^-[y_3][y_1]
\EEA

\Bigskip

Concentrations in terms of these variables are
\BEA\label{eq:2-2}
&&[x_1] = \frac{ a_{x_2} (a_{y_3}+k_{y_3}^+)\delta  + k_{y_3}^+( a_{x_2}+k_{x_1}^-) \delta' }{(a_{y _3} (a_{x_2} + k_{x_1}^-) k_{y _3}^- + (a_{x_1} a_{x_2} + a_{x_1} k_{x_1}^- + a_{x_2} k_{x_1}^+) a_{y _3} + (a_{x_1} a_{x_2} + a_{x_1} k_{x_1}^- + a_{x_2} k_{x_1}^+) k_{y _3}^+)}\\ \nonumber
&&[x_2] = \frac{k_{x_1}^+ k_{y_3}^+\delta' -(a_{x_1} a_{y_3} + a_{y_3} k_{y_3}^- + a_{x_1}k_{y_3}^+) \delta}{(a_{y _3} (a_{x_2} + k_{x_1}^-) k_{y _3}^- + (a_{x_1} a_{x_2} + a_{x_1} k_{x_1}^- + a_{x_2} k_{x_1}^+) a_{y _3} + (a_{x_1} a_{x_2} + a_{x_1} k_{x_1}^- + a_{x_2} k_{x_1}^+) k_{y _3}^+)}\qquad \quad\\ \nonumber
&&[x_3] = \frac{a_{y_2} k_{x_3}^- \delta'  + (a_{y_1} a_{y_2} + (a_{y_2} + k_{y_1}^-) k_{x_3}^- + a_{y_1} k_{y_1}^- + a_{y_2} k_{y_1}^+) \delta}{(a_{x_3} (a_{y_2} + k_{y_1}^-) k_{x_3}^- + (a_{y_1} a_{y_2} + a_{y_1} k_{y_1}^- + a_{y_2} k_{y_1}^+) a_{x_3} + (a_{y_1} a_{y_2} + a_{y_1} k_{y_1}^- + a_{y_2} k_{y_1}^+) k_{x_3}^+)}\\ \nonumber
&&[y_1] = \frac{(a_{x_3} a_{y_2} \delta' + (a_{y_2} + k_{y_1}^-) \delta k_{x_3}^+ + a_{y_2} \delta' k_{x_3}^+)}{(a_{x_3} (a_{y_2} + k_{y_1}^-) k_{x_3}^- + (a_{y_1} a_{y_2} + a_{y_1} k_{y_1}^- + a_{y_2} k_{y_1}^+) a_{x_3} + (a_{y_1} a_{y_2} + a_{y_1} k_{y_1}^- + a_{y_2} k_{y_1}^+) k_{x_3}^+)}\\ \nonumber
&&[y_2] = \frac{-(a_{x_3} a_{y_1} \delta' + a_{x_3} \delta' k_{x_3}^- + a_{y_1} \delta' k_{x_3}^+ - \delta k_{x_3}^+ k_{y_1}^+)}{(a_{x_3} (a_{y_2} + k_{y_1}^-) k_{x_3}^- + (a_{y_1} a_{y_2} + a_{y_1} k_{y_1}^- + a_{y_2} k_{y_1}^+) a_{x_3} + (a_{y_1} a_{y_2} + a_{y_1} k_{y_1}^- + a_{y_2} k_{y_1}^+) k_{x_3}^+)}\\ \nonumber
&&[y_3] = \frac{(a_{x_1} a_{x_2} \delta' + a_{x_1} \delta' k_{x_1}^- + a_{x_2} \delta' k_{x_1}^+ + (a_{x_2} \delta + a_{x_2} \delta' + \delta' k_{x_1}^-) k_{y_3}^-)}{(a_{y _3} (a_{x_2} + k_{x_1}^-) k_{y _3}^- + (a_{x_1} a_{x_2} + a_{x_1} k_{x_1}^- + a_{x_2} k_{x_1}^+) a_{y _3} + (a_{x_1} a_{x_2} + a_{x_1} k_{x_1}^- + a_{x_2} k_{x_1}^+) k_{y _3}^+)}
\EEA

Again the condition that the rate constants are all positive $(>0)$ gives a strict sign to the coefficients of $\delta$ and $\delta'$ in the expressions of concentrations. We can then replace these expressions in equation (\ref{eq:2-0}) and taking care only of the sign, we get two new equations:
\medskip
\BEA\label{eq:2-3}
0=A\delta'-B\delta-C\delta\delta'-D\delta'^2-E\delta^2 \nonumber \\
0=-A'\delta'+B'\delta-C'\delta\delta'-D'\delta'^2-E'\delta^2 
\EEA
\medskip\\
where $A,B,C,D,E$ and $A',B',C',D',E'$ are functions of the rates, and for positive rate constants, all of them have positive values.\\
\medskip\\

The stationary states of the type II $\ell=2$ core correspond to the solutions of equation (\ref{eq:2-3}) in the positive quadrant. \\
\medskip

\begin{Lemma}
    There is no degenerate stationary state in the positive quadrant for (\ref{eq:2-3})
\end{Lemma}
\noindent {\bf Proof.} Assume that (\ref{eq:2-3}) has a degenerate stationary state in the positive quadrant (not at origin) satisfying the given inequality, then\\
\medskip

The Jacobian Matrix of the system at this point is 
$$
J=\left[ 
\begin{array}{cc} 
 -B-2E\delta-C\delta' & A-2D\delta'-C\delta  \\
 B'-2E'\delta-C'\delta' & -A'-2D'\delta'-C'\delta 
\end{array} 
\right]
$$
The state being degenerate implies that the columns of this matrix are linearly dependent, so assume there exists $\mu$ such that: Column 1 + Column 2 $\times \mu\frac{\delta'}{\delta}$ $=0$. In terms of expressions, this gives
\BEA
0= -B\delta-2E\delta^2-C\delta'\delta+\mu(A\delta'-2D\delta'^2-C\delta\delta') \nonumber \\
0=B'\delta-C'\delta\delta'-2E'\delta^2+\mu(-A'\delta'-2D'\delta'^2-C'\delta\delta') 
\EEA
Using equation (\ref{eq:2-3}) and replacing terms, these equations can be written as
\BEA\label{eq:2-4}
0=(\mu-1)B\delta+(\mu-2)E\delta^2-\mu D\delta'^2-C\delta\delta' \nonumber \\
0=(1-\mu)A'\delta'+(1-2\mu)D'\delta'^2-\mu C'\delta\delta'-E'\delta^2 
\EEA
In the second equation of (\ref{eq:2-4}), if $\mu>1$ all terms are negative and equality is not possible. Thus $\mu<1$. Also from the first equation of (\ref{eq:2-4}), $0<\mu<1$ is not possible. Thus $\mu<0$.\\
With a negative $\mu$, there is only one term in each equation of (\ref{eq:2-4}) different in sign from the rest. This results in the inequalities
\BEQ\label{eq:2-5}
|\mu|D\delta'^2>C\delta\delta' \qquad E'\delta^2>|\mu|C'\delta'\delta
\EEQ

Looking at equations (\ref{eq:2-2}), if we take them of the form
\BEA
[x_1]=b_{x_1}\delta+c_{x_1}\delta' \qquad [x_3]=b_{x_3}\delta+c_{x_3}\delta' \qquad [y_1]=b_{y_1}\delta+c_{y_1}\delta' \qquad[y_3]=b_{y_3}\delta+c_{y_3}\delta'  \qquad
\EEA
\BEQ
D=\nu_1^-c_{x_1}c_{x_3}, \quad C=\nu_1^-(c_{x_1}b_{x_3}+b_{x_1}c_{x_3}), \quad E'=\nu_2^-b_{y_1}b_{y_3}, \quad C'=\nu_2^-(c_{y_1}b_{y_3}+b_{y_1}c_{y_3}) \quad
\EEQ
with all the constants positive. (\ref{eq:2-5}) results in
\BEQ
|\mu|\nu_1^-c_{x_1}c_{x_3}\delta'>\nu_1^-c_{x_1}b_{x_3}\delta \qquad \nu_2^-b_{y_1}b_{y_3}\delta>|\mu|\nu_2^-b_{y_1}c_{y_3}\delta' \nonumber
\EEQ
\BEQ
|\mu|c_{x_3}\delta'>b_{x_3}\delta \qquad b_{y_3}\delta>|\mu|c_{y_3}\delta'
\EEQ
Looking back at (\ref{eq:2-2}), $b_{x_3}>c_{x_3}$ and $b_{y_3}<c_{y_3}$. This results in contradicting inequalities with $\delta,\delta'$ and $\mu$. Such a $\mu$ cannot exist.  

\Bigskip

\begin{Lemma}
    For degradation of the form $a'=\{\alpha a_{x_1},\alpha a_{x_2},\alpha a_{x_3},\alpha a_{y_1},\alpha a_{y_2},\alpha a_{y_3}\}$, the stationary state at the origin has Jacobian determinant 0 for only one value of $\alpha\in\mathbb{R}^+$ 
\end{Lemma} 
\noindent {\bf Proof.} Similar to the type IV case, the Jacobian matrix is of the form
$$
J=\left[ 
\begin{array}{cc} 
 -B & A  \\
 B' & -A' 
\end{array} 
\right]
$$
Assume fixed reaction constants ($\{$\textbf{$k$},\textbf{$\nu$}$\}$). $a=\{a_{x_1},a_{x_2},a_{x_3},a_{y_1},a_{y_2},a_{y_3}\}$, and let us consider the same system with degradation rates of the form ($a'=\{\alpha a_{x_1},\alpha a_{x_2},\alpha a_{x_3},\alpha a_{y_1},\alpha a_{y_2},\alpha a_{y_3}\}$) where $\alpha>0$.\\
And using equations (\ref{eq:2-2}), the Jacobian determinant can be written as 
\begin{equation}
    Det\: J= F(\alpha,k,\nu,a)-f(k,\nu)f'(k,\nu)
\end{equation}
where $F(\alpha,k,\nu,a)$ is, in fact, a degree 5 polynomial in $\alpha$ with all positive coefficients which is a strictly increasing function of $\alpha$. For fixed rates and varying $\alpha>0$, $Det\: J=0$ for only one value of $\alpha$.\\

\medskip

Lemmas 8.6-8 imply that the Type II$_2$ cycle also satisfies the assumptions of Theorem \ref{bigboy} if the zero stationary state is non-degenerate for vanishing degradation rates.

\subsubsection{Some degradations zero and some non-zero for Type II and Type IV}\label{app:some}
The above lemmas still hold if for the species in each equation of (\ref{eq:2-22}) and of (\ref{eq:4-11}), at least one has non-zero degradation. For Type IV, this corresponds to at least one of the species from $\{x',u,v\}$ and one from $\{x,w\}$ having non-zero degradation. For Type II the corresponding pair of sets are $\{x_3,y_1,y_2\}$ and $\{y_3,x_1,x_2\}$. \\

\medskip

For the case where there is no degradation for all the species for one set of the pair, (\ref{eq:2-2}) and (\ref{eq:4-00}) become invalid. But taking a look at \textbf{Lemma \ref{lemma:4.1}}, we note that these sets correspond to the linkage classes of each of these types without degradation. Thus if only one of these two sets of each type has non-zero degradation for some species but there is no degradation for any species of the other set, the deficiency is still 0 and we only have a unique stationary state. This encompasses all possible degradation sets.

\subsubsection{Type V}
\begin{center}
\begin{tikzpicture}
\draw(0,0) circle(2);  \draw(0,0) node {${\cal C}$};
\draw(-4*0.707,0) node {${\cal C}'$};
  \draw[<-,ultra thin](1.5,0) arc(0:30:1.5 and 1.5);
\draw(-4*0.707+1.414 ,1.414) arc(45:315:2 and 2);
\draw[<-,ultra thin] (-4*0.707-1.5,0) arc(180:150: 1.5 and 1.5);

\draw[ultra thick](-2*0.707,2*0.707) arc(135:113:2 and 2);
\draw[ultra thick](-2*0.707,2*0.707) arc(45:67:2 and 2);
\draw[ultra thick](-1.2,1.2) node {$\mathbf v$};
\draw(-2*0.707,2*0.707) node {\textbullet};
\draw(-1.3,-1.2) node {$\mathbf u$};
\draw[ultra thick](-0.8,1.6) node {$\mathbf x$};
\draw[ultra thick](-2.2,1.6) node {$\mathbf x'$};

\draw[ultra thick](-2*0.5,-2*0.866) arc(-120:-135:2 and 2);
\draw[ultra thick](-4*0.707 + 2*0.5,-2*0.866) arc(300:315:2 and 2);

\draw(-0.8,-1.5) node {$\mathbf w$};
\draw(-2*0.5,-2*0.866) node {\textbullet};
\draw[ultra thick](-2*0.5,-2*0.866) arc(150:270:1.2 and 0.68);
\draw[->,ultra thick](0,-2.75) arc(270:360:2.75 and 2.75);
\draw[ultra thick](2.75,0) arc(360:497:2.75 and 2.75);

\draw(-4*0.707 + 0.8,-1.5) node {$\mathbf w'$};
\draw(-4*0.707 + 2*0.5,-2*0.866) node {\textbullet};
\draw[ultra thick](-4*0.707 + 2*0.5,-2*0.866) arc(30:-90:1.2 and 0.68);
\draw[->,ultra thick](-4*0.707 + 0,-2.75) arc(-90:-180:2.75 and 2.75);
\draw[ultra thick](-4*0.707 - 2.75,0) arc(-180:180-497:2.75 and 2.75);

\draw(2.8,-1.5) node {\bf \tiny  back};
\draw(-4*0.707-2.8,-1.5) node {\bf \tiny  back};

\draw(4,0) node {Type V};
\end{tikzpicture}
\end{center}
$\mathcal{X}=\{x,x',u,v,w,w'\}$\\
This time we introduce three variables, $\delta$, $\delta'$ and $\delta''$. 
\begin{equation}\label{eq:5-0}
    \delta=\nu_{1}^+[v]-\nu_{1}^-[x][x'],  \quad \delta'=\nu_{2}^+[w]-\nu_{2}^-[u][x'],  \quad \delta''=\nu_{3}^+[w']-\nu_{3}^-[u][x]
\end{equation}
Removing non-fork species, the set of equations are
\BEA\label{eq:5-1}
-(k_x^++a_x)[x]+\delta+\delta''+k_x^-[w]=0\\ \nonumber
-(k_{x'}^++a_{x'})[x']+\delta+\delta'+k_{x'}^-[w']=0\\ \nonumber
-(k_u^++a_u)[u]+\delta'+\delta''+k_u^-[v]=0\\ \nonumber
-(k_u^-+a_v)[v]-\delta+k_u^+[u]=0\\ \nonumber
-(k_x^-+a_w)[w]-\delta'+k_x^+[x]=0\\ \nonumber
-(k_{x'}^-+a_{w'})[w']-\delta''+k_{x'}^+[x']=0
\EEA

Further, from (\ref{eq:5-1}) we also get that
\begin{equation}\label{eq:2-22}
\delta'+\delta''-\delta= a_{u}[u]+a_{v}[v]  \qquad \delta+\delta''-\delta'= a_{w}[w]+a_{x}[x] \qquad \delta+\delta'-\delta''= a_{w'}[w']+a_{x'}[x']    
\end{equation}
So we are restricted to the positive quadrant in the $\delta,\delta',\delta''$ space.\\

\Medskip

\begin{Lemma}
    For type V core with rate constants and degradation values as in (\ref{eq:5-1}), there exists a continuous function $m_b(a,k)$, such that any positive stationary state satisfies $m_b(a,k)\geq\max\{[z],z\in{\cal X}\}$
\end{Lemma}
\noindent {\bf Proof.} The inequalities after putting $\delta,\delta',\delta''\geq0$ in (\ref{eq:5-1}) are of the form
\BEA
\frac{k_{x}^++a_{x}}{k_{x}^-}[x]>[w] \qquad \frac{k_{u}^++a_{u}}{k_{u}^-}[u]>[v]  \qquad \frac{k_{x'}^++a_{x'}}{k_{x'}^-}[x']>[w'] 
\EEA
Further from \ref{eq:5-0} we get 
\BEA
   &&[v]>\frac{\nu_1^-}{\nu_1^+}[x][x'] \implies \frac{k_{x}^++a_{x}}{k_{x}^-}\frac{k_{u}^++a_{u}}{k_{u}^-}[u]>\frac{\nu_1^-}{\nu_1^+}[w][x'] \nonumber \\
  && [w]>\frac{\nu_2^-}{\nu_2^+}[u][x']
\EEA
Which on composing gives an upper bound for $[x']^2$ with degradation terms only in the numerator (with a positive coefficient), and similarly for $[x]$ and $[u]$ which bound everything.

\Bigskip

Concentrations in terms of these variables are
\BEA\label{eq:5-2}
&&[x]=\frac{(a_{w}+k_{x}^-)\delta+(a_{w}+k_{x}^-)\delta''-k_{x}^-\delta'}{a_{w}a_{x}+a_{x}k_{x}^-+a_{w}k_{x}^+} \qquad [w]=\frac{k_{x}^+\delta+k_{x}^+\delta''-(k_{x}^++a_{x})\delta'}{a_{w}a_{x}+a_{x}k_{x}^-+a_{w}k_{x}^+} \nonumber \\
&&[x']=\frac{(a_{w'}+k_{x'}^-)\delta+(a_{w'}+k_{x'}^-)\delta'-k_{x'}^-\delta''}{a_{w'}a_{x'}+a_{x'}k_{x'}^-+a_{w'}k_{x'}^+} \qquad [w']=\frac{k_{x'}^+\delta+k_{x'}^+\delta'-(k_{x'}^++a_{x'})\delta''}{a_{w'}a_{x'}+a_{x'}k_{x'}^-+a_{w'}k_{x'}^+} \nonumber \\
&&[u]=\frac{(a_{v}+k_{u}^-)\delta'+(a_{v}+k_{u}^-)\delta''-k_{u}^-\delta}{a_{v}a_{u}+a_{u}k_{u}^-+a_{v}k_{u}^+} \qquad [v]=\frac{k_{u}^+\delta'+k_{u}^+\delta''-(k_{u}^++a_{v})\delta}{a_{v}a_{u}+a_{u}k_{u}^-+a_{v}k_{u}^+} 
\EEA
The rate constants are all positive, and thus \ref{eq:5-0} gives rise to three equations 
\BEA\label{eq:5-3}
0=-A\delta+B\delta'+C\delta''-D\delta^2+E\delta'^2+F\delta''^2-G\delta\delta'-H\delta\delta''-I\delta'\delta'' \nonumber \\
0=A'\delta-B'\delta'+C'\delta''+D'\delta^2-E'\delta'^2+F'\delta''^2-G'\delta\delta'-H'\delta\delta''-I'\delta'\delta'' \nonumber \\
0=A''\delta+B''\delta'-C''\delta''+D''\delta^2+E''\delta'^2-F''\delta''^2-G''\delta\delta'-H''\delta\delta''-I''\delta'\delta'' 
\EEA
\begin{Lemma}
    There is no degenerate stationary state in the positive quadrant for (\ref{eq:5-3})
\end{Lemma}
\noindent {\bf Proof.} Assume that (\ref{eq:5-3}) has a degenerate stationary state in the positive quadrant (not at origin) satisfying the given inequality, then
\medskip

The Jacobian Matrix of the system at this point is 
$$
J=\left[ 
\begin{array}{ccc} 
-A-2D\delta-G\delta'-H\delta'' & B+2E\delta'-G\delta-I\delta'' & C+2F\delta''-H\delta-I\delta' \\
A'+2D'\delta-G'\delta'-H'\delta'' & -B'-2E'\delta'-G'\delta-I'\delta'' & C'+2F'\delta''-H'\delta-I'\delta' \\
A''+2D''\delta-G''\delta'-H''\delta'' & B''+2E''\delta'-G''\delta-I''\delta'' & -C''-2F''\delta''-H''\delta-I''\delta' 
\end{array} 
\right]
$$
Without loss of generality (utilising the system symmetry), assume $\delta\leq\delta',\delta''$\\
For it to be singular, there must exist real numbers of the form $(\mu+1)\frac{\delta'}{\delta}$ and $(\omega+1)\frac{\delta''}{\delta}$ which, on multiplying with Column 2 and 3 respectively (and adding the columns) give 0 to show the linear dependence of the columns of the matrix. Simplifications of these equations yield
\BEA\label{eq:5-4}
\mu(B\delta'+2E\delta'^2-G\delta\delta'-I\delta''\delta')+\omega(C\delta''+2F\delta''^2-H\delta\delta''-I\delta'\delta'')=B\delta'+C\delta''-A\delta>0\nonumber\\
\mu(-B'\delta'-2E'\delta'^2-G'\delta\delta'-I'\delta''\delta')+\omega(C'\delta''+2F'\delta''^2-H'\delta\delta''-I'\delta'\delta'')=-B'\delta'+C'\delta''+A'\delta>0\nonumber\\
\mu(B''\delta'+2E''\delta'^2-G''\delta\delta'-I''\delta''\delta')+\omega(-C''\delta''-2F''\delta''^2-H''\delta\delta''-I''\delta'\delta'')=B''\delta'-C''\delta''+A''\delta>0\nonumber\\
\EEA

From \ref{eq:5-2}, we also get the relations $B'>C'=A',E'>F',D'$. Take the second equation of \ref{eq:5-3}, we get the form
\begin{equation}
    C'\delta''+2F'\delta''^2-H'\delta\delta''-I'\delta'\delta''=(B'\delta'-A'\delta)+(E'\delta'^2-D'\delta^2)+G'\delta\delta'+F'\delta''^2
\end{equation}
This expression is thus $>0$ and $<B'\delta'+E'\delta'^2+G'\delta\delta'<B'\delta'+2E'\delta'^2+G'\delta\delta'+I'\delta''\delta'$\\
Thus the second equation of \ref{eq:5-4} gives $\omega>\mu$. A similar procedure for the third equation gives $\omega<\mu$, thus a contradiction.

\begin{Lemma}
    For degradation of the form $a'=\{\alpha a_{x},\alpha a_{x'},\alpha a_{u},\alpha a_{v},\alpha a_{w},\alpha a_{w'}\}$, the stationary state at the origin has Jacobian determinant 0 for only one value of $\alpha\in\mathbb{R}^+$ 
\end{Lemma} 
\noindent {\bf Proof.} The Jacobian matrix is of the form
$$
J=\left[ 
\begin{array}{ccc}
-A & B & C\\
A' & -B' & C'\\
A'' & B'' & -C''
\end{array} 
\right]
$$
Assume fixed reaction constants ($\{$\textbf{$k$},\textbf{$\nu$}$\}$). $a=\{a_{x},a_{x'},a_{u},a_{v},a_{w},a_{w'}\}$, and let us consider the same system with degradation rates of the form ($a'=\{\alpha a_{x},\alpha a_{x'},\alpha a_{u},\alpha a_{v},\alpha a_{w},\alpha a_{w'}\}$) where $\alpha>0$.\\
From \ref{eq:5-2}, we get $C=B=k_1$(constant,$=k_u^+$ in this case) and $A=k_1+f_1(\alpha)$ where $f_1$ is an increasing monotonic polynomial in $\alpha$, and similarly for other rows (with functions $f_2$ and $f_3$). The determinant takes the form\\
\BEA
   && Det\: J= -(k_1+f_1)(f_2f_3+f_2k_3+f_3k_2)+k_1(2k_2k_3+k_2f_3)+k_1(2k_2k_3+k_3f_2) \nonumber \\
   &&= F(k_1,k_2,k_3)-G(\alpha,k_1,k_2,k_3)
\EEA
Where $F$ is fixed for fixed kinetics and as a sum of products of monotonic polynomials, $G$ is also an increasing monotonic polynomial in $\alpha$. Thus this value can be zero only for one value of $\alpha$\\

\medskip

Lemmas 8.9-11 imply that the Type V cycle also satisfies the assumptions of Theorem \ref{bigboy} if the zero stationary state is non-degenerate for vanishing degradation rates.

\section*{References}
\bibliographystyle{abbrv}
\bibliography{mybib}

\end{document}